\documentclass[fleqn,usenatbib]{mnras}

\usepackage{newtxtext,newtxmath}

\usepackage[T1]{fontenc}

\DeclareRobustCommand{\VAN}[3]{#2}
\let\VANthebibliography\thebibliography
\def\thebibliography{\DeclareRobustCommand{\VAN}[3]{##3}\VANthebibliography}

\usepackage{graphicx}	
\usepackage{amsmath}	
\usepackage{multirow}
\usepackage{multicol}

\newcommand{\aperp}{\alpha_\perp}
\newcommand{\apar}{\alpha_\parallel}
\newcommand{\aperpm}{[\alpha_\perp-1]\%}
\newcommand{\aparm}{[\alpha_\parallel-1]\%}

\newcommand{\hompc}{$h\,\mathrm{Mpc}^{-1}$}
\newcommand{\mpcoh}{$h^{-1}\,\mathrm{Mpc}$}
\newcommand{\Plin}{P_{\rm lin}}

\newcommand{\bq}{\boldsymbol q}

\newcommand{\hq}{\hat{q}}
\newcommand{\hk}{\hat{k}}

\newcommand{\bx}{\boldsymbol x}

\newcommand{\bk}{\textbf{k}}
\newcommand{\bp}{\textbf{p}}

\newcommand{\bs}{\textbf{s}}
\newcommand{\hn}{\hat{n}}

\newcommand{\Mpc}{h^{-1}\ \text{Mpc}}
\newcommand{\Gpc}{h^{-1}\ \text{Gpc}}
\newcommand{\kMpc}{h\ \text{Mpc}^{-1}}

\newcommand{\edit}[1]{\textcolor{black}{{{#1}}}}
\newcommand{\editnew}[1]{\textcolor{black}{{{#1}}}}

\usepackage[dvipsnames]{xcolor}
\definecolor{aac}{rgb}{0.858, 0.188, 0.478}

\def\pvmhid#1{}

\title[BAO Theory and Modelling Systematics for DESI 2024]{Baryon Acoustic Oscillation Theory and Modelling Systematics for the DESI 2024 results}

\author[Chen et. al.]{Shi-Fan Chen$^{1}$\thanks{E-mail: sfschen@ias.edu}\footnotemark[3],
Cullan Howlett$^{2}$\thanks{E-mail: c.howlett@uq.edu.au}\thanks{Both authors contributed equally to this work.},
Martin White,$^{3,4}$
Patrick McDonald,$^{4}$
Ashley J. Ross,$^{5}$ \newauthor
Hee-Jong Seo,$^{6}$
Nikhil ~Padmanabhan,$^{7}$
J.~Aguilar,$^{4}$
S.~Ahlen,$^{8}$
S.~Alam,$^{9}$
O.~Alves,$^{10}$
U.~Andrade,$^{11,10}$ \newauthor
R.~Blum,$^{12}$
D.~Brooks,$^{13}$
X.~Chen,$^{7}$
S.~Cole,$^{14}$
K.~Dawson,$^{15}$
A.~de la Macorra,$^{16}$
Arjun~Dey,$^{12}$
Z.~Ding,$^{17}$ \newauthor
P.~Doel,$^{13}$
S.~Ferraro,$^{4,18}$
A.~Font-Ribera,$^{13,19}$
D.~Forero-S\'{a}nchez,$^{20}$
J.~E.~Forero-Romero,$^{21,22}$ \newauthor
C.~Garcia-Quintero,$^{23,24}$
E.~Gazta\~naga,$^{25,26,27}$
S.~Gontcho A Gontcho,$^{4}$
M.~M.~S~Hanif,$^{10}$
K.~Honscheid,$^{5,28,29}$ \newauthor
T.~Kisner,$^{4}$
A.~Kremin,$^{4}$
A.~Lambert,$^{4}$
M.~Landriau,$^{4}$
M.~E.~Levi,$^{4}$
M.~Manera,$^{30,19}$
A.~Meisner,$^{12}$ \newauthor
J.~Mena-Fern\'andez,$^{31}$
R.~Miquel,$^{32,19}$
A.~Munoz-Gutierrez,$^{16}$
E.~Paillas,$^{33,34}$
N.~Palanque-Delabrouille,$^{35,4}$ \newauthor
W.~J.~Percival,$^{33,36,34}$
A.~P\'{e}rez-Fern\'{a}ndez,$^{16,37}$
F.~Prada,$^{38}$
M.~Rashkovetskyi,$^{23}$
M.~Rezaie,$^{39}$ \newauthor
A.~Rosado-Marin,$^{6}$
G.~Rossi,$^{40}$
R.~Ruggeri,$^{41,2}$
E.~Sanchez,$^{42}$
D.~Schlegel,$^{4}$
J.~Silber,$^{4}$
G.~Tarl\'{e},$^{10}$ \newauthor
M.~Vargas-Maga\~na,$^{16}$
B.~A.~Weaver,$^{12}$
J.~Yu,$^{20}$
S.~Yuan,$^{43}$
R.~Zhou,$^{4}$
and Z.~Zhou$^{44}$ \vspace{0.4 cm}
\\
Affiliations listed in Appendix~\ref{app:affiliations}.
}

\date{Accepted 2024 September 3. Received 2024 September 3; in original form 2024 February 22}

\pubyear{2024}

\begin{document}
\label{firstpage}
\pagerange{\pageref{firstpage}--\pageref{lastpage}}
\maketitle

\begin{abstract}
This paper provides a comprehensive overview of how fitting of Baryon Acoustic Oscillations (BAO) is carried out within the upcoming Dark Energy Spectroscopic Instrument's (DESI) 2024 results using its DR1 dataset, and the associated systematic error budget from theory and modelling of the BAO. We derive new results showing how non-linearities in the clustering of galaxies can cause potential biases in measurements of the isotropic ($\alpha_{\mathrm{iso}}$) and anisotropic ($\alpha_{\mathrm{ap}}$) BAO distance scales, and how these can be effectively removed with an appropriate choice of reconstruction algorithm. We then demonstrate how theory leads to a clear choice for how to model the BAO and develop, implement and validate a new model for the remaining smooth-broadband (i.e., without BAO) component of the galaxy clustering. Finally, we explore the impact of all remaining modelling choices on the BAO constraints from DESI using a suite of high-precision simulations, arriving at a set of best-practices for DESI BAO fits, and an associated theory and modelling systematic error. Overall, our results demonstrate the remarkable robustness of the BAO to all our modelling choices and motivate a combined theory and modelling systematic error contribution to the post-reconstruction DESI BAO measurements of no more than $0.1\%$ ($0.2\%$) for its isotropic (anisotropic) distance measurements. We expect the theory and best-practices laid out to here to be applicable to other BAO experiments in the era of DESI and beyond.
\end{abstract}




\section{Introduction}
\label{sec:intro}

Constraining the expansion history of the Universe is important both for improving our empirical cosmographical knowledge and for the constraints it places on the constituents of the Universe and our theories of gravity.  Within the standard Friedmann–Lema\^{i}tre–Robertson–Walker (FLRW; \citealt{Friedmann1922, Lemaitre1931, Robertson1935, Walker1937})  model, the scale factor -- $a(t)$ -- is the only degree of freedom at the background level, determining distance relations over cosmic history. Constraining its behaviour has been a major goal of cosmology for close to a century.  One of the most robust and accurate ways of determining $a(t)$ is through measurements of baryon acoustic oscillations (BAO) in the distribution of galaxies and gas.  Oscillations in the baryon-photon fluid prior to decoupling leave an imprint in both the cosmic microwave background and the matter density with a characteristic length scale, related to the distance $r_d$ a sound wave can travel prior to the baryon-drag epoch, just after recombination.  Measurements of this scale at low redshift in large-scale structure provide a ``standard ruler'' for constraining the expansion history of the Universe, both by measuring the angular diameter distance ($D_{A}(z)$) and the Hubble parameter ($H(z)$) to a specific redshift. This ruler can be made ``absolute'' by connecting the BAO scale back to early universe physics, e.g. through constraints from the cosmic microwave background (CMB; \citealt{Peebles70,Sunyaev70,Dodelson20,Huterer23}) or big-bang nucleosynthesis (BBN; \citealt{Cooke2018}).

The BAO method has become a workhorse of modern cosmology and provides some of our tightest current constraints on the cosmological expansion history \citep{Cole05,Eisenstein05,Beutler11,Blake11,Alam21}. Relying as it does on a feature at large spatial separations that is little affected by non-linear evolution or the complex astrophysics of galaxy formation, it is robust and theoretically well understood.  For this reason, it is generally regarded as a low-systematics method for constraining cosmology. Indeed, significant theoretical effort over the past two decades has led to a fairly mature understanding of the effect of the small shifts and modulations to the BAO signal incurred by large-scale structure physics which we recap and build on in this work (\S\ref{sec:bao_overview}). 

The purpose of this paper is to use our theoretical understanding to generate a theory and modelling systematics budget, and suggest fitting best practices, for BAO measurements in the Dark Energy Spectroscopic Instrument (DESI). DESI is a Stage-IV spectroscopic instrument \citep{DESI2016a,Instrument_Overview_KP} that will deliver galaxy BAO measurements with unprecedented precision up to $z \approx 2$ from galaxy clustering alone, with even greater reach when Lyman-$\alpha$ data are included at higher redshift. At the end of its five-year run DESI is expected to yield measurements on the (isotropic) BAO scale at a cumulative $0.2\%$ precision \citep{DESI2016a,DESISV}.\edit{\footnote{Where the cumulative or "aggregate" BAO precision is the statistical uncertainty we would expect in the BAO scale if all DESI galaxies were put into the same redshift bin and used to produce a single BAO measurement. It is equivalent to an inverse sum of the BAO precision across all our independent measurement bins.}} Already, galaxies from the early data release (EDR) in the luminous red galaxy (LRG) sample and bright galaxy sample (BGS), representing a small fraction of the total DESI data, yield BAO detections at the few percent level \citep{DESIEDR,BAO_EDR}.

This paper is part of a set focused on the analysis of clustering in DESI Data Release 1 (DR1; \citealt{DESI2024.I.DR1,DESI2024.II.KP3}), and in support of the key paper presenting the main galaxy BAO measurements for that data set \citep{KP4}. The DR1 data further contains BAO information in the Lyman-$\alpha$ forest \citep{DESI2024.IV.KP6}, and the cosmological implications of the joint galaxy and Lyman-$\alpha$ BAO measurements are discussed in \citet{DESI2024.VI.KP7A}. Information beyond the BAO --- particularly the fullshape information in the galaxy 2-point function---are studied in \citet{DESI2024.V.KP5,DESI2024.VII.KP7b,DESI2024.VIII.KP7C}. While the aggregate isotropic BAO scale precision from analysis of DR1 is closer to $0.5\%$, our conclusions and recommendations are somewhat general and aimed towards the full DESI sample. Our understanding of galaxy clustering in the standard model of cosmology ($\Lambda$CDM) gives us a baseline of \textit{known} theoretical systematics that need to be considered in order to make robust measurements of the BAO scale, even before \textit{unknown} modifications due to nonstandard physics are considered.

The structure of the paper is as follows: In Section \ref{sec:bao_overview} we begin with a broad overview of the BAO signal as observed in large-scale structure simulations. In Sections \ref{sec:nonlinear} and \ref{sec:recon} we review the theory of the BAO and BAO reconstruction, computing error estimates due to theory systematics, filling in extant gaps in the literature where necessary, and motivating our baseline form for fitting the BAO signal. In Section \ref{sec:broadband} we introduce a cubic-spline based method for marginalizing over the non-BAO broadband of galaxy clustering and harmonize broadband models in Fourier and configuration space. Section~\ref{sec:numerics} introduces the N-body simulations and inference pipeline \textsc{Barry} implementing the suggested analysis choices in this paper, and the systematic effects of various numerical choices are examined in Section~\ref{sec:results}. We conclude in Section~\ref{sec:conclusions}.

\section{The Observed Baryon Acoustic Oscillation Signal: An Overview}
\label{sec:bao_overview}

Baryon acoustic oscillations (BAO) manifest as an oscillatory, or wiggly, imprint in the linear matter power spectrum $\Plin(k)$.  The physics is very well understood and can be computed to high accuracy by numerical evolution of the coupled Einstein, fluid and Boltzmann equations as is done e.g. by \texttt{CAMB} \citep{CAMB,Howlett2012} and \texttt{CLASS} \citep{CLASS}.  To recap the basic physics, prior to recombination, when the densities are high, baryon-photon scattering is rapid compared with the travel time across a wavelength. This allows the behaviour of the baryon and photons to be described as a single fluid, whose continuity and Euler equations reduce to a driven harmonic oscillator with slowly varying natural frequency.  During this `tight-coupling' phase the baryon perturbations hence also undergo harmonic motion with very slowly decaying amplitude \citep{Peebles70,Sunyaev70}.  These `baryon acoustic oscillations' become `frozen-in' at decoupling when the photons release their hold on the baryons, setting the oscillation frequency $r_d$, and the matter perturbations begin to grow as $\propto a$.  The combination of oscillatory densities and velocities which project onto the growing mode gives the final post-recombination spectrum.\footnote{At low $k$ this is a mixture of density and velocity modes but at higher $k$ it is primarily the velocity mode which is why the higher harmonics of the BAO are out of phase with the peaks in the CMB, the latter of which arise primarily from \textit{photon} density perturbations.}  This then grows in a scale-independent fashion until the present day, up to small corrections due to massive neutrinos which start off relativistic but transition to non-relativistic at some point post-recombination. The baryon acoustic oscillations themselves are superposed upon the smooth power spectrum, whose shape is primarily determined by the epoch of matter-radiation equality.  The amplitude of the oscillations depends upon the driving force (gravitational potentials) and the baryon-to-photon ratio.  Larger oscillations come from higher baryon density $\omega_b$ and/or smaller total-matter density $\omega_m$. Additionally, light species (e.g.\ neutrinos) which propagate ahead of the sound waves also exert a gravitational influence on the BAO which can shift their phase \citep{Bashinsky04}, and any isocurvature modes in the initial conditions can also alter the shape of the observed signal \citep{Zunckel11}; the impact of some these effects on DESI BAO measurements is discussed in \citet{cosmology_dependence}, but we note that they are also more directly constrained by the CMB itself.

Given a linear matter power spectrum $\Plin$, the redshift-space \textit{galaxy} power spectrum in linear theory and in the plane-parallel or distant-observer approximation, is given by \citep{Kaiser87} 
\begin{equation}
    P_{gg}(k,\mu) = (b_1 + f\mu^2)^2 \Plin(k).
\end{equation}
where $f=d\ln D/d\ln a\approx\Omega_m^{0.55}$ is the growth rate and $b_1$ is the linear (scale-independent, deterministic) bias.  Since the linear spectrum is otherwise smooth, and neither the nonlinearities of structure/galaxy formation nor survey systematics are expected to produce features with the same frequency, the aim of BAO measurements in galaxy surveys is to independently extract the oscillatory part of this linear signal from the overall clustering of galaxies and use it as a standard ruler. The BAO feature is a localized peak in the correlation function at a scale determined by the sound horizon at the baryon-drag epoch ($r_d$), or equivalently as a series of oscillations in the power spectrum with Fourier-frequency $2\pi/r_d$.  At a given redshift, $z$, the line-of-sight and transverse determinations of this ruler fix $H(z)\,r_d$ and $r_d/D_A(z)$, respectively. Modern spectroscopic surveys typically convert the measured redshifts and sky positions of galaxies into comoving coordinates using a fiducial cosmology (usually flat $\Lambda$CDM with a fiducial $\Omega_{m, \rm fid}\approx 0.3$), in which the multipoles of the power spectrum $P(k,\mu) = \sum_\ell P_\ell(k) \mathcal{L}_\ell(\mu)$ and correlation functions $\xi(r,\mu)$ can be computed in physical units. These 2-point functions are then fitted against a BAO ``template'' that is scaled in the line-of-sight and transverse directions \citep{Padmanabhan08} to extract the so-called dilation parameters
\begin{equation}
    \qquad \qquad \alpha_{||} = \frac{H^{\mathrm{fid}}(z)r^{\mathrm{tem}}_{s}}{H(z)r_{s}}, \qquad \qquad \alpha_{\perp} = \frac{D_{A}(z)r^{\mathrm{tem}}_{s}}{D^{\mathrm{fid}}_{A}(z)r_{s}},
    \label{eqn:alpha_defs}
\end{equation}
where `fid' and `tem' refer to quantities in the fiducial and template cosmologies respectively. \edit{ It is also common to see the BAO dilation parameters recast into a purely isotropic term $\alpha_{\mathrm{iso}}\equiv\alpha_{||}^{1/3}\alpha_{\perp}^{2/3}$ and an anisotropic term $\alpha_{\mathrm{ap}} \equiv (1+\epsilon)^{3} \equiv \alpha_{||}/\alpha_{\perp}$. We will use these all interchangeably in this work. }

Let us describe the observation and extraction of the BAO in galaxy surveys outlined above in more detail, beginning with the dynamical model. In order to do so it is helpful to split the power spectrum into a BAO template consisting of the wiggles alone $P_w$ and a ``no-wiggle'' portion $P_{nw}$ containing the broadband power which we shall discuss further below (\S\ref{ssec:pnw})
\begin{equation}
    \Plin(k) = P_{nw}(k) + f_b P_{w}(k).
    \label{eqn:Plin_split}
\end{equation}
In the definition of the wiggle/no-wiggle split above we have \edit{included in the wiggly component a pre-factor of the baryon fraction $f_b \approx 15\%$. This small parameter controls the size of the BAO signal relative to the overall clustering signal, and we will use it as a convenient power-counting parameter to denote the suppression of contributions involving higher powers of the wiggle component. However, since $f_b$ is degenerate with the amplitude of $P_w$ we will generally include it in $P_{w}$ for convenience in the rest of the paper, except when it is useful for book-keeping like here.} The measurement of the linear BAO signal can then be thought of as the measurement of the linear dependence of the observed galaxy power spectrum, $P_{gg}$, on $f_b$. \footnote{There is some ambiguity in this `split' as, in addition to generating the acoustic oscillations, the fact that the baryon perturbations oscillate rather than grow between matter-radiation equality and photon-baryon decoupling means the potentials evolve as they would in a universe with only dark matter $\Omega_{m} = \Omega_{c}$ (i.e., the baryons don't contribute to the gravitational collapse).  This alters the shape of $\Plin$ just to the right of the matter-radiation equality peak leading to a faster decrease than would otherwise occur. Ultimately, this means that $P_{nw}(k)$ in a Universe with baryons is not the same as the power spectrum of pure cold dark matter.} Schematically, for the observed 2-point function we can write
\begin{equation}
    P_{gg}(k,\mu) = (P_{gg})_{f_b = 0} + f_b \left( \frac{d P_{gg}}{d f_b} \right)_{f_b=0} + \mathcal{O}(f_b^2).
    \label{eqn:BAO_model_exp}
\end{equation}
Based on this structure models of the BAO typically take the form
\begin{equation}
    P_{gg}(k,\mu) =  \mathcal{B}(k,\mu)P_{nw}(k) + \mathcal{C}(k,\mu)P_{w}(k) + \mathcal{D}(k,\mu)
    \label{eqn:BAO_model}
\end{equation}
where the term proportional to $\mathcal{C}(k,\mu)$ is the leading BAO signal \edit{$dP_{gg}/d f_b = C(k,\mu) P_w$ we want to measure, with $C(k,\mu) = (b+f\mu^2)^2$, in linear theory}. The rough non-oscillatory broadband shape is contained in the $\mathcal{B}(k,\mu)$ term, given by the Kaiser formula in linear theory. The final term $\mathcal{D}(k,\mu)$ represents any remaining contributions to the observed power spectrum, which are largely due to nonlinear coupling of the smooth linear spectrum and observational systematics. It may also absorb higher order (e.g. $(f_b)^2$) terms in the BAO, though these will tend to be both rather suppressed and oscillatory. The systematics we will discuss in this paper largely have to do with the goodness of this approximation; specifically we will consider the extent to which nonlinear oscillatory terms modify the $\mathcal{C}(k,\mu)$ signal we wish to measure, and how well the remaining terms can be absorbed by $\mathcal{D}(k,\mu)$ under the assumption of smoothness. 

Given one adopts a fiducial cosmology for converting galaxy redshifts to distances, and a template linear spectrum with corresponding $r_d^{\rm tem}$ and $P_w^{\rm tem}$, the assumption is then that any of the scale information carried in the BAO wiggles can be accounted for with a simple scaling $P_w(k) \sim P_w^{\rm tem}(r_d k / r_d^{\rm tem})$ in Equation~\ref{eqn:BAO_model}. Allowing for $H(z)$ and $D_A(z)$ to vary independently away from the fiducial cosmology, such that 
\begin{equation}
    \bk_{\rm obs} = \left( \frac{D_A(z)}{D_A^{\rm fid}(z)} \bk_\perp^{\rm true}, \frac{H^{\rm fid}(z)}{H(z)}  k_\parallel^{\rm true} \right)
    \label{eqn:kalpha}
\end{equation}
leads to the conclusion that the free parameters required to ``match'' the template to the observed BAO are precisely the dilation parameters defined in Equation~\ref{eqn:alpha_defs}, with smooth changes to the broadband and the shape and normalization of the wiggles captured by marginalizing over $\mathcal{B,C,D}$. This borrows philosophically from the Alcock-Paczynski effect \citep{AP1979}, which describes the dilation of scales and angles when analysing galaxy positions assuming a fiducial cosmology that differs from the true underlying cosmology of the Universe.

It is worth noting that not all of the cosmological dependence of the BAO signal (e.g.\ its amplitude, damping or phase shifts) is captured by this rescaling --- indeed, some of these, such as the phase shift,cannot even be captured by other parameters like galaxy bias or broadband modelling in standard BAO fits. In general, the physical effects that change $\Plin$ also change the CMB anisotropies, often more significantly \citep{Eisenstein04}, and tend to be disfavored by existing measurements \citep[e.g.][]{PCP18}. The systematic errors incurred by these additional cosmology-dependent effects are characterized in an accompanying DESI paper \citep{cosmology_dependence}, so we will only note here that the extraction of $\alpha_{\parallel,\perp}$ in BAO fits has been shown to be rather robust to these \citep{VargasMagana2014,Bernal20,Carter2020}, while referring interested readers to that upcoming companion paper for further details in the context of DESI Y1. In addition, the definition of $P_w$ is not unique and several different template extraction methods exist in the literature. Properly constructed, these $P_w$ should only differ by smooth functions of $k$ that are degenerate with the marginalized broadband terms, and we will explore the extent to which this is true in \S\ref{ssec:pnw}.

Our interest in the rest of this paper will be in quantifying the effects of dynamical nonlinearities and modelling choices on measuring $\alpha_{\parallel,\perp}$ so as to form a theoretical error budget for DESI BAO analyses. In doing so we will work primarily in the plane-parallel approximation and validate our models against the DESI Y1 `cubic mocks' based on single-redshift snapshots of N-body simulations in periodic, cubic boxes.  This isolates the `theoretical systematics' from those depending upon observing geometry or observational non-idealities.  However, before moving into these investigations it is important to describe a few additional details necessary to make contact with observations. 

Firstly, galaxy surveys measure clustering not in translationally invariant cubic boxes but within defined angular and redshift windows. In the power spectrum this effect, along with wide-angle contributions to common power spectrum estimators, can be accounted for by convolving theory predictions with suitably computed window matrices \citep{FKP94,Blake2018,Beutler21}. Since these effects should ideally be exactly computable given the galaxy window function we will not consider their systematics here but relegate them to the DESI Y1 catalog papers. Secondly, the measured 2-point function is not sampled from a single point but rather a weighted sum across the redshift range $P = \sum_i w_i P(z_i)$; to leading order this is taken into account by taking the BAO measurement to be at the effective redshift, but in cosmological analyses it is also possible to take into account higher order terms e.g.
\begin{equation}
    \hat{\alpha}_{\parallel,\perp} = \alpha_{\parallel,\perp}(z_{\rm eff}) + \frac12 \left( \frac{d^2 \alpha_{\parallel,\perp}}{dz^2} \right)_{z_{\rm eff}} \sigma_z^2 + ...
\end{equation}
where $\sigma_z^2$ is the mean square width of the redshift bin.  We discuss this further\edit{, and define the effective redshift,} in Appendix \ref{app:zevol}. Finally, in our modeling of the BAO signal we have ignored various large-scale effects including wide-angle effects, unequal time effects and relativistic effects. These typically scale as $(H/k)^2$ in the power spectrum and so are extremely suppressed on the range of scales we fit, since our fiducial setup has $k_{\rm min} = 0.02 \kMpc$. Many of these effects can produce out-of-phase contributions but their effect on the inferred BAO scale is rather limited (typically less than $0.05\%$; Appendix~\ref{app:wide_angle}). As such, we will not further consider them in the rest of this work.

\section{Nonlinear Baryon Acoustic Oscillations in Galaxy Clustering}

\label{sec:nonlinear}

Nonlinear structure formation and galaxy bias significantly modify the observed shape and phase of the linear BAO signal, especially pre-reconstruction \citep{Bharadwaj96,Meiksin99,ESW07,Seo08,Smith08,Crocce08,Matsubara08,Padmanabhan09b,Sherwin12,Carlson13}. Broadly speaking, we can split the effects of these nonlinearities into two categories: (1) the damping, or modulation, of the amplitude of the linear wiggles and (2) the shift in the observed BAO scale through \textit{out-of-phase} contributions to galaxy clustering.

The most important contributions to both of these effects come from the displacements $\Psi$ of galaxies on large scales, which are well-described by perturbation theory. Within Lagrangian perturbation theory (LPT), which models structure formation via the trajectories of galaxies, $\bx(\bq,t) = \bq + \Psi(\bq,t)$ where $\bq$ is the initial Lagrangian position of the galaxy, the galaxy power spectrum in real space can be written as \citep{Matsubara08,Carlson13}
\begin{align}
    P(k) = \int d^3\bq\ & e^{i\bk\cdot\bq - \frac12 k_i k_j A_{ij}(\bq)} \big( 1 + 2 b_1^L i k_i \langle \delta(\bq) \Delta_i \rangle \nonumber \\
    &+ (b_1^{L})^2 \xi_{\rm lin}(\bq) + 2 b_1^L b_2^L i k_i \langle \delta(\bq) \Delta_i \rangle \xi_{\rm lin}(\bq) + ... \big)
    \label{eqn:lpt_rs}
\end{align}
where $\Delta = \Psi(\bq) - \Psi(\textbf{0})$ is the pairwise displacement between two galaxies separated by $\bq$, $A_{ij} = \langle \Delta_i \Delta_j \rangle$ is its variance, and $\xi_{\rm lin}$ is the linear matter correlation function. The numbers $b_n^L$ are the free parameters in the Lagrangian bias expansion of galaxies, related to the canonical Eulerian ones through simple linear transformations e.g. $b_1 = 1 + b_1^L$ \citep{bias_review}. Throughout this work we will operate under the assumption that galaxies are biased tracers of cold dark matter and baryons but not (massive) neutrinos \citep{Castorina15}. This has been shown to be an excellent approximation in simulations---close to the $0.1\%$ level for the galaxy power spectrum in redshift space on the scales we are interested in---and as such we will operate within this regime throughout the text \citep{Bayer22}. We note that the discussion in this section is not meant to constitute an exhaustive treatment of galaxy clustering in perturbation theory, and nor is Equation~\ref{eqn:lpt_rs} meant to be the complete expression for the 1-loop galaxy power spectrum; the intent is rather to pick out the pieces of physics most relevant to the BAO signal in order to estimate their systematic effect on BAO measurements.\footnote{See Appendix~\ref{app:rsd_bao_shift} and onwards for a more complete accounting for galaxies in redshift space and post-reconstruction.}

\subsection{Real Space}
\label{sec:nlpre}

Let us begin with modifications to the linear BAO signal in real space due to only one power of the baryon fraction, i.e. which involves contributions of order $\mathcal{O}(P_w \times P_{nw}^n)$. Physically we can think of these as distortions to the BAO feature due to gravitational nonlinearities of the smooth component.

We can see the origin of both the `in-phase' BAO damping and `out-of-phase BAO' shift directly in Equation~\ref{eqn:lpt_rs}. In order to do so it is again useful to split the linear correlation function $\xi_{\rm lin}$ into its contributions from the wiggle and no-wiggle components $\xi_{w(nw)}$---the former has a sharp peak at $r_d$. Focusing on the $b_1^2$ piece we can see that the wiggle-component can therefore be approximated as (\citealt{Vlah16}, see also e.g. \citealt{Baldauf15,Senatore15,Blas16} for discussions outside of LPT)
\begin{align*}
    (b_1^L)^2 &\int d^3\bq\ e^{i\bk\cdot\bq - \frac12 k_i k_j A_{ij}(\bq)} \xi_w(\bq) \nonumber \\
    &\approx (b_1^L)^2  e^{-\frac12 k^2 \Sigma^2_{\rm NL}} \int d^3\bq\  e^{i\bk\cdot\bq} \xi_w(\bq) = b_1^2 e^{-\frac12 k^2 \Sigma^2_{\rm NL}} P_w(k)
\end{align*}
where $\Sigma^2_{\rm NL}$ is the angular average of $A_{ij}(\bq)$ for $q = r_d$. At first-order in LPT, i.e.\ in the Zeldovich approximation, this is given by\footnote{The saddle-point approximation here involves an angular integral dependent on the relative orientations of $\hq$ and $\hk$. The approximation we have made here is to only resum the isotropic component, but other resummation choices also exist in the literature, e.g. \citet{Baldauf15,Blas16} have an integrand proportional to $1 - j_0 + 2 j_2$ instead, corresponding to resumming the $\hq = \hk$ component. The differences in the BAO amplitude in 1-loop perturbation theory using these two resummation choices are well below the percent level (see e.g.\ Figs.~3 and 5 of \citealt{Chen20b}) while in the BAO fitting only case their difference should be well within the priors of $\Sigma^2_{\rm NL}$ set in the analysis (see Section~\ref{sec:damping}). For a more detailed exploration of the angular integral in LPT see \citet{bispectrum_in_prep}.}
\begin{equation}
    \Sigma^2_{\rm NL} = \frac23 \int \frac{dk}{2\pi^2} P(k) \left[ 1 - j_0(k r_d) \right] \quad .
    \label{eqn:zel_damping}
\end{equation}
This gives us the damping of the BAO by the bulk displacements of galaxies. Similar reasoning shows that any ``wiggle'' component of the power spectrum is similarly damped \citep{Beutler19b,Vasudevan19,Chen20b}. An important feature of Equation~\ref{eqn:zel_damping} is that the damping comes from modes with wavelengths \textit{smaller} that the BAO, i.e.\ with $k \gtrsim 1/r_d$; physically, this reflects that galaxies separated by $r_d$ are moved coherently by modes larger than the BAO scale, but incoherently by those smaller. In $\Lambda$CDM universes similar to our own the square displacement $\Sigma^2_{\rm NL}$ is dominated by linear displacements close to the $1/r_d$ cutoff, whose effects are sufficiently large that they have to be kept in the exponential, or resummed, instead of being perturbatively expanded. However, we stress that while BAO damping in $\Lambda$CDM universes similar to our own is dominated by contributions from relatively long-wavelength modes as described above, small-scale, nonlinear, displacements can also contribute. We discuss the importance of these effects when fitting BAO, especially considering redshift-space distortions and reconstruction, further below.

The BAO shift effect can be similarly derived by approximating the nonlinear effect of no-wiggle contributions by their values near $q = r_d$ when they multiply well-localized wiggly components such as $\xi_w$. Let us consider the $b_1 b_2$ term in Equation~\ref{eqn:lpt_rs}. In this case, writing to linear order
\begin{equation}
    \langle \delta(\bq) \Delta_i \rangle = -\frac{1}{3} q_i \sigma^2(q) 
\end{equation}
where $\sigma^2(q)$ is the mean-square linear density fluctuation in a sphere of radius $q$,\footnote{\edit{This correlator, $\langle \delta \Delta_i \rangle$, describes the mean inflow of matter towards an overdensity and is more commonly written in the LPT literature as \citep{Carlson13}
\begin{equation}
U_i(\bq) = -\hat{q}_i \int \frac{dk\ k}{2\pi^2} \Plin(k) j_1(kq). \nonumber
\end{equation}
To relate the two expressions we note that
\begin{equation}
    j_1(kq) = \frac{1}{3} (kq) W_q(k) \nonumber 
\end{equation}
where $W_q = 3 (\sin(kq)/(kq) - \cos(kq))/(kq)^2$ is the Fourier transform of the spherical top hat window. This is a consequence of Birkhoff's theorem, i.e. that the inflow velocity depends only on the enclosed mass at radius $q$.}} we have
\begin{align}
    &2i b_1^L b_2^L  i k_i \int d^3\bq\ e^{i\bk\cdot\bq}\ \langle \delta(\bq) \Delta_i \rangle\ \xi_w(\bq) \nonumber \\
    &\approx -\frac23 b_1^L b_2^L \sigma^2_s k_i \nabla_{k_i} \int d^3\bq\ e^{i\bk\cdot\bq}\  \xi_w(\bq) = -\frac23 \sigma^2_s b_1^L b_2^L \frac{d P_w}{d\ln k}.
    \label{eqn:realbaoshift1}
\end{align}
In the above we have defined $\sigma^2_s = \sigma^2(r_d)$ to be the mean square density at the peak of the wiggly correlation function. This directly translates into a physical shift in $\alpha_{\mathrm{iso}}$ and, sign-included, has the interpretation that galaxies with positive $b_2$ are more likely to form in overdense regions, and galaxies sitting in an overdensity will tend to flow towards each other, shrinking the average observed BAO \citep{Crocce08,Padmanabhan09b,Sherwin12}. Importantly, this effect is caused by overdensities due to modes with wavelengths \textit{longer} than $r_d$. While we have focused on the phase shift due to $b_2$, all other quadratic contributions to clustering will produce similar shifts; working in Eulerian perturbation theory (EPT), \citet{Sherwin12} showed that the real-space shift for a biased tracer with quadratic density bias is\footnote{We note that \citet{Sherwin12} presented two different expressions for the BAO phase shift, the one we have quoted here and a larger one in the correlation function. However, the two expressions are in fact equivalent under Fourier transform and, since the covariance is (more) diagonal in Fourier space and therefore corrections therein more directly translated into systematic shifts in parameters, we use the former expression throughout this text. In addition, we note that the precise weighting of the $k$ modes in $\sigma^2_s$ stems from our LPT calculation.}
\begin{equation}
    \Delta \alpha_{\rm iso, NL} = \frac{47}{105} \left(1 + \frac{70}{47} \frac{b_2}{b_1} \right) \sigma^2_s \approx 0.3 \% \left(1 + \frac{70}{47} \frac{b_2}{b_1} \right) \left( \frac{D(z)}{D(0)} \right)^2
    \label{eqn:realbaoshift2}
\end{equation}
where $D(z)$ is the linear growth factor and we have evaluated the shift in the fiducial DESI cosmology.

\begin{figure}
 \includegraphics[width=0.48\textwidth]{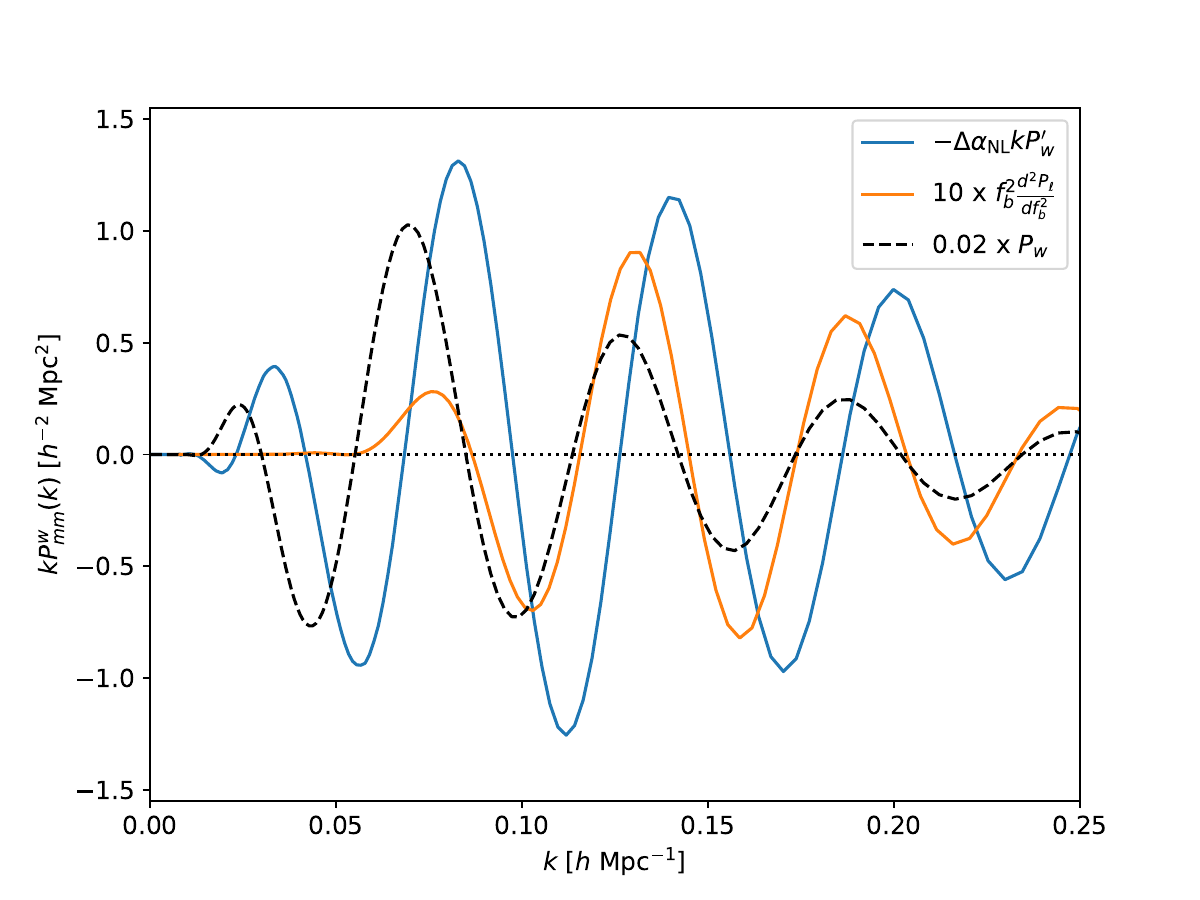}
 \caption{A summary of oscillatory, non-linear, effects that lead to shifts in the measured BAO position in the real-space 1-loop matter power spectrum. Exact expressions are given in Eqs.~\ref{eqn:realbaoshift1}-\ref{eqn:mode_coupling}. For reference, the dashed line is the linear theory wiggle power spectrum $P_w$ scaled down to $2\%$ of its nominal amplitude, and the $f_b^2$ contribution amplified by a factor of $10$, for ease of comparison.}
 \label{fig:wiggly_effects}
\end{figure}

The BAO feature can also couple with itself to produce additional out-of-phase contributions to the signal. These are contributions of the order $\mathcal{O}(P_{w}^n)$ and hence $f^{n}_{b}$. For example, at quadratic order we can write the contribution in Fourier space
\begin{equation}
    \frac{d}{d f_b^2} P(k) = \int \frac{d^3\bp}{(2\pi)^3} K_2(\bp, \bk - \bp)\ P_w(p) P_w(|\bk-\bp|)
    \label{eqn:mode_coupling}
\end{equation}
where $K_2$ is the quadratic perturbation theory kernel for galaxy clustering. Since each power of $P_w$ is sinusoidal with frequency $r_d$, their convolution in Fourier space can produce a frequency dependence distinct from that in linear theory. In the case of nonlinear matter clustering or linear bias we have that $K_2$ is the quadratic matter kernel $F_2(\bp_1,\bp_2)$ which peaks when 
$\bp_1 = \bp_2 = \bk/2$, producing an out-of-phase oscillatory signal $\sin(r_d k/2)^2 \sim \cos(r_d k)$ \citep{Padmanabhan09b}. Since this signal only has support within one BAO wiggle of $\bk/2$ it scales roughly as $f_b^2 (\pi / r_d)^2 k P_w(k/2)^2$ , compared to the nonlinear shift derived from Eqs.~\ref{eqn:realbaoshift1} and~\ref{eqn:realbaoshift2} which scales as $\propto \sigma^2_s \times k P_w'$ --- it is thus subdominant since $k/2$ will typically appear after the peak of the power spectrum and because it contains an extra power of the baryon fraction $f_b$. 

A comparison of both of the BAO shift contributions we have discussed, in the case of matter clustering, is shown in Figure~\ref{fig:wiggly_effects}. Note that unlike in the case of a pure sine wave the actual BAO wiggles in the linear power spectrum lead to a $d^2 P/df_b^2$ that also contains a significant in-phase component, further suppressing the contribution of this effect to the total BAO shift. 

In addition to the above effects, which describe the clustering of galaxies assuming dark matter and baryons can be treated as a single matter fluid on cosmological scales, the very same physics that generates the BAO produces residual differences in dark matter and baryon clustering that remain even after decoupling. The dominant effects can be written as \citep{Schmidt16,Chen19}
\begin{equation}
    \delta_g \ni b_{\delta_{bc}} \delta_{bc} + b_{\theta_{bc}} \theta_{bc} + b_{v^2} v^2_{bc}
\end{equation}
where we have defined the relative linear perturbations as $X_{bc} = X_b - X_c$, with $\delta$ and $\theta$ the linear densities and velocity divergences of the baryons and dark matter, respectively. The linear relative densities and biases are constant and decay with redshift as $a^{-1}(z)$, respectively. The third term is the so-called ``relative velocity effect'' which was first described in \citet{Tseliakhovich10} as the effect of supersonic baryon-dark matter velocities on gravitational collapse. Unlike the first two terms, it appears only at one-loop order in perturbation theory, but can be enhanced as the relative velocity squared is a dimensionful parameter --- studies of the relative velocity effect have \edit{often used the value} $b_{v^2} \sim 0.01 \sigma_{bc}^{-2}(z)$ where $\sigma_{bc}^2$ is the mean square fluctation of relative velocity perturbations, which correspond to roughly $30$ km s$^{-1}$ flows at recombination\edit{, based on early investigations into this effect.}

Compared to the usual galaxy biases, $b_1, b_2$ etc., the sizes of these relative bias parameters are somewhat less well understood, though some recent progress has been made. \cite{Barreira20} measure the relative density bias $b_{\delta_{bc}}$ using the separate-universe technique in hydrodynamical simulations and find distinct trends for dark matter halos (with halo mass) and galaxies (with stellar mass). While the trend in halos can be well-modelled by the change in amplitude of local fluctuations due to baryon density, as can halo-mass selected galaxies, the trend in galaxies selected by stellar mass is sensitive to other baryonic physics and exhibits an opposite (positive) trend with mass, particularly at the redshifts $z<2$ that DESI will observe. Nonetheless for all of these competing effects, the inferred bias values for both relevant halos and galaxies are well described by $|b_{\delta_{bc}}| \lesssim 1$ for the galaxies relevant to DESI. The linear relative velocity divergence bias is less well understood but physical estimates through excursion sets and other methods \citep{Schmidt16,Blazek16} suggest $|b_{\theta_{bc}}| \lesssim 7 (1+z)^{-1}$, where the redshift dependence cancels the decay of peculiar velocities. The size of the relative velocity effect at DESI redshifts is fairly poorly understood, with estimates as high as $b_{v^2} \sim 0.01 \sigma_{bc}^{-2}$ or as low as $10^{-5} \sigma_{bc}^{-2}$ for DESI-like galaxies \citep{Yoo11,Blazek16,Schmidt16}. \edit{A direct estimate of the effect of streaming velocities on baryonic infall into halos gives an order $(10^5 M_{\odot} / M)^{2/3}$ effect \citep{Blazek16}, which yields $b_{v^2}$ much closer to the lower estimate for DESI galaxies, while reaching the higher value often used in the literature requires that galaxies today retain significant memory of star formation at early times \citep{Yoo13}.}

\begin{figure}
 \includegraphics[width=0.48\textwidth]{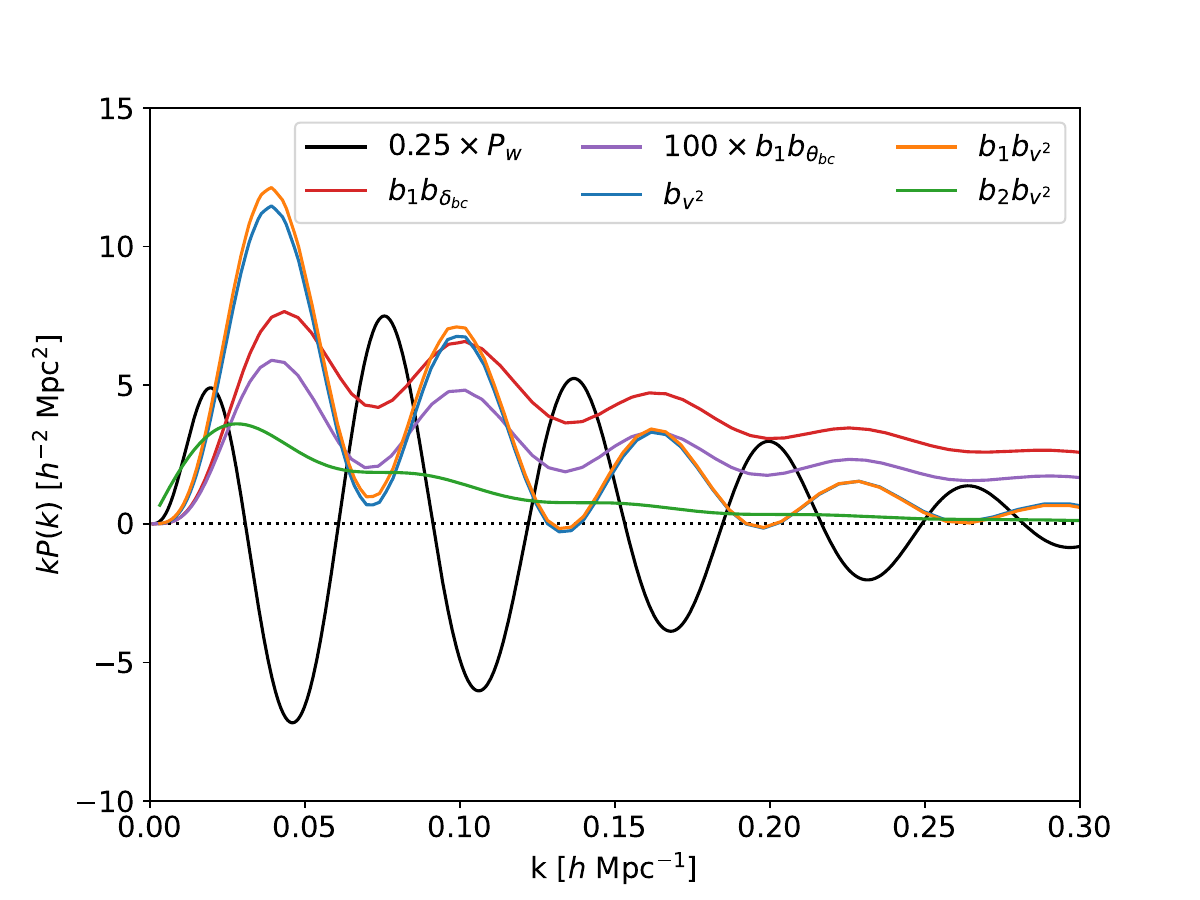}
 \caption{Contributions to the real-space galaxy power spectrum at $z=0$ from relative baryon-dark matter perturbations, compared to the matter $P_w$ (black). All three leading effects show prominent oscillatory features independent from the $P_w$, with the largest contribution from the relative density perturbation for values $b_{\delta_{bc}} \sim 1$ \edit{(red)}. The contributions due to the relative velocity divergence \edit{($b_{\theta_{bc}}$)} \edit{(purple)} are quite small and are scaled here by a factor of $100$ for visibility. The relative velocity effect terms \edit{($\propto b_{v^2}$)} assume an effect of order $0.01 \sigma_{v}^{-2}$, a very pessimistic choice at the upper end of the range of estimates. \edit{Contributions due to the cross correlation of the $b_{v^2}$ term with matter, linear and quadratic bias are shown in blue, orange and green, respectively, with the former two showing particularly prominent oscillations at this value of $b_{v^2}$.}}
 \label{fig:relative_velocity_rs}
\end{figure}

Figure~\ref{fig:relative_velocity_rs} shows the leading contributions to the galaxy power spectrum of each of these terms in real space at $z=0$ \citep{Yoo11,Schmidt16,Blazek16,Chen19}, compared to linear theory wiggles. When the relative velocity effect \edit{($b_{v^2}$; blue, orange and green curves)} is the size of the most generous estimates quoted above it is the leading oscillatory contribution, with anti-correlated wiggles up to one quarter the size of the matter oscillations . The leading relative density contribution \edit{($b_{\delta_{bc}}$; red)} is also quite substantial, and its importance grows as a function of redshift since $\delta_{bc}$ is stationary while $\delta_m$ decays as $D(z)$ with redshift. The relative velocity \textit{divergence} \edit{($b_{\theta_{bc}}$; purple)} on the other hand is suppressed by up to two orders of magnitude compared to these other contributions and thus its contribution to any BAO shift should be negligible. It is worth noting that all of the shown relative bias contributions contain significant or even dominant \edit{in-, or exactly out-of-phase contributions with oscillations with the opposite sign to the BAO. These oscillations mainly modify the amplitude ofthe BAO rather than shift it}; if these are present at a significant level in DESI galaxies they would lead to modifications of their clustering (e.g. inconsistencies between the BAO and broadband amplitudes) which could be constrained in full-shape fits~\citep{Blazek16,Beutler2016:1612.04720v3,Slepian2016:1607.06098v1}. These constraints could then be used to reduce the theoretical error on the BAO. 

Overall, the unknown amplitude of these effects means they must be accounted for, conservatively, in our theoretical systematic error budget. However, this requires further consideration of these effects in redshift-space, which will be carried out next.

\subsection{Redshift Space}
\label{sec:rsd}

We now extend the discussion in the previous section to include redshift-space distortions. In LPT, redshift-space distortions can be modelled by including an extra displacement along the line of sight $\hn$ proportional to the peculiar velocity, i.e.\  $\Psi_s = \Psi + \hn \left( \hn \cdot \dot{\Psi} \right)/(aH)$ \citep{Matsubara08,Carlson13}. Within linear theory this is equivalent to taking $\Psi_{s,i} = R_{ij} \Psi_j$, where $R_{ij} = \delta_{ij} + f\hn_i \hn_j$ and $f(z)$ is the linear growth rate.

The effect of redshift-space distortions on the BAO damping is straightforward. Since the exponentiated displacements all get multiplied by $R_{ij}$, the damping factor acquires an anisotropy given by \citep{ESW07,Matsubara08,Ivanov18}
\begin{equation}
    -\frac12 k_i k_j R_{ia} R_{jb} \Sigma^2_{\rm NL} \delta_{ab} = -\frac12 K^2 \Sigma^2_{\rm NL}\, , \quad K_i = R_{ij} k_j\, ,
\end{equation}
such that $K^2 = (1 + f(2+f)\mu^2)k^2$. As in real space, small-scale nonlinear effects can also affect the damping of the BAO peak. In practice, the most significant contribution will be from the so-called ``Fingers of God'' (FoG) --- random motions, and therefore redshift-space displacements --- due to virial motions of galaxies within halos \citep{Jackson72}. To leading order, these contribute an additional $-\frac12 k^2 \mu^2 \Sigma^2_{\rm FoG}$ to the exponential damping.   Since virial velocities come from a range of halo masses with different dispersions  \citep{Sheth96,Diaferio96,White01} their distribution is not Gaussian \citep{Peebles76,Davis83}. Their overall effect may be better described by other probability distributions \citep{Seljak11,Chen21}, such as the e.g.\ Lorentzian form sometimes used in the literature \citep{Peacock92,Park94,Peacock94,Ballinger96}. We will quantify the impact of these choices in \S\ref{sec:damping}.

The effect on the nonlinear BAO shift is somewhat more complicated. Following the same reasoning as in real space, we can easily come to the conclusion that the BAO shift should become anisotropic and enhanced along the line of sight. This is because redshift-space distortions from peculiar velocities streaming towards overdensities will cause galaxies separated by an overdensity along the line of sight to appear closer to each other than those separated perpendicular to such an overdensity. An exact calculation of all the anisotropic nonlinearities proportional to the large-scale overdensity $\sigma^2_s$ yields (Appendix~\ref{app:rsd_bao_shift})
\begin{align}
    P_{\rm shift} = -& (1/315)  \sigma^2_s k P_w'(k) (b_1 +  f \mu^2) (141 b_1 + 210 b_2 + 112 b_s \nonumber \\
    &+ 
   f (93 + 330 b_1 + 210 b_2 + 112 b_s - 42 (1 + b_1) f) \mu^2 \nonumber\\
   &+ 3 (80 + 77 b_1 - 28 f) f^2 \mu^4 + 189 f^3 \mu^6)\, ,
   \label{eqn:bao_shift}
\end{align}
where we have also included the possible effect of a tidal bias $b_s$. Evidently, the shift can no longer be mapped into a single $\Delta \alpha_{\mathrm{iso}}$ as it could in real space, though it is worth noting that the total shift for modes perpendicular to the line of sight ($\mu=0$) remains the same, while the shift along the line of sight becomes significantly enhanced.

In order to translate Equation~\ref{eqn:bao_shift} into systematic shifts in $\apar, \aperp$ we conduct a simple Fisher analysis assuming a cosmic-variance limited survey and Gaussian covariances with a fiducial model given by the Kaiser form plus a polynomial broadband model as described in \S\ref{sec:broadband}. Defining the systematic error vector $
\epsilon = (P^0_{\rm shift}, P^2_{\rm shift})$ to be the multipoles of Equation~\ref{eqn:bao_shift} we have that the shift in $\alpha$'s is given by
\begin{equation}
    \Delta \alpha^{\rm NL}_{\parallel,\perp} = F^{-1}_{\alpha_{\parallel,\perp}, b} t^b_i C^{-1}_{ij} \epsilon_j\, ,
    \label{eqn:fisher_shifts}
 \end{equation}
where $C^{-1}_{ij}$ is the inverse covariance,  $t^b_i$ are the linear templates for the free parameters $p_b$ and $F_{ab}$ is the Fisher matrix. For simplicity we assume the coevolution relations $b_2 = b_2^L + \frac{8}{21} b_1^L$ and $b_s = -\frac{2}{7} b_1^L$, where the Lagrangian $b_2^L$ is fixed to be a function of $b_1^L$ according to the Sheth-Tormen mass function \citep{Sheth01}.

\begin{figure}
 \includegraphics[width=0.48\textwidth]{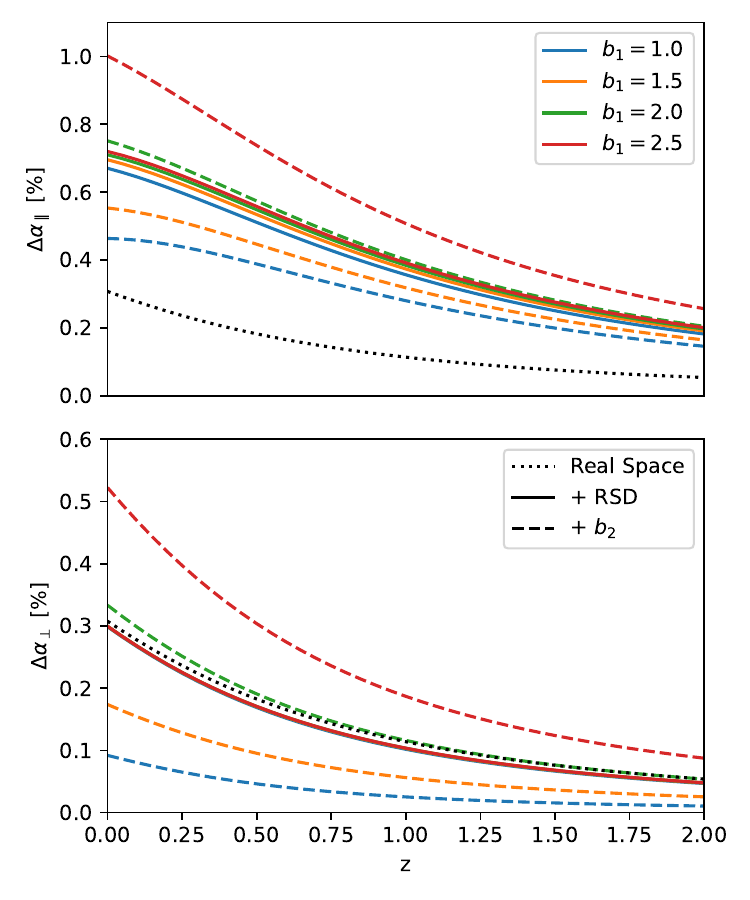}
 \caption{Fisher forecasts for the BAO shift in redshift space due to nonlinear clustering as a function of redshift $z$ and linear bias $b_1$, assuming a signal-dominated survey. Solid lines show results for linear bias only while dashed lines show the shift when a quadratic bias parameter, predicted from the peak-background split, is included, showing that the scatter in potential shifts due to the unknown nonlinear bias parameters is substantial. The dotted black line shows the shift in real space, which is similar in size to $\Delta \alpha_\perp$ in most cases but substantially smaller than $\Delta \alpha_\parallel$.}
 \label{fig:bao_shifts}
\end{figure}

Figure~\ref{fig:bao_shifts} shows the expected shifts in the BAO scale in the above setup as a function of $b_1$ and redshift. The solid lines show the expected BAO shift in the absence of any nonlinear bias: they roughly reproduce the real-space prediction (black dotted) in the case of $\aperp$ but are up to 3 to 4 times larger for $\apar$. This makes sense since the perpendicular part of Equation~\ref{eqn:bao_shift} is identical to the real-space prediction.

The dashed lines in Figure~\ref{fig:bao_shifts} show the result of adding in nonzero quadratic density and shear biases using peak-background split predictions. These show significant deviations away from the linear bias-only predictions; roughly, for $b_1^L < 1$ the quadratic bias becomes negative, reducing the effect, and vice versa for more biased tracers, which cluster more strongly around overdensities. The bias relations used here are known to qualitatively reproduce the behaviour of dark matter halos in N-body simulations \citep{Lazeyras18,Abidi18}; however, they are not guaranteed to hold for the galaxy samples observed in DESI, and as such the scatter between the different lines in the plot should be thought of as an indicator of the possible systematic scatter due to the unknown size of nonlinear biasing. Conversely, a \textit{full-shape} fit of galaxy clustering aimed not just at extracting the BAO signal and where all perturbation theory parameters are varied automatically incorporates this uncertainty into its parameter constraints. At least in the case of pre-reconstruction, where perturbation theory models are quite mature, an alternative procedure to fitting the BAO may be to employ such a model, as one would do in a ``full'' fit of the galaxy power spectrum, while simultaneously marginalizing over broadband terms (\S\ref{sec:bao_overview}) as is typically done in BAO fits. Beyond standard BAO measurements, we note that these shifts have a larger effect on $\alpha_{\parallel,\perp}$ than the expected shift due to light species such as neutrinos, which has emerged as a topic of recent interest (e.g. \cite{Baumann19}), and therefore must be carefully marginalized over when looking for ``exotic'' effects in BAO physics; on the other hand, as we will see, they are very well mitigated by reconstruction.

Converting the error estimates above into $\alpha_{\rm iso}$ and $\alpha_{\rm ap}$, the former unsurprisingly is subject to nonlinear shifts spanning the geometric average $\Delta \alpha_\parallel/3 + 2 \Delta \alpha_\perp/3$, such that the range of scatter due to the unknown size of nonlinear biases is directly translated over from Figure~\ref{fig:bao_shifts}. In the case that the nonlinear biases are given by the peak background split and Lagrangian coevolution we have, roughly, $\Delta \alpha_{\rm iso} \approx 0.5 \sigma^2_s ( 1 + (b_1-1)) D(z)^2$, or roughly $0.3\%$ for a bias one tracer at redshift zero, while neglecting any nonlinear Eulerian bias gives $\approx 0.9 \sigma^2_s D(z)^2$. The scatter between these two again is indicative of the uncertainty induced in pre-reconstruction measurements by nonlinearities. On the other hand, it is interesting to note that this scatter essentially cancels in $\alpha_{\rm ap}$ such that we have $\Delta \alpha_{\rm ap} \approx 1.2 \sigma^2_{s} f(z) D(z)^2$, corresponding to approximately a $0.4\%$ shift at $z=0$, to within $0.1\%$ accuracy over the entire range of biases and redshifts shown, with only minor variations increasing with nonlinear bias. This suggests that a direct correction based on this estimate can be performed for $\alpha_{\rm ap}$ when measuring it from pre-reconstruction BAO. We caution that the fitting forms above are approximate and intended for error estimation.

We can also apply the same Fisher formalism to extract the BAO shifts due to the quadratic coupling of $P_w$ with itself and the relative velocity effect. Redshift-space distortions similarly enhance the size of the former along the line of sight; however, since this effect ultimately has a rather distinguishable scale dependence compared to the BAO wiggles and their derivatives the contribution to the measured BAO scale is extremely suppressed --- for a bias $b=2$ tracer at $z = 0.8$ the effect is of order $f_b^2 \times 0.1\%$, which is entirely negligible. This result justifies neglecting any order $f_b^2$ effects in BAO fits even though they are nominally only suppressed by one power of the baryon fraction $f_b \approx 15\%$.

Finally, we compute the redshift-space power spectrum contributions of the relative velocity effect using the Lagrangian formalism of \citet{Chen19}, converting their real-space expressions into redshift-space as in \cite{Vlah19,Chen21} (see also \citealt{Schmidt16}). By operating within LPT we avoid having to perform wiggle/no-wiggle splits on the relative velocity spectra to model their nonlinear damping; the standard algorithms to perform this split are optimized for the matter power spectrum and cannot typically be used ``out of the box'' in this case where the wiggles are also more pronounced. Keeping corrections only to first order in the relative bias parameters we can express the resulting BAO shifts from Equation~\ref{eqn:fisher_shifts} as $\Delta \alpha_{\parallel,\perp} = b_i (d\alpha_{\parallel,\perp}/db_i)$ for $b_i = \{ b_{\delta_{bc}}, b_{\theta_{bc}}, b_{v^2} \}$. In addition to the shift in best-fit parameters, the presence of these relative bias terms can also change the strength of BAO constraints by modifying the BAO signal amplitude with their in-phase components. However, this effect will be avoided so long as the covariance matrices and model can match the amplitude of the observed BAO.

\begin{figure}
 \includegraphics[width=0.48\textwidth]{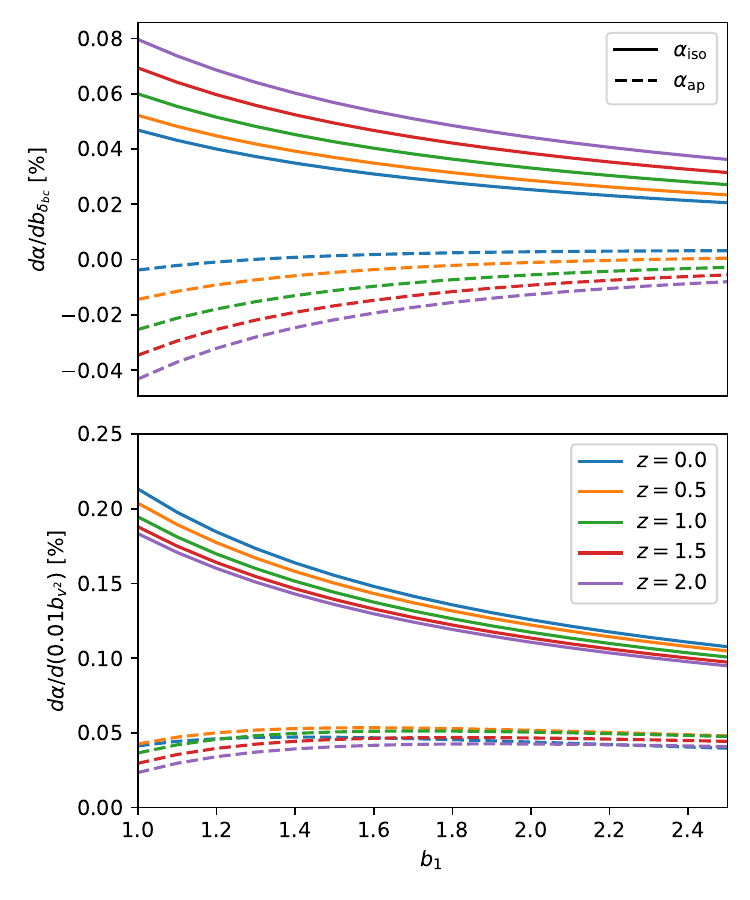}
 \caption{BAO shifts from relative velocity effect as a function of linear bias $b_1$ and redshift $z$, computed for (top) the relative density bias $b_{\delta_{bc}}=1$ and (bottom) $b_{v^2}=0.01 \sigma_{bc}^{-2}$. The normalization for $b_{v^2}$ here is on the upper range of estimates in the literature, with estimates in the literature spanning three orders of magnitude smaller---we adopt $10^{-3} \sigma_{bc}^{-2}$ as the fiducial mean in this work.  We have omitted the shift due to the relative velocity divergence since its effect is expected to be much smaller than the two shown.}
 \label{fig:relative_bias_shifts}
\end{figure}

Figure~\ref{fig:relative_bias_shifts} shows the resulting redshift-space BAO shift per unit relative bias. The top panel shows the effect of the relative overdensity bias $\delta_{bc}$. The magnitude of the shift diminishes towards lower redshift and higher biases and is opposite in sign for $\alpha_{\rm iso, ap}$. This is as expected, since the relative component stays constant with redshift unlike the growing adiabatic component, so perpendicular to the line of sight the relative size of the effect scales as $b_{\delta_{bc}} / (b_1 D(z))$, while along the line of sight, the shift is somewhat reduced by the presence of redshift-space distortions sourced by matter, such that the shift in $\alpha_{\rm ap}$ is negative. We note that while \cite{Barreira20} find that $b_{\delta_{bc}}$ is typically of order one for both galaxies and halos and the expected shifts are therefore of the same order of magnitude as the derivatives shown in the top panel, they also find that $b_{\delta_{bc}}$ tends to increase with halo mass. Given that the exact value of this bias is unknown as a conservative estimate we should therefore expect a $\sim0.05 D^{-1}(z) \%$ shift in the BAO from this effect. 

The bottom panel similarly shows the expected shift per $0.01 b_{v^2}$ as a function of bias and redshift. Since these terms exist at 1-loop order it is necessary to make some assumptions about nonlinear biasing; we have therefore adopted the peak-background split values of $b_2$ like in Figure~\ref{fig:bao_shifts}, though our results in this case are not very sensitive to this choice as the relative velocity effect is mostly sourced by long-wavelength modes. Here again the signal decreases as a function of linear bias, but unlike the relative density bias, it stays roughly constant with redshift since the 1-loop terms contributing to it scale similarly to the linear adiabatic modes. Given the lack  \edit{of a clear measurement in either data or simulations of the value of $b_{v^2}$, or a clear theoretical consensus in the literature about its expected size,} the derivatives plotted can be thought of as a worst-case scenario for this bias. If we take a log-uniform prior on $b_{v^2}$ spanning the range $10^{-5}$ to $10^{-2} \sigma^{-2}_{bc}$ the mean shift is roughly $0.02 b_1^{-1} \%$ on $\alpha_{\rm iso}$ with negligible redshift dependence and a significantly smaller anisotropic shift. \edit{However, since the straightforward estimate based on halo mass yields of a few times $10^{-5}$, absent discovering an astrophysical mechanism that preserves the relative velocity signal in early star formation in DESI target galaxies it seems safe to assume this is also a conservative estimate. Further, while the shift due to a $b_{v^2}$ in the upper range of reported values in the literature would be comparable to the statistical precision of the full Y5 DESI data, we expect that full-shape analyses in the meantime will be able to place stronger bounds on its value, since data from the BOSS survey already come quite close to constraining these values \citep{Yoo13,Beutler2016:1612.04720v3}.}

\section{Reconstruction}
\label{sec:recon}
\subsection{The Standard Reconstruction Algorithm}
\label{sec:reconstandard}

In \S\ref{sec:nonlinear}, we considered the effects of unadulterated nonlinearities in structure formation on the BAO signal in galaxy clustering. However, as described in the introduction (\S\ref{sec:intro}), the bread-and-butter of BAO measurements in spectroscopic surveys like DESI concerns the BAO signal in galaxy clustering \textit{post-reconstruction}. The reconstruction algorithm presented in \cite{Eisenstein07} amounts to a nonlinear transformation of the observed galaxy density field to enhance \textit{and unbias} the BAO signal --- the purpose of this subsection is to review these effects in some detail using perturbation theory \cite[see][for earlier discussions]{Padmanabhan09a,Noh09,White15,Schmittfull15,Hikage17,Ding18,Chen19b,Hikage20}.

The first step of the reconstruction algorithm is to generate the reconstructed displacement $\Psi_r(\bs)$ at a given redshift space position $\bs$. Schematically, this is done by dividing the galaxy density field by $b + f\mu^2$ after smoothing the observed galaxy density field by a Gaussian filter $\mathcal{S}(k) = e^{-k^2 R^2/2}$ with $R$ sufficiently large that the smoothed field is well-described by linear theory. In this case, the linear Zeldovich displacement can be reconstructed as 
\begin{equation}
    \Psi_r(\bk) =  \frac{i \bk}{k^2}  \left( \frac{\mathcal{S}(k) \delta_g(\bk)}{b + f\mu^2} \right) \approx \Psi_{\rm Zel}(\bk).
    \label{eqn:recon_disp}
\end{equation}
In practice, survey geometry and other effects mean that the simple procedure above cannot be followed literally (i.e.\ the un-windowed $\delta_g(\bk)$ cannot be directly observed and divided by the Kaiser factor). In the below, however, we will mostly use the approximation in Equation~\ref{eqn:recon_disp} except when describing potential biases due to imperfect reconstruction, leaving the details of the reconstruction algorithm itself to our companion ``optimal reconstruction'' papers \citep{OptimalRecon1,OptimalRecon2}.

Given $\Psi_r$, the reconstruction algorithm proceeds to shift both the observed galaxies and randoms by the \textit{negative} reconstructed displacement in order to cancel the nonlinear damping due to long-wavelength displacements, and the reconstructed density field is defined to be the difference between the \textit{displaced} galaxies and \textit{shifted} randoms, $\delta_r = \delta_d - \delta_s$. Subtracting off the overdensity of the shifted randoms restores the linear clustering power removed by displacing the galaxies. Using number conservation, the overdensities of the galaxies and randoms post-reconstruction can be expressed as \citep{Schmittfull15}
\begin{equation}
    1 + \delta_{d,s}(\bx_r) = \int d^3 \bx\ \delta_D \left(\bx_r - \bx - \Psi_r(\bx) \right) \left( 1 + \delta_{d,s}(\bx) \right).
\end{equation}
Here $\bx_r$ is the position of galaxies and randoms post-reconstruction, and the densities $\delta_{d,s}(\bx)$ refer to their pre-reconstruction densities, i.e. $\delta_d(\bx) = \delta_g(\bx)$ and $\delta_s(\bx) = 0$. Equivalently we have in Fourier space
\begin{equation}
    \delta_{d,s}(\bk) = \int d^3\bx\ e^{-i\bk \cdot \bx} \left( e^{-i\bk \cdot \Psi_r(\bx)} \left( 1 + \delta_{d,s}(\bx) \right) - 1 \right).
    \label{eqn:deltak_rec}
\end{equation}
In order to make a connection with Lagrangian perturbation theory we can further substitute $\bx = \bq + \Psi(\bq)$, such that the integral over $\bx$ is replaced with $\bq$ and $\bx_r = \bq + \Psi(\bq) + \Psi_r(\bq+\Psi)$ \citep{White15,Chen19}.

In order to include redshift-space distortions the reconstructed displacement is multiplied by the matrix $R_{ij}$ (\S\ref{sec:rsd}). Where this boost factor is applied varies across the literature; in this paper, and generally in DESI Y1, we will focus on two conventions:
\begin{itemize}
    \item \textbf{RecIso}: galaxies shifted by $-\mathbf{R} \Psi_r$, randoms shifted by $-\Psi_r$
    \begin{align}
        \delta_d &= (b_1 + f\mu^2 - (1+f\mu^2) \mathcal{S}) \delta, \; \delta_s = - \mathcal{S} \delta\, , \nonumber \\
        \delta_r &= (b_1 + (1 - \mathcal{S}) f \mu^2) \delta\, , \nonumber
    \end{align}
    \item \textbf{RecSym}: galaxies shifted by $-\mathbf{R} \Psi_r$, randoms shifted by $-\mathbf{R} \Psi_r$. 
    \begin{align}
        \delta_d &= (b_1 + f\mu^2 - (1+f\mu^2) \mathcal{S}) \delta, \; \delta_s = - (1+f\mu^2) \mathcal{S} \delta\, , \nonumber \\
        \delta_r &= (b_1 + f \mu^2) \delta\, , \nonumber
    \end{align}
\end{itemize}
where we have also written down the linear theory predictions for each scheme. The former convention, \textbf{RecIso}, was chosen by the SDSS, BOSS and eBOSS collaborations \citep{Padmanabhan12,Alam17,Alam21}; since linear redshift space distortions are sourced by the same linear velocities as in $\mathbf{R}\Psi_r$, removing them from the displaced galaxies without doing the same to the randoms leads to the removal of linear redshift space distortions on large scales and ``isotropizes'' the clustering.  The second convention, \textbf{RecSym}, was introduced in \citet{White15} and treats randoms and galaxies ``symmetrically'' such that the linear theory broadband is preserved. It is important to note that (a)  the anisotropic clustering still contains BAO information, so subtracting it may result in weaker BAO constraints for a fixed shot noise and (b) the cancellation of RSD depends on the extent to which the true and fiducial cosmologies match, such that even given the true Zeldovich displacement the anisotropic term would be more like $ (f - \mathcal{S} f_{\rm fid}) \mu^2$, leaving a residual that could be confused for anisotropic BAO from the Alcock-Paczynski effect. \textbf{RecSym} is on the other hand robust against any such mistakes to leading order since both particles and randoms are moved identically.

\subsection{Nonlinear Modeling of the Reconstructed Field}
\label{sec:nlrecon}

The reconstructed field $\delta_r$ carries a somewhat different set of nonlinearities than the original galaxy density $\delta_g$, as expected since the reconstruction algorithm was designed specifically to reduce the nonlinear damping of the BAO signal. The nonlinear modelling of the reconstructed BAO is rather nontrivial and, while much progress has been made in recent years including efforts to model the post-reconstruction matter field \citep{Hikage17,Hikage20} and galaxies in the Zeldovich approximation \citep{White15,Chen19b}, to the best of the authors' knowledge no complete model taking into account both nonlinear dynamics and bias as well as dependences on the reconstruction algorithm itself (e.g. due to the fiducial cosmology assumed) has yet been presented. Nonetheless, it is possible to understand the effect of nonlinearities post-reconstruction using the same methods as the pre-reconstruction case. In the below, we will outline general arguments. More detailed calculations, particularly in the more tractable case of matter, will be presented in Appendix~\ref{app:ept_recon}.

\subsubsection{Real Space}

Let us first examine the damping of the BAO peak post-reconstruction. For the displaced galaxies we can write the following version of Equation~\ref{eqn:deltak_rec}:
\begin{equation}
    \delta_d(\bk) = \int d^3\bq\ e^{i\bk \cdot (\bq + \Psi(\bq) + \Psi_r(\bq+\Psi))} \left(1 + \delta_g(\bq) \right)\, ,
\end{equation}
where we have dropped delta functions in $\bk$ for brevity. Compared to the pre-reconstruction case we can see that the featureless displacement that smooths the power spectrum is now given by $\Psi(\bq) + \Psi_r(\bq) \approx (1 - S) \ast \Psi$ to leading order, i.e. the damping parameter is now
\begin{equation}
    \Sigma^2_{\rm rec} = \frac23 \int \frac{dk}{2\pi^2}\ \left(1 - \mathcal{S} \right)^2 P(k) (1 - j_0(k r_d))\, .
    \label{eqn:zel_damping_recon}
\end{equation}
The case of the shifted field is similar: At first sight, it looks like the shifted field starts out un-displaced so that the only BAO damping comes from the shift we apply to the randoms; however, since the shift field is computed from the \textit{Eulerian} density field it already comes pre-displaced, such that its ultimate damping scale is the same as in the displaced field (Appendix~\ref{app:recon_damping}). Another way to see this is that the full reconstructed density field is given via Equation~\ref{eqn:deltak_rec} as (see also \citealt{Shirasaki21,Sugiyama21})
\begin{equation}
    \delta_{\rm rec}(\bk) = \int d^3\bx\ e^{-i\bk\cdot\bx - i\bk\cdot\Psi_r(\bx)} \delta_g(\bx)
    \label{eqn:magic_recon_eqn}
\end{equation}
and, since $\delta_g(\bx)$ can be expressed as a sum of bias operators at the Lagrangian position $\bq$ \citep{Mirbabayi15} it follows that the total shift is given in \edit{Equation~\ref{eqn:zel_damping_recon}. This result that the BAO post-reconstruction is still damped by a simple single-exponential form, which we derive for the first time, is in contrast to previous derivations in the literature using perturbation theory and EFT \cite{Ding18,Chen19b}.}

\begin{figure}
    \centering
    \includegraphics[width=0.45\textwidth]{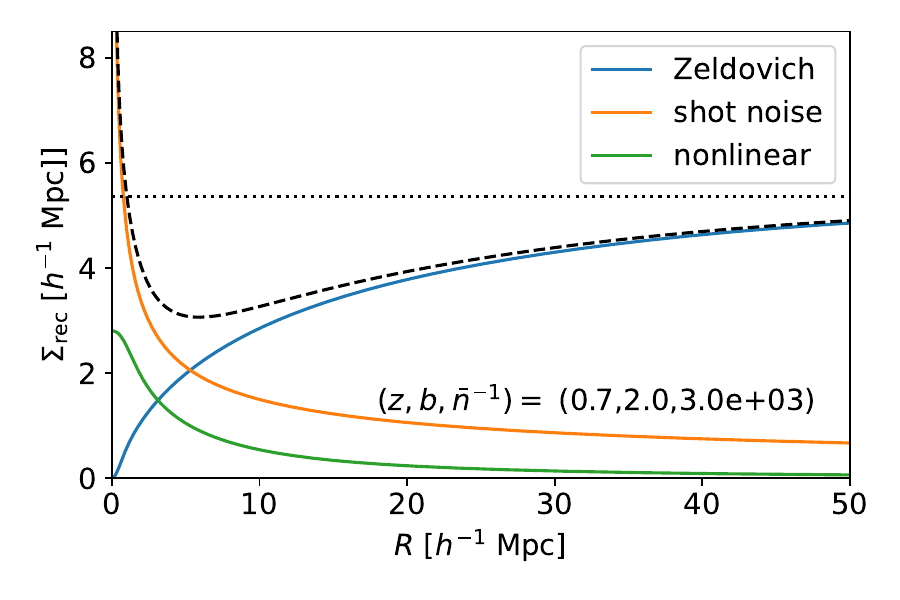}
    \includegraphics[width=0.45\textwidth]{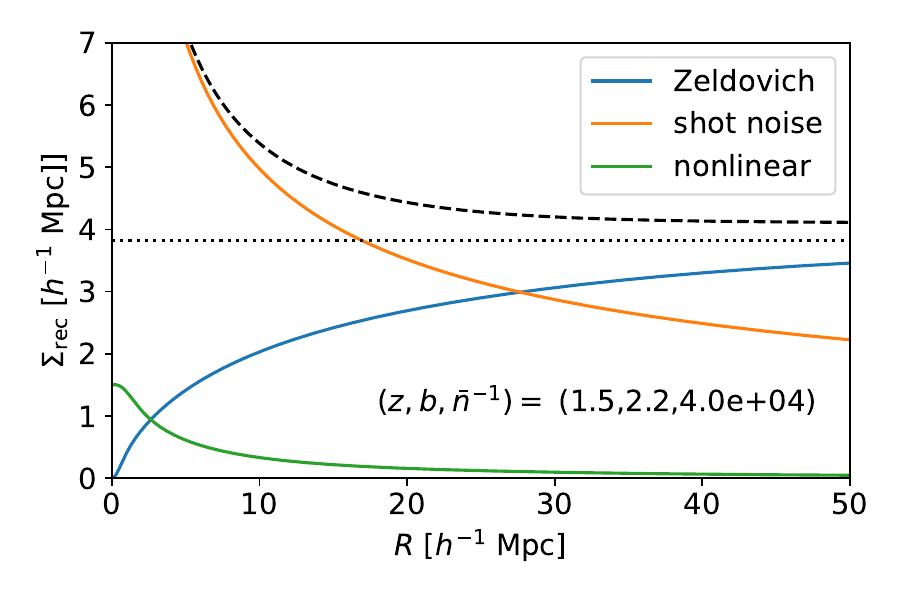}
    \caption{Contributions to the post-reconstruction BAO damping due to Zeldovich displacements (blue), shot noise (orange) and nonlinearities (green) as a function of smoothing scale $R$, with the total shown in the black dashed line. The black dotted line shows the full Zeldovich displacement at the BAO scale\edit{, i.e. the pre-reconstruction damping scale}. The green curve is estimated by computing the mean square reconstructed displacement $\langle \Psi^2_r \rangle$ using the 1-loop power spectrum for the value of bias shown, with nonlinear Eulerian biases assumed to be zero. The top and bottom panels show representative dampings for the DESI LRG and QSO samples, respectively, with the shot noise value reported in [$h^{-3}$ Mpc$^3$].}
    \label{fig:damping_scale_nonlins}
\end{figure}

We thus have that in real space the damping form is equivalent to the pre-reconstruction case except with a modified damping parameter, i.e.
\begin{equation}
    P(k) = b^2 \left(P_{nw}(k) + P_w(k) e^{-\frac12 k^2 \Sigma^2_{\rm rec}} \right)\, .
    \label{eqn:recsym}
\end{equation}
In addition, since the shift field is reconstructed from noisy galaxy densities it will add an additional uncorrelated damping proportional to the shot noise $P_{\rm shot} / \bar{n}$ \citep{White10}
\begin{equation}
    \Sigma^2_{\rm shot} = \frac23 \frac{P_{\rm shot}}{b^2 \bar{n}} \int \frac{dk}{2\pi^2} \mathcal{S}(k)^2\ (1 - j_0(k r_d)) \approx \frac{1}{6\pi^{3/2}} \frac{P_{\rm shot}}{b^2 \bar{n}R}\, ,
    \label{eqn:sigma2shot}
\end{equation}
where we have assumed that $R \ll r_d$. Here $P_{\rm shot}$ describes the deviation of the galaxy stochasticity away from a Poisson distribution, where we would have $P_{\rm shot} = 1$. For galaxies with inverse comoving density $3 \times 10^3 h^{-3} \textrm{Mpc}^3$ this is equal to $\Sigma^2_{\rm shot} = 6 b^{-2} R_{15}^{-1} P_{\rm shot} h^{-2} \textrm{Mpc}^2$ where $R_{15}$ is the reconstruction smoothing in units of $15\ \Mpc$. Again, it is important to note that this will not in general be the \textit{only} additional contribution to the post-reconstruction damping, but its presence is a good gauge of the size of nonlinear corrections to the Zeldovich damping. Some illustrative values of the expected damping as a function and smoothing scale are shown in Figure~\ref{fig:damping_scale_nonlins} --- while in general the damping can be somewhat reduced by decreasing $R$, other nonlinearities eventually make this choice sub-optimal. In the case of bias values and redshifts representative of the DESI QSO sample, we can see that the extra damping incurred by the shot noise brings the total post-reconstruction damping above the pre-reconstruction one for essentially all smoothing scales. However, for a suitably large $R$ the smoothing scales are rather comparable while the nonlinear BAO shift is still cancelled post-reconstruction such that performing reconstruction may still be preferred.

The effect of reconstruction on the nonlinear BAO shift follows a similar logic. Reconstruction explicitly removes, up to the smoothing scale $R$, the Zeldovich displacement terms that produce the shift; since the shift depends on modes $k \lesssim 1/r_d \ll 1/R$ the cancellation is essentially perfect if the linear displacement is properly estimated (see Appendix~\ref{app:recon_shift} for an explicit calculation). However, mis-estimating the linear bias $b_1$ can lead to residual displacements post-reconstruction, leading to errors roughly of order $(\Delta b_1 / b_1) \Delta \alpha_{\rm iso, NL}$, where $\Delta b_1$ is the difference between the true linear bias and that assumed during the application of the reconstruction algorithm. It is important to note that the relevant quantity here is the error on the linear bias and not its product with $\sigma_8$. The latter is typically extremely well constrained by the data while, absent external priors, the constraints on the former are essentially dictated by the errors on $\sigma_8$ itself. Conceptually, this is because we can only cancel the shift as well as we can know the amplitude of matter fluctuations within spheres of the BAO radius $r_d$.

Reconstruction also has a mixed effect on the relative biases. The contributions due to $b_{\delta_{bc}, \theta_{bc}}$ are essentially linear BAO's that are out-of-phase with the matter one, and reconstruction thus simply sharpens them in the same way, taking the damping from $\Sigma^2_{\rm NL} \rightarrow \Sigma^2_{\rm rec}$. Since both signals are identically strengthened, reconstruction is not expected to reduce the theoretical error due to this effect. On the other hand, the $b_{v^2}$ contributions are due to its coupling with quadratic nonlinearities in the galaxy density, which is in fact dominated by the large-scale displacements cancelled by reconstruction as $v_{bc}^2$ is dominated by long-wavelength modes \citep{Blazek16}. While we thus expect reconstruction to reduce the contributions of the relative velocity effect to the BAO---and indeed calculations within the Zeldovich approximation show that its oscillatory components are reduced by roughly a factor of two post-reconstruction---a complete calculation of the post-reconstruction relative velocity effect is outside the scope of this work. A conservative estimate is therefore to use the pre-reconstruction shifts as shown in Figure~\ref{fig:relative_bias_shifts} in our error budget.

\subsubsection{Redshift Space: Fitting Form and Algorithm Choice}

In the case of \textbf{RecSym} the arguments in real space carry over to redshift space essentially identically; since both the reconstructed displacement and the Zeldovich displacements responsible for the nonlinear BAO shift and BAO damping are multiplied by the same matrix $\textbf{R}$, reconstruction simply cancels the shift equally in redshift space and we get $k^2 \Sigma^2_{\rm rec} \rightarrow K^2(\mu) \Sigma^2_{\rm rec}$ as in the pre-reconstruction case. This motivates the fitting form
\begin{equation}
    \mathcal{C}(k,\mu) = (b + f\mu^2)^2 \exp\left[-\frac12 k^2 \left(\mu^2 \Sigma_\parallel^2 + (1-\mu^2) \Sigma_\perp^2 \right) \right]
\label{eqn:propagator}
\end{equation}
both pre- and post-reconstruction, with the understanding that the phase shift on top of this will be larger pre-reconstruction, and that $\Sigma^2_{\parallel,\perp}$ are roughly given by the Zeldovich prediction with uncertainty due to nonlinearities and imperfect knowledge of the true cosmology. In particular, as with the linear bias, the extent to which these cancellations happen depends on the fidelity with which we can constrain the linear growth rate $f$, such that we should expect a remaining theoretical error of size $(\Delta f / f) \Delta \alpha_{\perp,\parallel}$. Both of these errors are related to our certainty on the underlying matter clustering amplitude $\sigma_8$, which we can conservatively place a 5$\%$ uncertainty on given current discrepancies between the primary CMB and late-universe probes, though we note that differences up to $10\%$ with the primary CMB exist e.g. from galaxy redshift-space clustering \citep{Abdallah22}. We remind the reader that, while Zeldovich displacements form the dominant contribution to BAO damping, small-scale nonlinearities can also contribute and will tend to be equivalent to Equation~\ref{eqn:propagator} only at leading order. We investigate the effect of including such effects (e.g. FoG) into BAO fits in \S\ref{sec:damping}, finding that they have negligible impact even at DESI precision. More generally, these effects beyond leading order can be mitigated by limiting the range of scales fit, especially in the power spectrum.

In the case of \textbf{RecIso} however some care must be taken, because only the real-space portion of the Zeldovich displacement is cancelled for the shifted randoms. In this case, it is reasonable to assume that much (though not all) of the contributions due to the anisotropic BAO shift will remain, though the $\mu = 0$ contribution should still be cancelled. We show this explicitly for matter in Appendix~\ref{app:recon_rsd}. Similarly, while the displaced galaxies will still have their nonlinear damping reduced to $K^2 \Sigma^2_{\rm rec}$ the shifted fields and their cross spectra with galaxies will now be damped according to
\begin{align}
    \Sigma^2_{ss}(\mu) &= \Sigma^2_{\rm rec} + f^2 \mu^2 \Sigma^2_{\rm NL} + 2 f \mu^2 \Sigma^2_x\, , \nonumber \\
    \Sigma^2_{ds}(\mu) &= (1 + f\mu^2) \Sigma^2_{\rm rec}  + f(1+f) \mu^2 \Sigma^2_x + f^2 \mu^2 \Sigma^2_{\mathcal{S}}\, 
    \label{eqn:reciso}
\end{align}
where we have defined $\Sigma^2_x = \frac23 \int \frac{dk}{2\pi^2} \left( 1 - \mathcal{S} \right) P_{\rm lin}(k) \left( 1-j_0(r_d k)\right)$
as the cross-correlation between the pre- and post-reconstruction Zeldovich displacements and $\Sigma^2_{\mathcal{S}} = \frac13 \int \frac{dk}{2\pi^2} \mathcal{S}^2 P_{\rm lin}(k)$ is the mean-square Zeldovich displacement after smoothing by $\mathcal{S}$. \edit{The distinct pairwise displacements above between the shifted galaxies and randoms in \textbf{RecIso} mean that the reconstructed power spectrum in this case does not have BAO which are damped by a single exponential, but rather piecewise according to Equation~\ref{eqn:reciso}, which can be also seen from the fact that Equation~\ref{eqn:magic_recon_eqn} does not hold if the galaxies and randoms are not moved by the same amount.} The above expressions reduce to the pre-reconstruction displacement when $\mathcal{S} \rightarrow 0$, and the post-reconstruction displacement in real space if $f \rightarrow 0$, but in general the BAO signal in \textbf{RecIso} will be more damped than it is in \textbf{RecSym}, and the nonlinear shift larger, because of the incomplete cancellation of displacements along the line of sight. We thus see that \textbf{RecIso} suffers from (a) a partially uncancelled nonlinear BAO shift (b) more damping (c) less signal and (d) greater sensitivity to systematics in reconstruction than \textbf{RecSym}.

\section{Broadband Models in Fourier and Configuration Space}
\label{sec:broadband}

Thus far, we have developed the theory for the effects of nonlinear structure formation on the oscillatory BAO signal. The main contributions come from large-scale displacements which smoothly alter the envelope of the damped BAO wiggles and shift the observed BAO. These can be effectively removed through reconstruction, leading to a robust BAO signal that can be only weakly biased through second-order bias effects and \edit{(for pre-reconstruction or under \textbf{RecSym} convention)} damped by a single exponential damping term. However, nonlinearities\edit{---which are modified, but not removed, by reconstruction--- not to mention systematics in the data and differences in the fiducial and true cosmologies, also contribute to the \textit{broadband} shape of the observed galaxy 2-point function. Since the express purpose of BAO fits is to isolate the geometric information in the oscillatory part of the power spectrum only, the ability to marginalize over any broadband differences is an integral part of any BAO modeling.} 

\edit{An essential feature of such a broadband model is that it must fit any feature in the power spectrum with frequency in Fourier space lower than that of the BAO, thereby isolating the BAO information from the rest of the clustering signal. This is equivalent in configuration space to saying that it must remove any possible features localized at radii smaller than the BAO scale, as we will discuss in detail in \S\ref{ssec:bb_config}. It is important to emphasize that this is done regardless of the source (fiducial cosmology, observational systematic or nonlinearity) of the broadband contribution and, conversely, if a systematic contributes to clustering in a way that is \textit{not} sufficiently smooth it would need to be specifically modeled in order to mitigate its effect on the BAO.}

The broadband terms in our fiducial BAO model (Equation~\ref{eqn:BAO_model}) are contained in $\mathcal{B}(k,\mu) P_{nw} + \mathcal{D}(k,\mu)$. In principle $\mathcal{D}$ can be any sufficiently flexible fitting form non-degenerate with the BAO wiggles, though in practice it is useful to isolate a ``baseline'' contribution expected from linear theory alone described by $\mathcal{B}(k,\mu)$. A further role of $\mathcal{D}(k,\mu)$ is to account for any residual broadband term in $P_{w}(k)$ --- ideally this would represent \textit{only} the BAO wiggles, but in practice this separation cannot be performed on the linear power spectrum exactly (leading to, i.e., differences in the BAO templates using different methods, see \S\ref{ssec:pnw} and Fig.~\ref{fig:template_test}). In this case, a good choice of $\mathcal{D}(k,\mu)$ ensures consistent results regardless of the method used to generate  $P_{w}(k)$.  A particularly simple, well-motivated choice for $\mathcal{B}(k,\mu)$ is \citep{Kaiser87,Peacock94,Cole95}
\begin{equation}
\mathcal{B}(k,\mu) = (b+f\mu^{2})^{2}\biggl(1+\frac{1}{2}k^{2}\mu^{2}\Sigma^{2}_{s}\biggl)^{-2}\, ,
\label{eqn:lorentzian}
\end{equation}
where in addition to the linear theory prediction we have included a rough ``Fingers of God'' (FoG) parametrization through a single free velocity dispersion parameter $\Sigma_{s}$ in order to capture the rapid damping of the power spectrum along the line-of-sight due to small-scale virial motions. 

Past BAO analyses (e.g., \citealt{Anderson14,Beutler17, Bautista21}) have tended to rely on parametrizations of $\mathcal{D}(k,\mu)$ with angular multipoles expanded in polynomials of the wavenumber $k$ in powers between some chosen $n_{\rm min,max}$. The logic is that polynomials of sufficiently low order cannot mimic the oscillatory BAO signal. In configuration space, past analyses have also employed polynomials in $r$ of sufficiently low degree with the expectation that the BAO peak in the correlation function cannot be reproduced by polynomials \citep{Ross17,Bautista21}. While historically demonstrated to be robust, this approach suffers from a few limitations: (1) there is no clear criteria for what orders of polynomial terms to include --- which is a function of both the scale cuts and noise of the analysis (the polynomial cannot have more critical points than the number of BAO wiggles detected) --- and even if there were, there is no clear relation between the order of polynomial and smoothness of the function; and (2) while low-order polynomials cannot mimic \textit{wiggles} in Fourier space, it is easy for them to reproduce a \textit{peak} in configuration space. Due to these limitations, the stability of any particular prescription for the broadband given a particular set of data was only empirical. In fact, the earliest broadband models used to fit BAO in the Sloan Digital Sky Survey were based on roughly corresponding powers of $r^n$ and $k^{-3-n}$ to fit the correlation function and power spectrum, respectively \citep{Xu12,Anderson12} but as the data improved the number of polynomial terms in the power spectrum increased while those in the correlation function stayed roughly constant.  In this work we instead develop and validate a new unified treatment of the broadband $\mathcal{D}(k,\mu)$ for the power spectrum (and correlation function) that avoids these limitations. \edit{While the arguments in this section are quite general, we note that their assumptions are implicity and explictly tested in many accompanying DESI 2024 BAO papers---we refer readers interested in the effects of fiducial cosmology and observational systematics on the broadband model developed to ~\cite{cosmology_dependence,KP4}. }

\begin{figure}
    \includegraphics[width=0.45\textwidth]{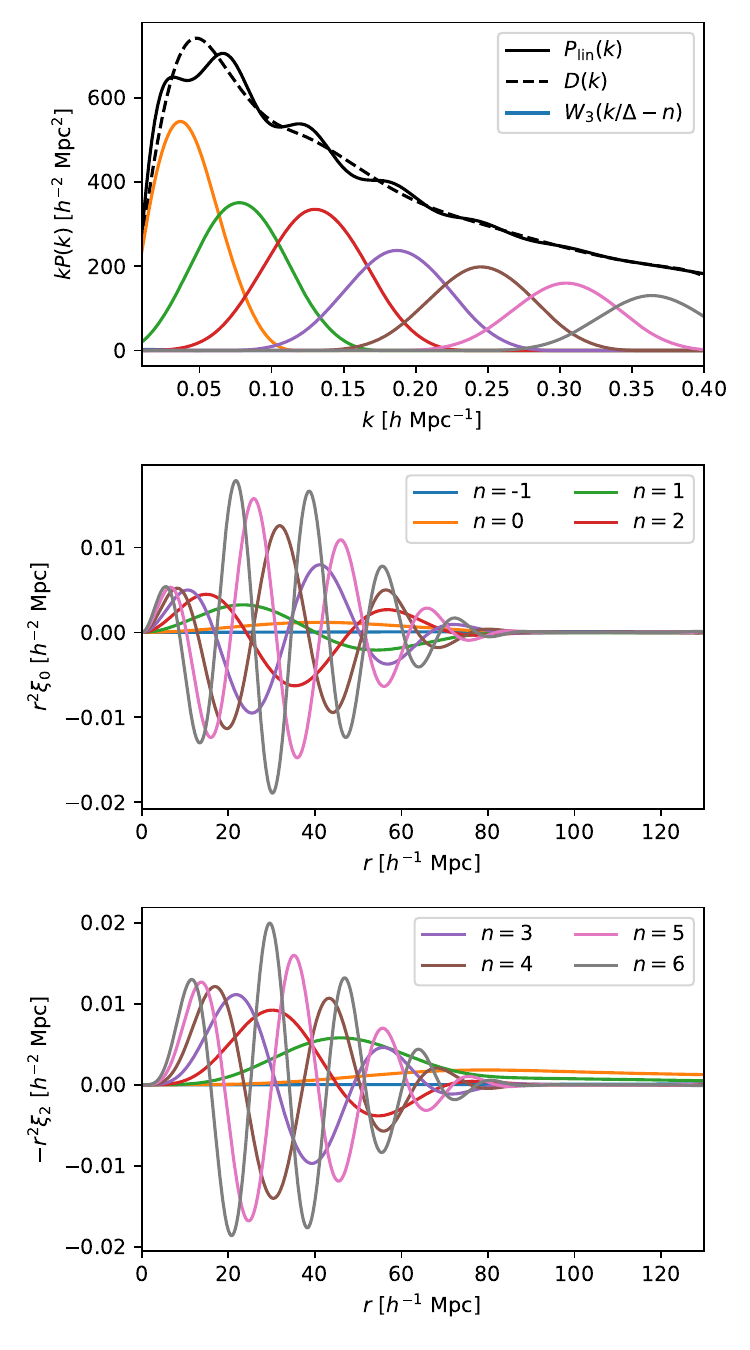}
    \caption{Smooth spline basis for broadband fitting. (Top) The basis in Fourier space, \edit{multiplied by $a_n$} coefficients fit to the linear power spectrum (black solid). The resulting broadband fit, which broadly matches the power spectrum but doesn't contain BAO wiggles, is shown in the black-dashed line. (Middle) The same spline functions \edit{without the coefficients $a_n$} Hankel-transformed to give the correlation function monopole. The first two basis functions have smooth large-scale contributions while the rest are oscillatory and confined to $r < 2\pi/\Delta$. The small wiggles at the largest $r$ are due to ringing from the discontinuous higher derivatives of the spline basis. (Bottom) Same as the middle plot but for the quadrupole. The first three basis functions in this case have smooth large-scale contributions.}
    \label{fig:spline_basis}
\end{figure}

\subsection{Fourier Space}

Our starting point will be to develop a broadband model in Fourier space $\mathcal{D}_\ell(k)$, where we allow for distinct terms for each clustering multipole $\ell$. The requirements on $\mathcal{D}_\ell$ are that they should not be degenerate with the BAO signal itself; this is really a requirement on the data that any theoretical or observational systematics not have features that can ``look'' like the BAO. In practice this requires that $\mathcal{D}_\ell$ be smoother than the BAO scale, or rather that its values be uncorrelated beyond some scale $\Delta > \pi/r_d$, which is half the wavelength of the BAO by the Nyquist sampling theorem.

A convenient way to do this is to parameterize the broadband using a cubic spline basis with bases separated by \edit{a suitably chosen} $\Delta$, i.e.
\begin{equation}
    \mathcal{D}_\ell(k) = \sum_{n=-1}^\infty a_{\ell,n} W_3\left(\frac{k}{\Delta} - n\right).
\end{equation}
Here $W_3$ is the `piecewise cubic spline' (4th-order) extension of the counts-in-cell interpolation kernel \citep{Hockney88,Jeong10}
\begin{equation}
W_{3}(x)=\frac{1}{6}
\begin{cases}
4 - 6|x|^{2} + 3|x|^{3} & |x| \le 1, \\
8-12|x|+6|x|^{2}-|x|^{3} & 1 < |x| \le 2, \\
0 & |x| > 2,
\end{cases}
\end{equation}
which has the convenient property that $\mathcal{D}_\ell(k>0) = 1$ when $a_{\ell,n} = 1$. The first few kernels in this basis fit to the linear power spectrum, \edit{where we have chosen to use $\Delta = 0.06\ h$ Mpc$^{-1} > \pi/r_d$,} are shown in Figure~\ref{fig:spline_basis}. The model is able to match the broadband of the power spectrum without reproducing the BAO signal itself. 

A nice feature of this parametrization of the broadband is that the free coefficients $a_{\ell,n}$ in this model are by construction never degenerate with the BAO signal, and thus cannot lead to numerical instability when fitting the BAO scale. As an extreme example, even if $n \Delta > k_{\rm max}$ such that $a_{\ell,n}$ is unconstrained, it will not interfere with the BAO signal within the range of fit where BAO are detected, but rather simply sample its prior. As such there is no fine-tuning required for the number of free parameters unlike in the polynomial case.

Finally, we note that in the Gaussian power spectrum likelihood typically used to constrain the BAO, any linear parameter in the power spectrum model, e.g. the $a_{\ell,n}$'s, can be analytically marginalized instead of being sampled directly. This is equivalent to adding an additional theoretical contribution to the covariance matrix if one does not care about the actual value of the marginalized-over parameters (\citealt{Taylor10}, see also Section~\ref{sec:barry}). Since we allow $a_{\ell,n}$ to take on arbitrary values, our prescription is equivalent to allowing infinite error at each wavenumber $P_\ell(k)$, with the proviso that the error across $k$ points are correlated when they are within $\Delta$ of each other, and the error on different multipoles is uncorrelated. A related theoretical-error based approach to fitting the BAO was proposed in \cite{Philcox20}, who impose an exponentially decaying correlation with $\Delta$ of the error but more narrowly estimate its size according to 1-loop perturbation theory. As we will now see, while the spline broadband induces a slightly counterintuitive theoretical covariance that is technically infinite at each $k$ point, by performing a simple coordinate (Fourier) transform the theoretical error can instead be strongly localized below some scale in configuration space.
 
\subsection{Configuration Space}
\label{ssec:bb_config}

\edit{The spline basis introduced in Fourier space is composed of smooth and well-localized functions which can be readily Hankel transformed into correlation functions. Figure~\ref{fig:spline_basis} shows their $j_{0,2}$ transforms: these basis functions turn into a basis of oscillatory functions in configuration space with zeros around $r = \pi/\Delta$, after which they rapidly decay. These functions have arbitrarily high frequencies below this radius as a function of $n$. This is to be expected since the functions in Fourier space have widths given by $\Delta$ and are centered at $n \Delta$.  These functions are given by 
\begin{equation}
    \Delta^3 B_{\ell,n}(r\Delta) = i^2 \int \frac{dk\ k^2}{2\pi^2} \ W_3(\frac{k}{\Delta} - n)\ j_\ell(k r)
    \label{eqn:corrspline}
\end{equation}
and we give their exact form in Appendix~\ref{app:corrspline}. Fitting these functions with free amplitudes in configuration space, i.e.
\begin{equation}
    \tilde{D}_\ell^{\rm spline}(r)  = \Delta^3 \sum_{n=-1}^\infty a_{\ell,n} B_{\ell,n}(r\Delta),
    \label{eq:corrspline2}
\end{equation}
nulls any signal in the correlation function, BAO or otherwise, up to $r_{\rm min} \approx \pi/\Delta$, since the functions sample all available Fourier frequencies up to that scale. While there are, in principle, an infinite number of broadband parameters in this expansion, their effects are localized to small $r$; we therefore have that the spline broadband model in Fourier space is equivalent to setting an $r_{\rm min}$ when performing the fit in configuration space rather than including any smooth broadband (e.g. polynomial) model. The two exceptions to this rule are the two basis functions $n=[0,1]$ in the quadrupole, which have some smooth dependence out to large $r$. Note that the $B_{2,-1}$ term also has nontrivial support at large $k$ but it is degenerate with the $k_{\rm min}$ contributions discussed below due to being confined to low $k$ only. We test the equivalence of these Fourier and configuration space configurations in \S\ref{sec:results} (see in particular the discussion around Fig.~\ref{fig:config_spline_test}). In particular, we note that while there is some ``spillover'' due to the finite localization of these functions to small $r$, the equivalence at the level of posteriors is quite insensitive to these numerical choices.}

Perhaps surprisingly, however, another (implicit) component of the Fourier-space broadband model does make it necessary to introduce a smooth polynomial broadband model in configuration space. In order to protect against unknown large-scale data systematics we typically only include scales above a certain $k_{\rm min}$ in the power spectrum when fitting the BAO. The Fourier transforms of these uncontrolled large-scale multipoles can be written as an expansion in even powers of the radius $r$ (odd powers are cancelled when computing even multipoles), such that
\begin{equation}
    \tilde{\mathcal{D}}_\ell(r > r_{\rm min}) = \tilde{a}_{0} + \sum_{n = 1}^{\infty} \tilde{a}_n \left( \frac{r k_{\rm min}}{2\pi}\right)^{2n}\, .
\end{equation}
For typical $k_{\rm min}$, e.g. $k_{\rm min} = 0.02\ h$ Mpc$^{-1}$, then $2\pi/k_{\rm min} \approx 300\ h^{-1}$ Mpc and the higher order terms in this series get progressively smaller at the BAO scale. We can hence truncate this to first order, $n=1$.

Overall, both the considerations above lead to a model for the broadband of the monopole and quadrupole of the correlation function given by
\begin{align}
    \tilde{\mathcal{D}}_{0}(r) &= \tilde{a}_{0,0} + \tilde{a}_{0,1} \left( \frac{r k_{\rm min}}{2\pi}\right)^{2}\, , \\
    \tilde{\mathcal{D}}_{2}(r) &= \tilde{a}_{2,0} + \tilde{a}_{2,1} \left( \frac{r k_{\rm min}}{2\pi}\right)^{2} + \Delta^3 \left( a_{2,0}B_{2,0}(r\Delta) + a_{2,1} B_{2,1}(r\Delta) \right)
\end{align}
with $B_{2,0}(r\Delta)$ and $B_{2,1}(r\Delta)$ given by Eq.~\ref{eqn:corrspline}.

\section{Numerical validation with Control Variate simulations}
\label{sec:numerics}

Having provided a comprehensive theoretical motivation for how the BAO can be modelled and fit given a set of power spectrum or correlation function measurements, we now turn to validating this model using a set of precise simulated measurements with a known cosmological model. In addition to what has been presented so far, there are a number of choices of how exactly one builds the full model that we will also test, many of which are required due to subtle differences between the model we have presented here and what has been done in previous works. The full extent of the things we aim to test/understand are:
\begin{itemize}
    \item{How do the BAO constraints compare between the previous polynomial-based broadband model and the new spline-based model?}
    \item{What is the impact of different prior choices for the BAO/FoG damping parameters $\Sigma_{\perp}$, $\Sigma_{||}$ and $\Sigma_{s}$?}
    \item{What is the impact of including the FoG term in the wiggle component of the model as well as the smooth component?}
    \item{What are the optimum scale cuts to use for fitting the power spectrum and correlation function?}
    \item{How does the choice of method used to perform the wiggle/no-wiggle split impact the BAO constraints?}
    \item{Does dilating the full power spectrum model or just the wiggle component make a difference?}
    \item{Does our updated model of the correlation function --- consistently treating it as the Hankel transform of the dilated power spectrum model with the same bias and RSD parameters; and accounting for the measurement binning --- change the BAO constraints compared to previous methods?}
    \item{Does the choice of MCMC sampler used to perform the fit impact the constraints?}
\end{itemize}
Combining all of these effects together, we will arrive at a single consensus value for the modelling systematic error within the DESI BAO analyses, to be combined with the theoretical systematic error from earlier.

\subsection{Simulations}
For all the numerical tests in this work, aimed towards motivating a modelling systematic error budget for DESI, we use a suite of 25 high precision measurements of the power spectrum and correlation function. These are obtained from \textit{control variate} realisations of a set of cubic simulations populated with a set of galaxies mimicking the clustering of the DESI LRG sample. Here, we give a brief summary of the relevant details of these simulations, and refer the reader to our companion papers, \citet{ELG_HOD_Systematics} and \citet{LRG_HOD_Systematics} for a more complete description.

Our base cubic simulations, the AbacusSummit\footnote{\url{https://abacussummit.readthedocs.io/en/latest/}} suite \citep{Maksimova2021} are a set of over 140 simulations across 97 different cosmological models, each with a box size $(2\Gpc)^{3}$ containing $6192^{3}$ particles; for an effective particle mass of $2\times10^{9}h^{-1}\,M_{\odot}$. In this work, we restrict to the `cosm000', \citealt{PCP18} $\Lambda$CDM-based cosmological model, \footnote{Specifically, with $\omega_{\mathrm{b}}=0.02237$, $\omega_{\mathrm{c}}=0.12$, $h=0.6736$, $A_{\mathrm{s}}=2.083\times10^{-9}$, $n_{\mathrm{s}}=0.9649$ and a single massive neutrino with mass $m_{\nu}=0.06\,\mathrm{eV}$.} and a single simulation snapshot with redshift $z=0.8$, which corresponds closely to the peak of the DESI LRG sample number density. The simulations are populated with the `vanilla' Halo Occupation Distribution model described in \citet{LRG_HOD_Systematics}, leading to a sample with constant number density $\sim1\times10^{-3}h^{3}\,\mathrm{Mpc}^{-3}$. Finally, we apply a reconstruction algorithm to these data, following \textbf{RecSym} convention, and other algorithmic choices presented in \citep{OptimalRecon1, OptimalRecon2}.

From these, a set of more precise clustering measurements is obtained using the \textit{control variate} (CV) technique \citep{Chartier21,Ding2022,Kokron22,DeRose23}. The CV technique works by subtracting off the intrinsic sample variance in simulated cosmological observables shared with highly correlated but more cheaply computable surrogate observables---in the case of reconstruction, \citet{Hadzhiyska23} showed that a reasonable such surrogate is the linear 2-point function computed from the same initial conditions.\footnote{That this is a good choice can be further understood from \S\ref{sec:recon} that shows that the reconstructed density, particularly in \textbf{RecSym}, has its displacement terms reduced to $(1 - \mathcal{S})\ast\Psi^{(1)}$. This displacement term is what leads to most of the decorrelation with the initial conditions.} We refer the interested reader to that work for further details, particularly as pertains to the implementation of CV in the context of DESI and AbacusSummit.

The uncertainties on our clustering measurements are obtained using a set of 1000 approximate EZmock simulations \citep{Chuang2015} with the same volume, particle number density, and HOD as the AbacusSummit simulations. Reconstruction is applied to these approximate simulations in the same way as for the AbacusSummit realisations. Validation of these covariance matrices and comparisons with other approaches for estimating them can be found in \citet{analytic_covariances,covariance_comparison,semianalytic_covariance}. Because we use a covariance matrix estimated from a finite number of simulations, we also de-bias the inverse covariance matrix and account for noise in its estimation following \cite{Hartlap2007,Percival2014,Percival2022}.

\begin{figure*}
    \centering
    \includegraphics[width=\textwidth]{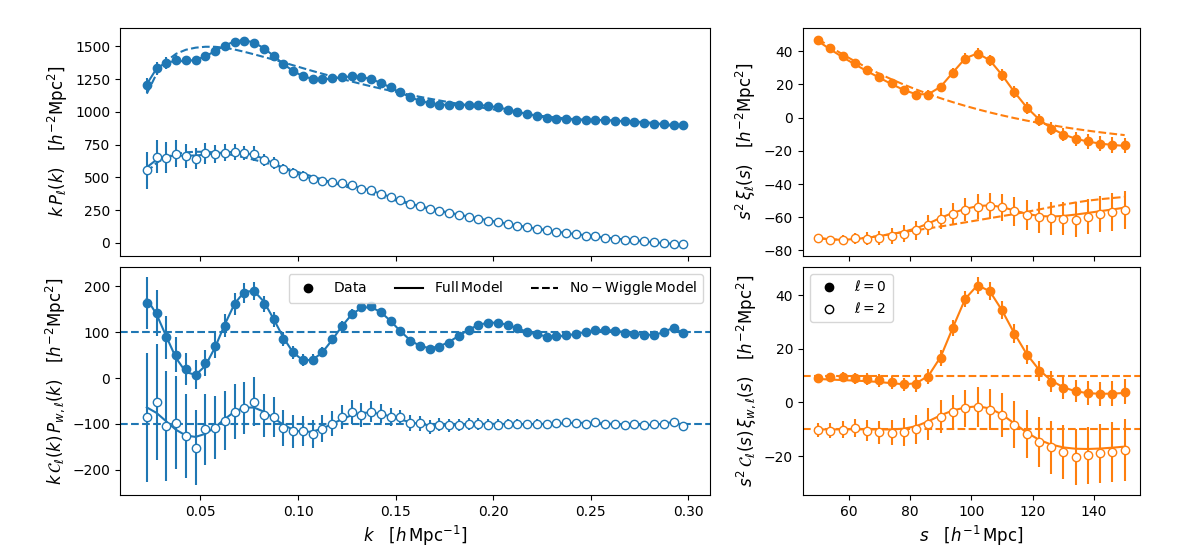}
    \caption{The mean clustering signal, taken from 25 control variate simulations of DESI's LRG sample, used for all the modelling tests in this work. In all panels closed/open points show the mean measured monopole and quadrupole clustering respectively, and error bars are those for a single realisation (i.e., not the error on the mean). Left panels are the power spectrum and right are the correlation function. For both, the top panels show the full signal, including the best-fitting full- and no-wiggle models (solid/dashed lines respectively) using our fiducial fitting methodology (see Section~\ref{sec:results}). The bottom panels show only the remaining wiggle component of the \edit{multipole} data/models after subtracting the no-wiggle model \edit{(i.e., showing only the second term on the right-hand side of Eq.~\ref{eqn:BAO_model}), multiplied by $k$, and arbitrarily offset by $\pm100$ ($\pm10$) for the power spectrum (correlation function) for visualisation purposes}.}
    \label{fig:simulations}
\end{figure*}

Figure~\ref{fig:simulations} shows the mean power spectrum and correlation function of the 25 realisations along with errorbars for a single realisation and our best fit fiducial BAO model (see \S\ref{sec:results}). This figure shows the extremely precise set of measurements available for us to test our modelling assumptions --- and when fitting the mean of 25 realisations, we further reduce the covariance matrix by a factor 25 compared to that shown in the figure to represent the error on the mean. When fitting a single of the 25 individual LRG realisations, the resulting precision on the BAO constraints is $\sim 0.3\%$ on $\alpha_{\mathrm{iso}}$ and $\sim 1\%$ on $\alpha_{\mathrm{ap}}$. Fitting the mock mean, with reduced covariance, gives a precision of $\sim 0.1\%$ on $\alpha_{\mathrm{iso}}$ and $\sim 0.5\%$ on $\alpha_{\mathrm{ap}}$.  For comparison, the expected aggregate precision on the BAO constraints for $0.0 < z < 1.1$ for DESI Y1 and the full DESI sample are $0.51\%$ ($1.78\%$) and $0.24\%$ ($0.68\%$) for $\alpha_{\mathrm{iso}}$ ($\alpha_{\mathrm{ap}}$) respectively \citep{DESISV}. Hence, a single one of our simulations is sufficient to investigate systematics at the level of $\sim2\times$ the precision of DESIs Y1 measurements, while averaging over all 25 realisations is sufficient to determine systematics for the full DESI sample. 

\subsection{Numerical implementation in Barry}
\label{sec:barry}

To test the various theoretical systematics and modelling choices identified above, we use the modular BAO-fitting code \textsc{barry} \citep{Hinton2020}. This code is designed to be fully modular, easily enabling comparative fits to the same data when different modelling choices are made. This is done by specifying a series of unique model and dataset pairs in one script, which are then submitted automatically as unique jobs to a supercomputing cluster. \cite{Hinton2020} used this architecture to study the impact of different perturbation-theory inspired models on the isotropic BAO signal \citep{Seo2016, Beutler17, Ding18, Noda2020, Chen21}, finding excellent agreement between the different models. However, as demonstrated in Section~\ref{sec:recon}, we find that a more nuanced and complete treatment of theory leads to a single clear choice of reconstruction convention (\textbf{RecSym}), and a single BAO damping term. We hence fix our model to the one presented in Section~\ref{sec:broadband}. All of the remaining modelling choices outlined at the start of this Section can be investigated by simple changes to the input parameters within \textsc{barry}.

To facilitate this, significant improvements have been made to the code relative to that presented in \cite{Hinton2020} which we will now describe. For all other base aspects of the code, we refer the reader to this reference. In contrast to \cite{Hinton2020}, the latest version of \textsc{barry} implements anisotropic BAO fitting as default, reverting back to a standard isotropic method only when specified for particular model-dataset pairs.

It is important to note that an `isotropic' BAO fit, both within \textsc{barry} and in general, is not the same as running an `anisotropic' fit to only the monopole with fixed $\epsilon=0$. The reason for this is due to the impact of the survey window function. Both the window function convolution and modelling of wide-angle effects are implemented via matrix multiplications as described in \cite{Beutler21}, which also allows the fitting of the odd-multipoles of the power spectrum if they are provided. Even when fitting only the monopole, the presence of a survey window function means that power leaks from higher-order multipoles into the monopole and vice versa, requiring the full set of multipoles to be modelled and the window function for all even and odd multipoles to be used. When an isotropic fit is performed, the higher-order multipoles are assumed to be zero and the part of the window function responsible for describing the leakage of power between these and the monopole is ignored. This is a fundamental difference in the modelling that is retained in \textsc{barry} for historical reasons and comparison to results from previous surveys, but in this work, we only consider the correct `anisotropic' approach of modelling all higher-order multipoles regardless of the exact parameters, scales or data multipoles being fit.

A second feature that is important to highlight is the ability to analytically marginalise over the broadband terms during the BAO fitting. As these enter at linear order, such marginalisation can be done fully (by modifying the likelihood used during MCMC fitting or optimisation) or partially (by using a standard Gaussian likelihood and finding the best-fit broadband terms at each step of the fitting). Appendix~\ref{app:marg} discusses this further and demonstrates the consistency between these approaches. All results presented hereon use full analytic marginalisation, which we argue is the formally correct choice, although partial analytic marginalisation is retained as an option in the code for comparison to previous works which used this approach (i.e., \citealt{Ross17, Bautista21}).

The full code used to perform all the tests in this work is available at \url{https://github.com/Samreay/Barry}.

\section{results}
\label{sec:results}

This section presents the results of our various systematic modelling tests. Unless otherwise stated, we fit the post-reconstruction data with Gaussian priors on the BAO and FoG damping parameters, $\Sigma_{\perp}=\mathcal{N}(2,1)$,  $\Sigma_{||}=\mathcal{N}(5,2)$,
$\Sigma_{s}=\mathcal{N}(2,2)$ on scales $0.02\kMpc \le k \le 0.30 \kMpc$ and $50\Mpc < s < 150\Mpc$ for the power spectrum and correlation function respectively. Both the linear galaxy bias and RSD parameters are allowed to vary within wide flat prior ranges. We dilate only the wiggle component of the model, and the FoG damping applies only to the smooth component. The wiggle/no-wiggle split is performed using the method of \cite{Wallisch2018} and all our fits use the Nautilus sampler \citep{Lange2023}.

\subsection{Validating the spline-based broadband model}
\label{sec:spline}

Our first set of tests validates the new spline-based broadband method. First, we test the dependence of our fit results to the smoothing scale $\Delta$. Intuitively, splines with too small a $\Delta$ will be able to reproduce the BAO signal, while splines with too large a $\Delta$ will lack sufficient freedom to fit the broadband. This is especially clear in configuration space where the former corresponds with $r_{\rm min}$ overlapping the BAO peak, while the latter corresponds with $r_{\rm min}$ extending too far out from the BAO peak. Fortunately, no fine tuning in $\Delta$ is required as fits using the spline method are very stable over a range of $\Delta$ until the BAO scale is reached.

\begin{figure}
    \centering
    \includegraphics[width=0.45\textwidth]{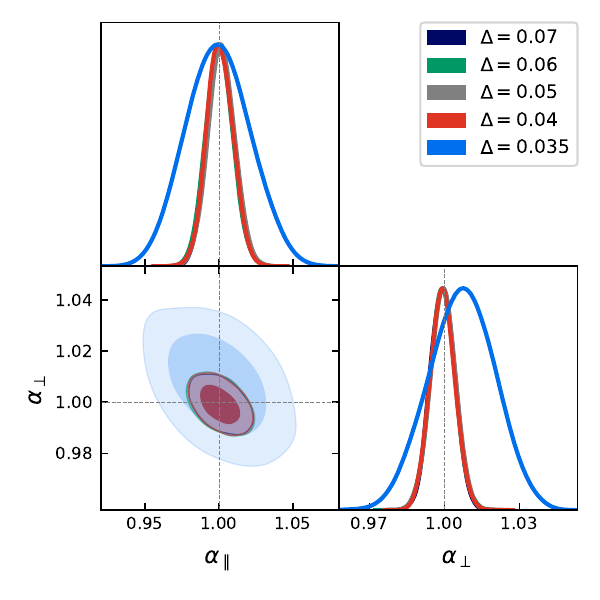}
    \caption{Fits to the mean power spectrum with varying levels of $\Delta$. The constraints on the BAO parameters $\alpha_{\parallel,\perp}$ are remarkably stable to $\Delta$ up to $k \sim 0.04 \kMpc$; at around $\Delta \sim 0.035 \kMpc$, just shy of $10\%$ above the Nyquist frequency, the constraints rapidly degrade.}
    \label{fig:delta_test}
\end{figure}

Figure~\ref{fig:delta_test} shows fits to the mean LRG power spectrum with varying values of $\Delta$. The constraints on $\alpha_{\parallel,\perp}$ are extremely stable to a broad range of $\Delta$, and only begin to rapidly degrade around $\Delta \sim 0.035 \kMpc$ or roughly $10\%$ above the Nyquist frequency of the BAO wiggles. The central values of the fits shift by less than $0.1\%$ in all cases and the best-fit $\chi^2$ changes only by around the amount implied by the increase in free spline parameters with support within the fitting range ($k_{\rm max} = 0.3 \kMpc$). Since there is no apparent bias due to using a slightly wider spacing $\Delta$ but a clear degradation of the overall signal well before the Nyquist frequency is reached we pick roughly twice this frequency, $\Delta = 0.06 \kMpc$, as our fiducial choice. At $k_{\rm max} = 0.3 \kMpc$ this corresponds to 7 free parameters per multipole.

\begin{figure}
    \centering
    \includegraphics[width=0.45\textwidth]{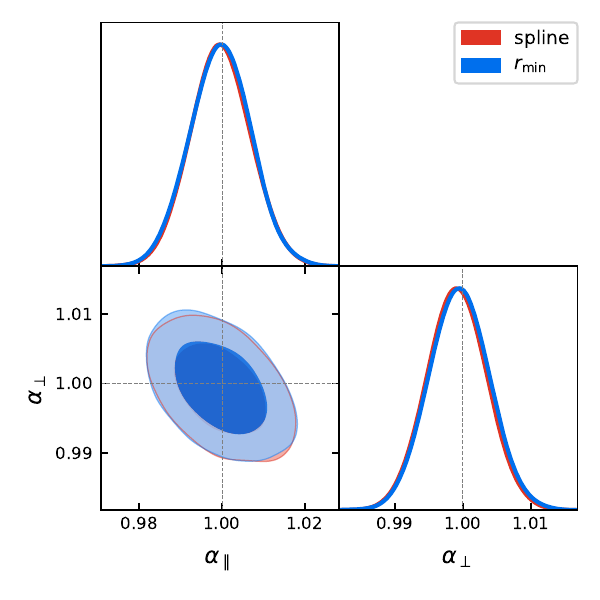}
    \caption{Fits to the mean correlation function using a wide scale range $0\ \Mpc < r < 130\ \Mpc$ and the full set of Hankel-transformed spline coefficients compared to when an $r_{\mathrm{min}}$ cut of $80 \Mpc$ is applied and only the $n=[0,1]$ spline bases in the quadruopole are used. The excellent consistency between the two demonstrates that our default method of fitting the BAO removing small scales and unnecessary spline components is sufficient.}
    \label{fig:config_spline_test}
\end{figure}

Next, we confirm that the Fourier-space broadband model is equivalent to setting $r_{\rm min} = \pi/\Delta$ with a minimal set of spline coefficients and large-scale polynomial terms. To do so we perform two fits: one where the spline broadband including $20$ bases with $\Delta = 0.06 \kMpc$ is Hankel transformed and the correlation function fit from $0 \Mpc < r < 130 \Mpc$ \edit{(summing from $n=-1$ to $n=19$ in Eq.~\ref{eq:corrspline2})}, and another where only the first two spline bases in the quadrupole with large-scale support in $\xi_2$ are kept \edit{(using only the $n=0$ and $n=1$ terms in Eq.~\ref{eq:corrspline2})}, fitting to $80 \Mpc < r < 130 \Mpc$, where $80 \Mpc$ is chosen as the $r_{\rm min}$ where the spline basis has essentially no support. The two constraints are shown in Figure~\ref{fig:config_spline_test}. The BAO constraints are essentially identical, up to percent level differences in constraining power, despite the radial fitting range differing by close to a factor of two. Both the means and variances of the posteriors on the linear theory amplitudes $b$ and $f$ also agree at the roughly $5\%$ level, which is notable given the very different radial ranges over which the data are fit. This suggests that these linear-theory amplitude parameters are mostly fitting the amplitude of the BAO only while sensitivity to the amplitude of the smooth component is removed either by $r_{\rm min}$ or by fitting the spline basis.

\begin{figure}
    \centering
    \includegraphics[width=0.45\textwidth]{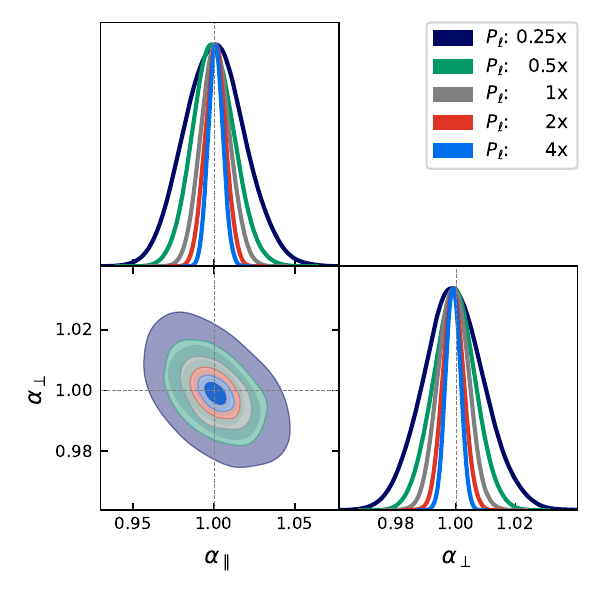}
    \includegraphics[width=0.45\textwidth]{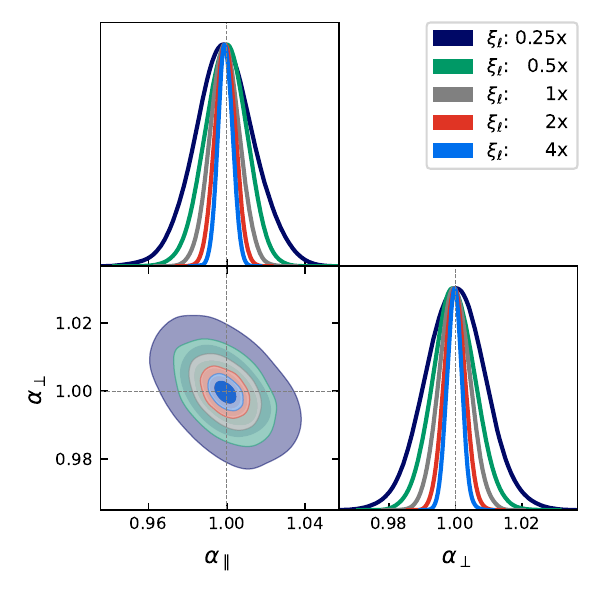}
    \caption{Fits to the mean power spectrum (top) and correlation function (bottom) when the covariance matrix is re-scaled between 0.25x and 4x. The excellent consistency of the constraints demonstrates that our default spline-basis never removes BAO information regardless of the SNR level.}
    \label{fig:cov_test}
\end{figure}

Finally, we would like to check that the fits are robust against various levels of noise. In order to perform this check we fit the data with the covariance re-scaled by factors of $A = 1/4, 1/2, 1, 2, 4$. This corresponds approximately to rescaling the volume by the inverse factor --- at $A=1/4$ and $A=4$ these rescalings correspond to more volume than all the DESI Y1 samples combined and less volume than BOSS CMASS. The results in Fourier and configuration space are shown in Figure~\ref{fig:cov_test}. The fits are remarkably stable across these rescalings, with the error bars on $\alpha_{\parallel,\perp}$ scaling as $\sqrt{A}$ to a very good approximation, demonstrating that the spline broadband does not sap BAO information out of the signal at any SNR level. Indeed, we have tested that this remains true even for a $16\times$ rescaling, beyond which the BAO scale can essentially only be constrained to order unity.

\subsection{Impact of BAO/FoG damping parameter choices}
\label{sec:damping}

The BAO damping parameters $\Sigma_{||}$ and $\Sigma_{\perp}$ play a key role in describing the shape of the measured BAO feature, which in turn impacts how well its peak position/wavelength can be determined. Furthermore, $\Sigma_{s}$ is designed to account for non-linear RSD in the broadband shape of the clustering but may interact with the BAO signal depending on how it is implemented. Previous studies \citep{Alam17,  Gil-Marin2020, Bautista21} have typically fixed the values of these parameters based on fits to a set of mocks, additionally demonstrating that fits to the data remain robust if different fixed values are used, or the parameters are freed up entirely. However, the treatment of $\Sigma_{s}$ has not always been consistent, and it is not clear to what extent the priors on the three $\Sigma$ parameters impact BAO constraints at DESI precision.

\begin{figure}
    \centering
    \includegraphics[width=0.5\textwidth]{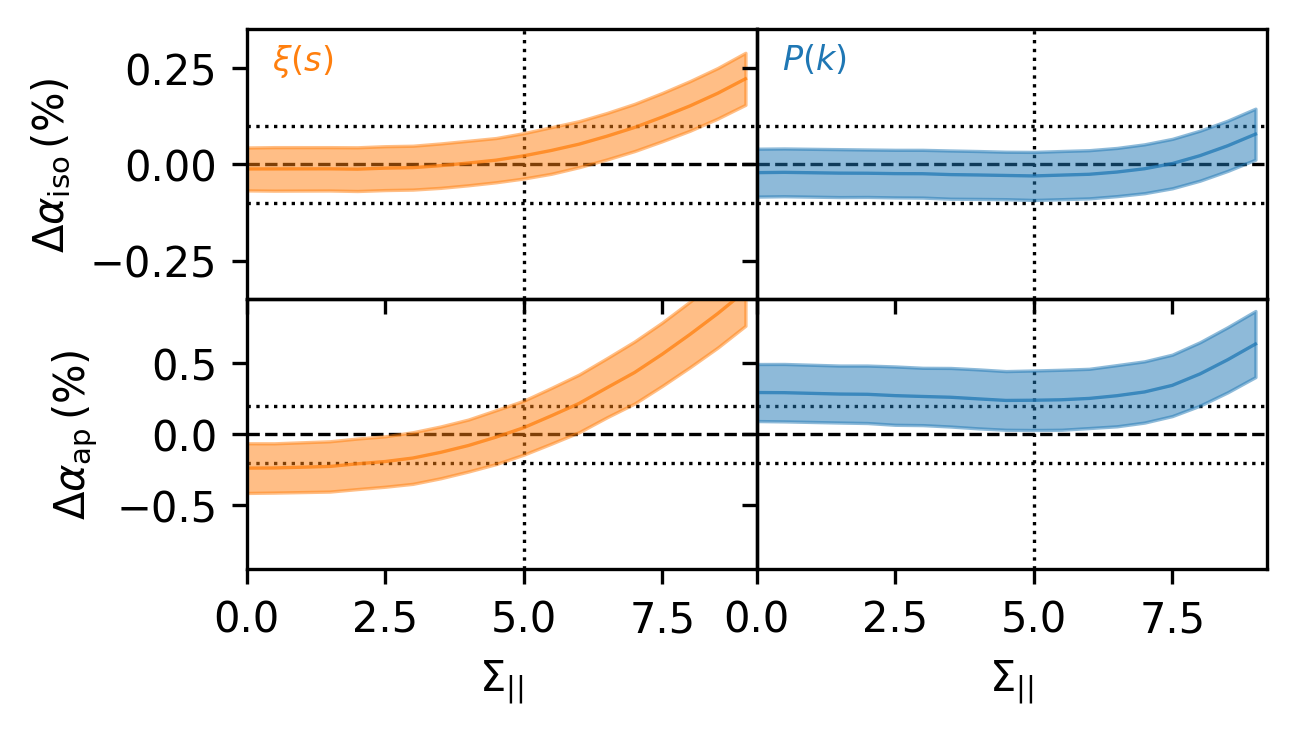}\\
    \includegraphics[width=0.5\textwidth]{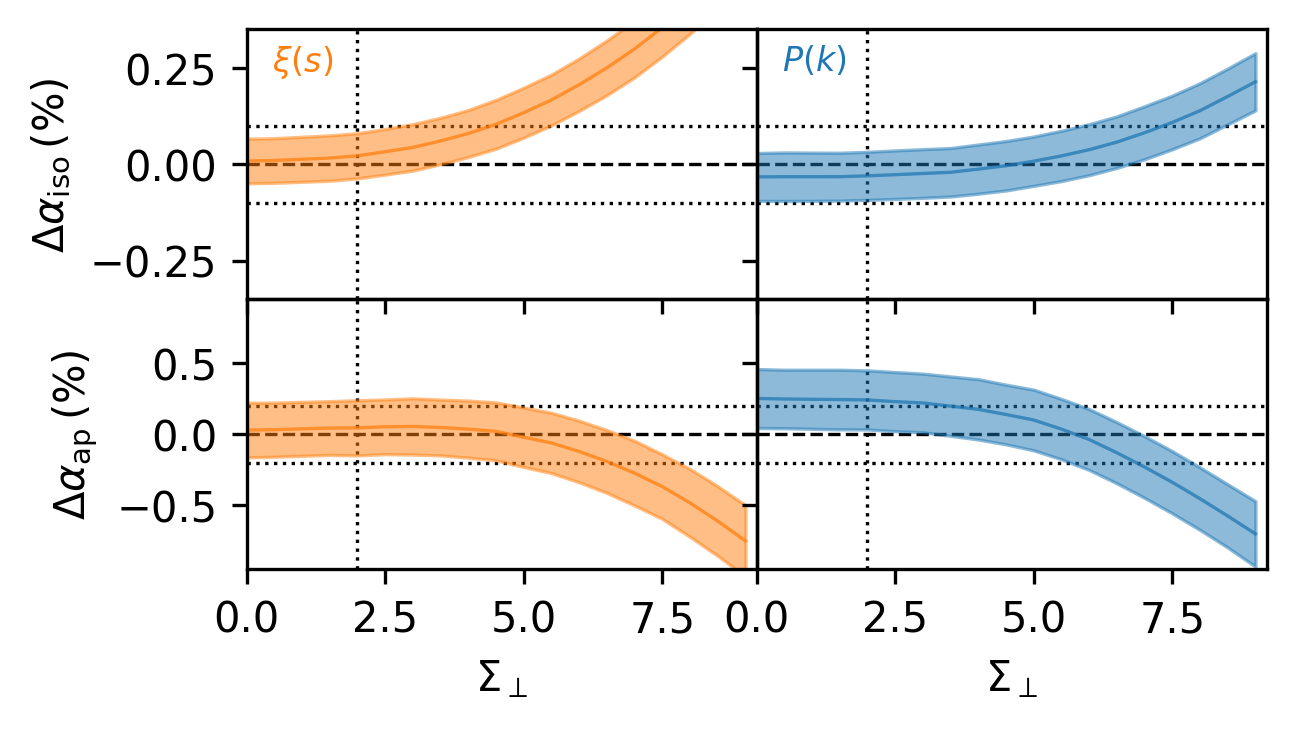}\\
    \includegraphics[width=0.5\textwidth]{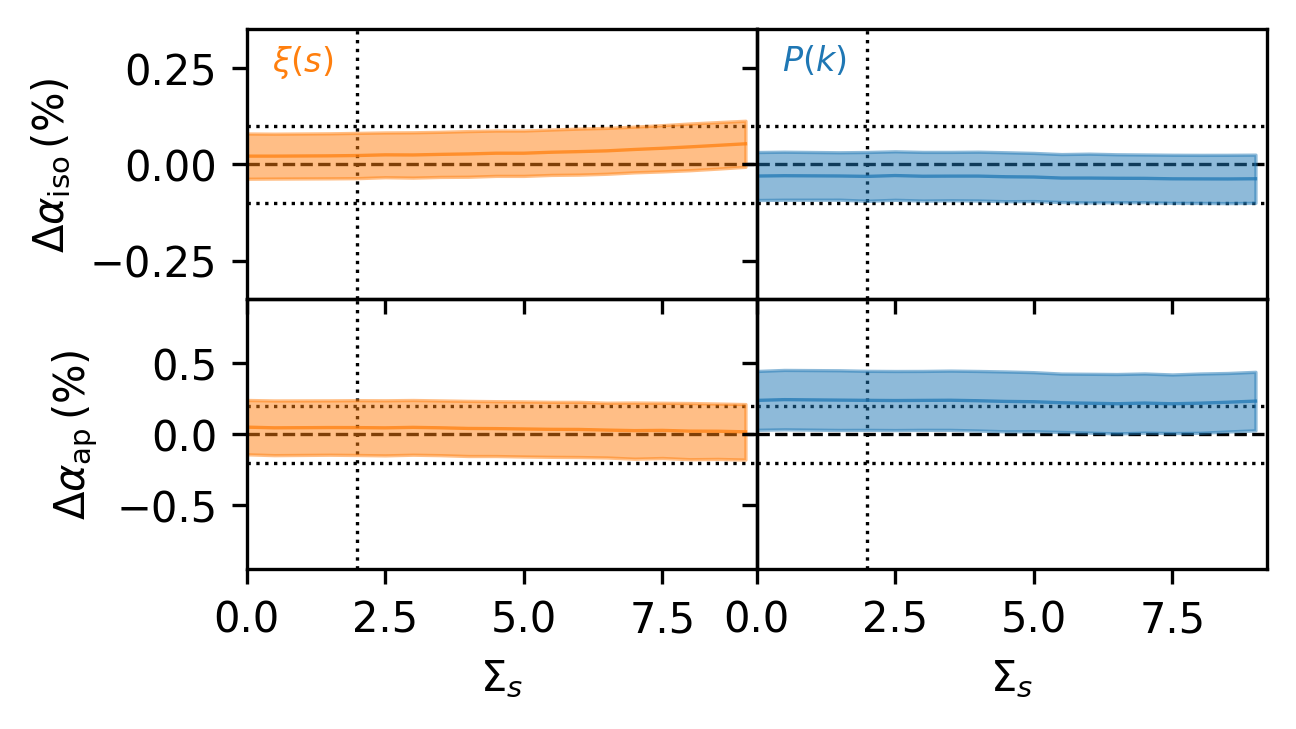}
    \caption{Percentage bias in the BAO dilation parameters as a function of the fixed damping scale used in the model when fitting the mean of the 25 LRG simulations. \edit{Each band shows the mean and standard deviation in the BAO parameters (relative to the expected value of 1) from fitting our mock mean measurements}. Left panels show fits to the correlation function, while right panels are fits to the power spectrum. In all cases, the vertical lines show our fiducial choice of damping scales, while the horizontal lines are placed at characteristic values of $0.1\%$ for $\alpha_{\mathrm{iso}}$ and $0.2\%$ for $\alpha_{\mathrm{ap}}$ to guide the eye. These results suggest that the BAO model can accommodate small inaccuracies in the fiducial choice when fixing the damping scales, but that larger-than-ideal biases occur if the fiducial value is more than $\sim 2\Mpc$ from the correct choice. This motivates allowing for some freedom in the damping scale, i.e., a Gaussian prior centred on the fiducial value with width $1-2\Mpc$.}
    \label{fig:sigmatests}
\end{figure}

To test, this we first investigate how fixing these to different values changes the BAO constraints. Figure~\ref{fig:sigmatests} shows the bias $\Delta \alpha_{\mathrm{iso}} = \alpha_{\mathrm{iso}}-1$ and similar for $\alpha_{\mathrm{ap}}$ when we fit the mock mean clustering. Our results show that fixing the choice of damping parameters is somewhat risky. While the constraints on the BAO dilation parameters are remarkably robust to moderate changes in the choices of damping scales, both the power spectrum and correlation function show systematic variation of increasing bias in the $\alpha$'s when the BAO damping $\Sigma$'s are increased significantly --- biases of order $0.1\%$ and $0.2\%$ for $\alpha_{\mathrm{iso}}$ and $\alpha_{\mathrm{ap}}$ respectively occur when the $\Sigma$ values are changed by more than $\sim 2\Mpc$ from their fiducial values. Similar trends have been identified in previous works, see e.g., Table B2 in \cite{Ross17}. We see no notable impact from varying $\Sigma_{s}$ on either the correlation function or power spectrum, as expected given our default model decouples it from the wiggle power spectrum.

The bias is more pronounced for the combination of $\alpha_{\mathrm{iso}}$ with $\Sigma_{\perp}$ than with $\Sigma_{||}$, while $\alpha_{\mathrm{ap}}$ shows a complex behaviour from variations in either damping parameter. This can be understood from theoretical considerations --- Taylor expanding the full BAO damping term in Eq.~\ref{eqn:propagator} and integrating with respect to the $\ell=0$ and $\ell=2$ Legendre polynomials allows us to write effective BAO damping scales for the monopole and quadrupole as 
\begin{align}
\Sigma^{2}_{\ell=0} &= \frac{1}{1+2\beta+\frac{3}{5}\beta^{2}}\biggl[\left(1+\frac{6}{5}\beta + \frac{15}{35}\beta^{2}\right)\Sigma^{2}_{||} + \left(2+\frac{4}{5}\beta+\frac{6}{35}\beta^{2}\right)\Sigma^{2}_{\perp}\biggl], \\
\Sigma^{2}_{\ell=2} &= \frac{1}{2\beta+\frac{6}{7}\beta^{2}}\biggl[\left(1+\frac{12}{7}\beta-\frac{5}{7}\beta^{2}\right)\Sigma^{2}_{||} - \left(1-\frac{2}{7}\beta-\frac{1}{7}\beta^{2}\right)\Sigma^{2}_{\perp}\biggl],
\end{align}
where $\beta=f/b$ is the linear RSD parameter. Information on $\alpha_{\mathrm{iso}}$ arises primarily from the monopole and so bias in this parameter actually occurs when variations in $\Sigma_{||}$ or $\Sigma_{\perp}$ cause a change in $\Sigma_{\ell=0}$. For the LRG mocks used here, where $\beta\approx0.35$, the same constant change applied to $\Sigma^{2}_{\perp}$ has a $\sim1.5\times$ greater impact on the damping scale in the monopole, compared to the same change applied to $\Sigma^{2}_{||}$. This is consistent with the larger bias seen in $\alpha_{\mathrm{iso}}$ when $\Sigma_{\perp}$ is varied compared to $\Sigma_{||}$. Conversely, the damping scale in the quadrupole is more affected by changes in $\Sigma_{||}$, but can also become too small if $\Sigma_{\perp}$ is increased enough at fixed $\Sigma_{||}$. As the quadrupole is the primary source of information on $\alpha_{\mathrm{ap}}$, considering variations in $\Sigma^{2}_{\ell=2}$ can explain the more nuanced behaviour of the $\alpha_{\mathrm{ap}}$ as a function of both $\Sigma_{||,\perp}$. We note that the dependence on $\beta$ means that the picture will be slightly quantitatively different for other tracers, though qualitatively similar.

Overall, this suggests that \textit{some} freedom should be given to the BAO damping parameters. We recommend using a Gaussian prior, centred on theory calculations such as in Fig.~\ref{fig:damping_scale_nonlins} or from fitting simulations, but with width $1-2 \Mpc$ to allow for the presence of non-linearities not captured by theory or deviations between the mocks and data which may otherwise cause small $0.1-0.2\%$ biases. When fitting the damping scale to simulations we advocate for fitting the BAO in high-precision simulated mocks (e.g. control-variate DESI mocks) using Equation~\ref{eqn:propagator} with the cosmology set to ``truth'' (i.e. $\alpha_{\parallel,\perp}=1$) --- previous works often instead fit to the ``propagator'', i.e. the correlation between the initial conditions and (reconstructed) galaxy density, but we note that this quantity is not strictly the same as the BAO damping since it is not cutoff at the BAO scale as in Equation~\ref{eqn:zel_damping}.

An alternative is to allow the BAO damping parameters to freely vary between wide flat priors such that the values are purely informed by the data. However, in other tests as part of this work, on more weakly constraining simulations than the control variates used here, we found that this generally reduces constraining power and has the potential to bias the BAO constraints. In the literature, nearly all previous BAO studies have elected to fix the damping parameters; of the studies we referred to only \cite{Gil-Marin2020} tested varying these, finding consistent BAO constraints, but we caution that the combined BOSS+eBOSS LRG dataset used there is particularly well constraining. We also note that future studies could also investigate whether it is more useful to place priors on $\Sigma_{\ell=0}$ and $\Sigma_{\ell=2}$ rather than $\Sigma_{\perp}$ and $\Sigma_{||}$ as our results suggest that it is changing along these degeneracy directions that ultimately cause biases in the BAO scale.

\begin{figure}
    \centering
    \includegraphics[width=0.5\textwidth]{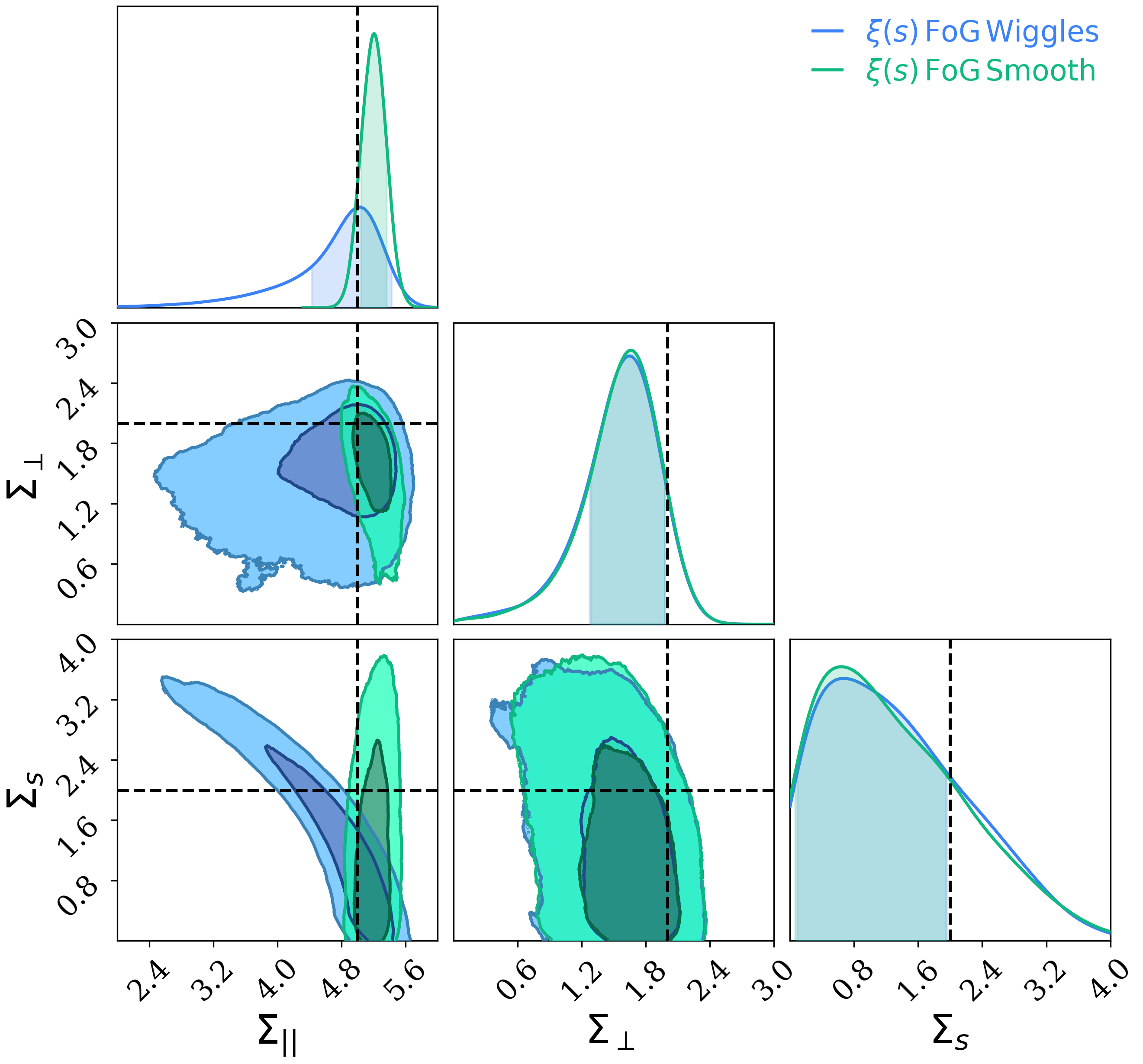}
    \caption{Constraints on the BAO and FoG damping parameters when fitting the mock mean correlation function with fixed $\alpha_{\mathrm{iso}}=\alpha_{\mathrm{ap}}=1$ and including the FoG damping in only the smooth model component ($\xi(s)$ FoG Smooth), or allowing it to multiply both the smooth and wiggle components ($\xi(s)$ FoG Wiggles). The dashed lines represent the central values we use for the priors on these parameters in our other fits. 
    Including the FoG damping in only the smooth model component is preferable as it removes a clear degeneracy between the BAO and FoG damping scales.}
    \label{fig:sigmascontour}
\end{figure}

Having determined a reasonable choice of prior for the damping parameters, we then test for any impact of including the FoG damping term in the wiggle component of the model as well as the smooth component, i.e., including the Lorentzian term from Eq.~\ref{eqn:lorentzian} in Eq.~\ref{eqn:propagator}. As a first demonstration of why we advocate for decoupling the FoG damping from the wiggle component of the model, we show in Fig.~\ref{fig:sigmascontour} that doing so effectively removes the degeneracy between $\Sigma_{||}$ and $\Sigma_{s}$  when fitting the mean of our mocks. This makes it far easier to enforce or determine suitable priors on these parameters in the manner described above and improves the convergence of the fitting. This breaking of the degeneracy occurs even when the $\alpha$'s are allowed to vary in concert with the damping parameters.

\begin{figure}
    \centering
    \includegraphics[width=0.5\textwidth]{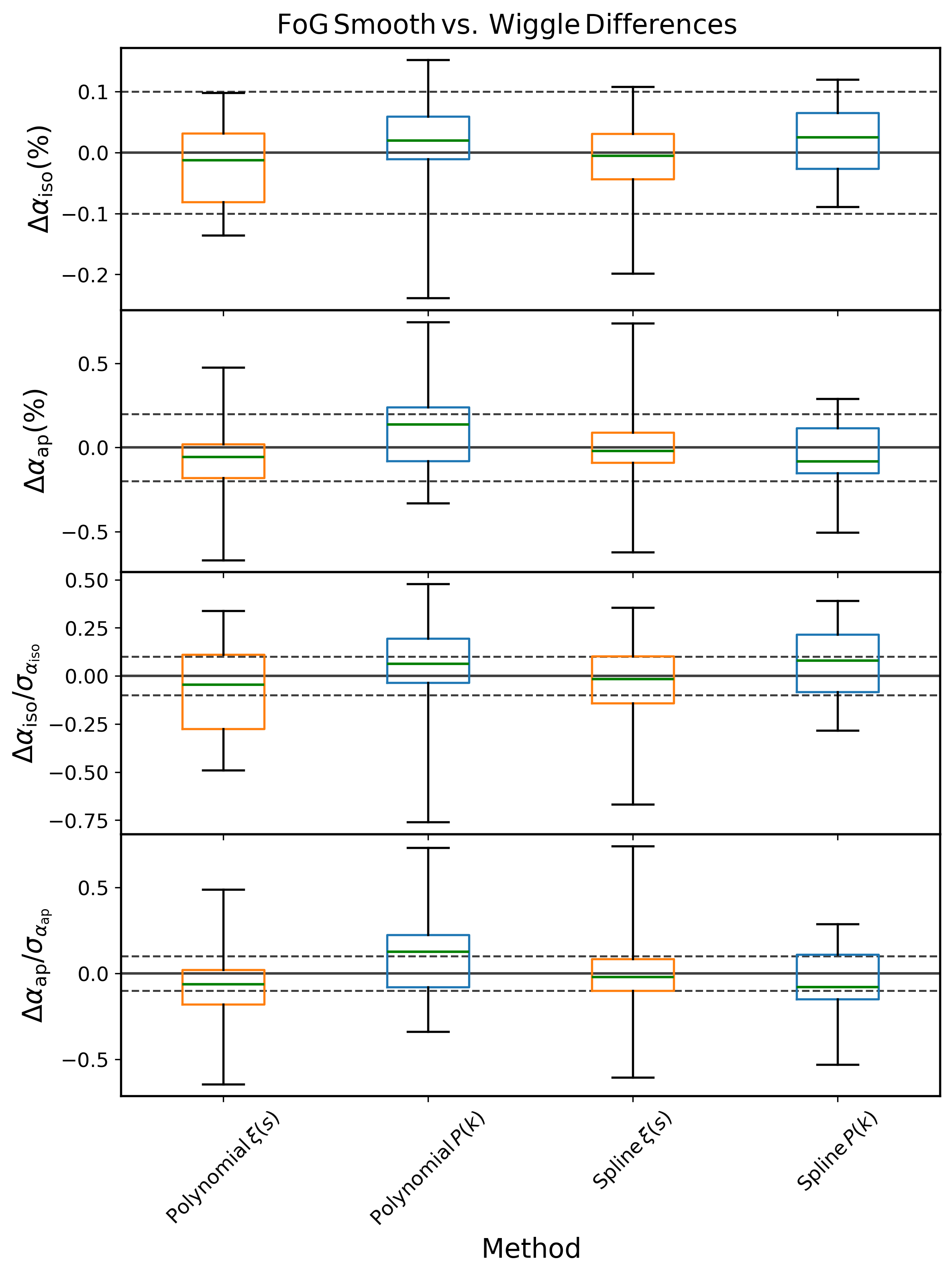}
    \caption{Difference in the BAO constraints between our 25 mock realisations fit with and without the FoG damping applied to the wiggle component in addition to the smooth component (without is our default). Each box-and-whisker plot shows the median difference over the individual mocks (\edit{green} line) the interquartile range (blue/orange boxes for power spectrum/correlation function respectively) and the largest individual differences (black error bars). These plots are shown for both the polynomial and spline-based broadband methods. The top two panels show the percentage differences between the same realisation, while the bottom two panels show the difference relative to the \textit{average} statistical uncertainty on the BAO parameter \edit{for a single realisation}. The horizontal lines are to guide the eye and are placed at 0.1 for $\Delta \alpha_{\mathrm{iso}}$ and the error-normalised differences, and $0.2\%$ for $\Delta \alpha_{\mathrm{ap}}$.}
    \label{fig:fog_test}
\end{figure}

Secondly, we look at the impact of this on the BAO constraints themselves by fitting each of our 25 individual mock realisations and then looking at the distribution of mock-to-mock differences in the $\alpha$'s for the two different modelling configurations. The results are shown in Fig.~\ref{fig:fog_test}, where we find that regardless of the choice of clustering statistic, or whether a polynomial-based or our new spline-based broadband model is used, the largest average difference is $\le 0.02\%$ and $0.11\%$ for $\alpha_{\mathrm{iso}}$ and $\alpha_{\mathrm{ap}}$ respectively. Albeit small, this latter difference is detected with $~2\sigma$ significance and so should be accounted for in our final systematic error budget. That said, the largest difference seen in any individual realisation is $\sim0.25\%$ ($0.75\sigma$) for $\alpha_{\mathrm{iso}}$ and $0.75\%$ ($0.75\sigma$) for $\alpha_{\mathrm{ap}}$, indicating that while on average the results are robust to the choice of where the FoG damping is included, this choice could lead to statistical fluctuations on individual datasets dependent on the noise in the data.

\subsection{Impact of different scale cuts}

\begin{figure}
    \centering
    \includegraphics[width=0.5\textwidth]{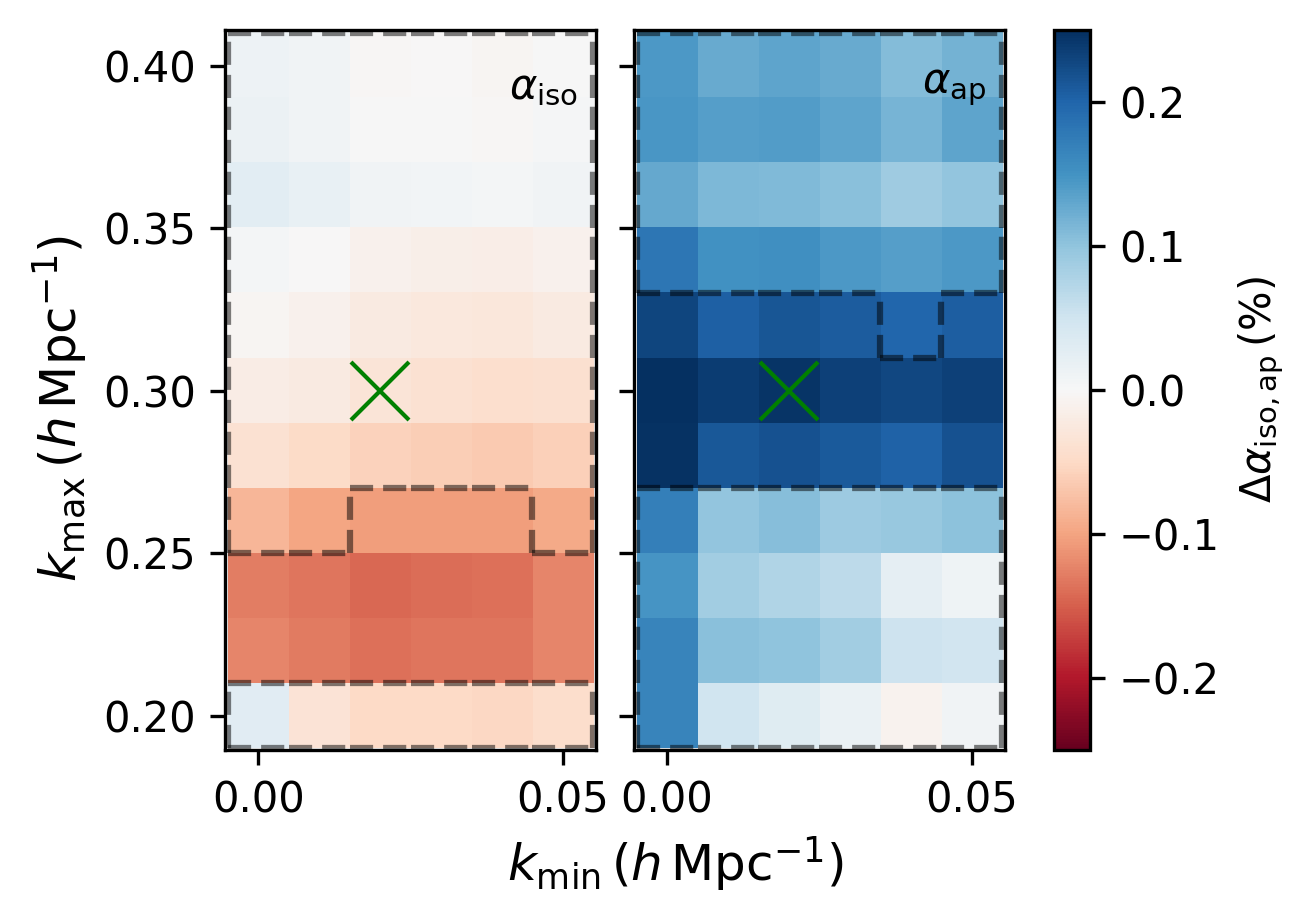}\\
    \includegraphics[width=0.5\textwidth]{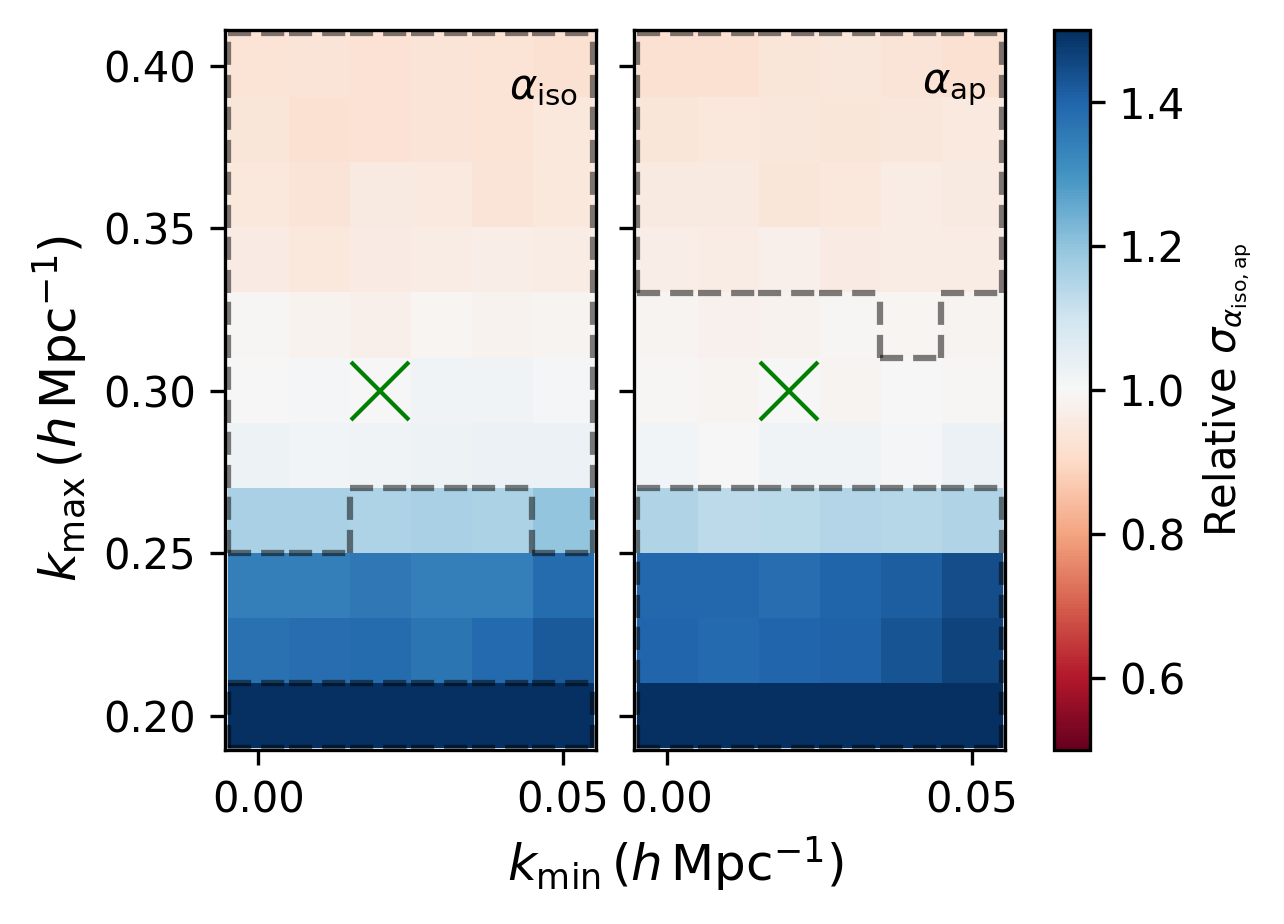}\\
    \caption{The impact of varying minimum and maximum scale cuts ($k_{\mathrm{min}}$ and $k_{\mathrm{max}}$) on BAO dilation parameters extracted from the mock mean LRG power spectrum. The top panel is coloured according to the percentage deviation of the fitted $\alpha$'s from unity. The bottom panel is the uncertainty of the $\alpha$'s relative to our fiducial fitting range (indicated by the green cross). In both cases, the left (right) panel is for $\alpha_{\mathrm{iso}}$ ($\alpha_{\mathrm{ap}}$). The regions enclosed by the dashed line in both of the left panels are where the deviation in $\alpha_{\mathrm{iso}}$ from unity is $<0.1\%$. For the right panels, it denotes a deviation in $\alpha_{\mathrm{ap}}$ of $<0.2\%$.}
    \label{fig:kminmax_test}
\end{figure}

\begin{figure}
    \centering
    \includegraphics[width=0.5\textwidth]{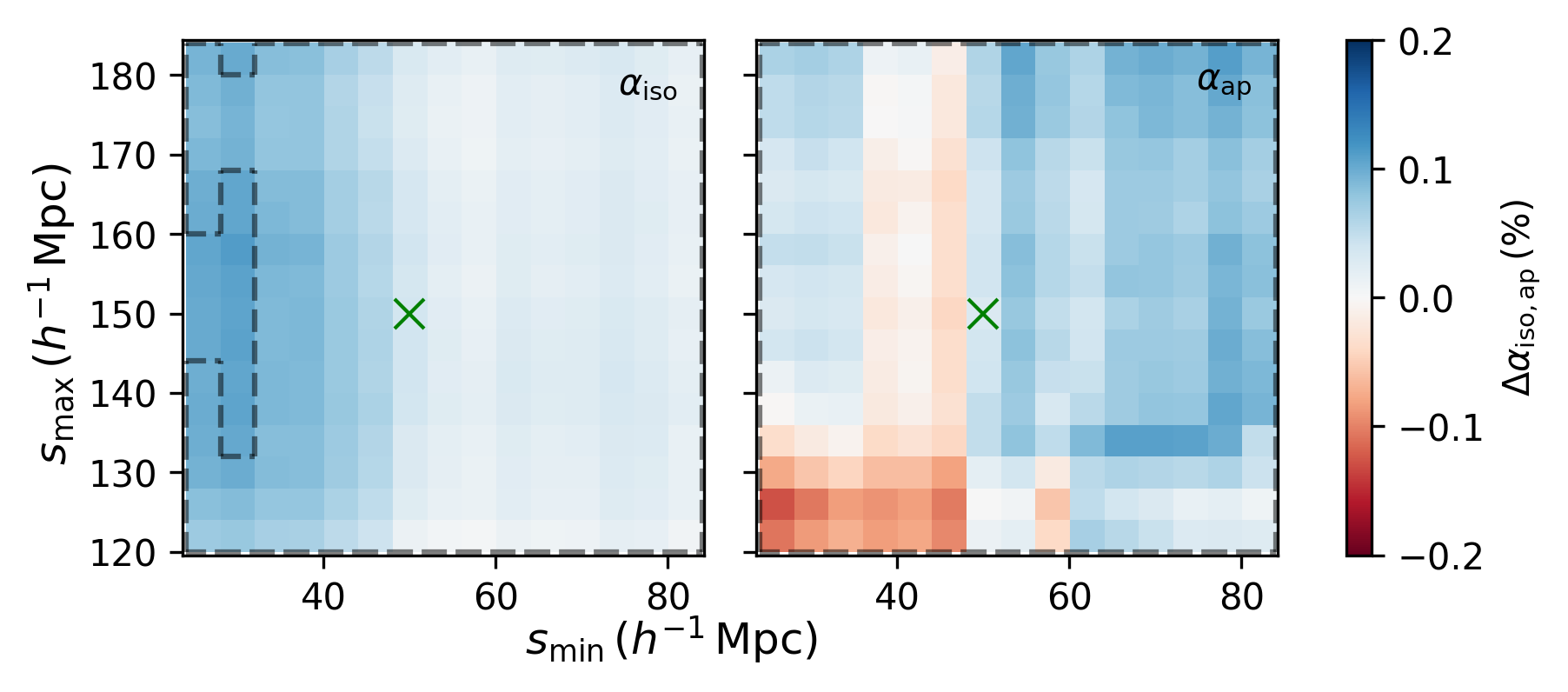}\\
    \includegraphics[width=0.5\textwidth]{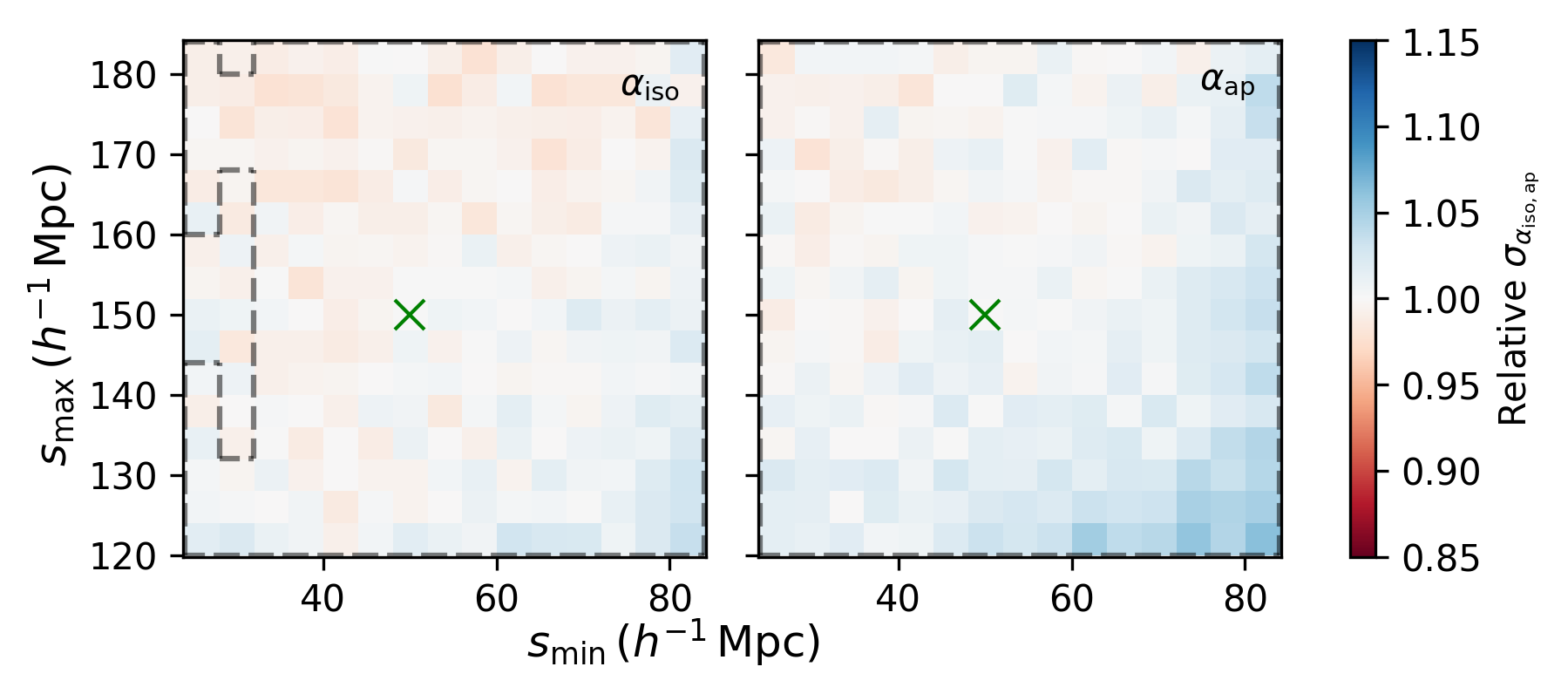}\\
    \caption{As for Fig.~\ref{fig:kminmax_test}, but showing BAO constraints from the correlation function when varying the minimum and maximum fitting scales $s_{\mathrm{min}}$ and $s_{\mathrm{max}}$.}
    \label{fig:sminmax_test}
\end{figure}

We next turn to verify the appropriate range of scales to include in the BAO fit. For both the power spectrum and correlation function we fit the mock mean clustering with a variety of choices for the minimum and maximum scale cuts and measure the bias in the dilation parameters, and their uncertainties relative to our default. These are shown in Figures~\ref{fig:kminmax_test} and~\ref{fig:sminmax_test} for the power spectrum and correlation function respectively.

For the power spectrum, we find that the difference from unity exceeds $0.1\%$ and $0.2\%$ for $\alpha_{\mathrm{iso}}$ and $\alpha_{\mathrm{ap}}$ respectively for a couple of narrow ranges of $k_{\mathrm{max}}$. The $k_{\mathrm{max}}$ at which this occurs is not the same for the two dilation parameters, and, more importantly, including even more non-linear scales does not make the bias worse. There is no significant trend of increasing bias when increasing/decreasing either $k_{\mathrm{min}}$ or $k_{\mathrm{max}}$. As such, we attribute these regions with slight deviations from unity to noise in the simulations, rather than a bias in the model --- looking at the measurements in Fig.~\ref{fig:simulations}, one can see a slight excess of power at around $k=0.29\kMpc$ which inconveniently coincides with our fiducial choice of scales. 

Turning to the uncertainty on the BAO parameters using the power spectrum, we do see a clear weakening of the constraints by up to $50\%$ when restricting the fitting to larger scales $k_{\mathrm{max}}=0.2 \kMpc$ and some small $10\%$ improvement when including smaller scales $k_{\mathrm{max}}=0.4 \kMpc$. There is also perhaps a very weak trend as $k_{\mathrm{min}}$ is increased and starts to encroach on the first BAO wiggle. It is worth reminding that as this is from a fit to the mean mock measurements, the uncertainties on the clustering measurements are significantly smaller than we expect for DESI DR1 and comparable to the aggregated BAO across $0.0 < z < 1.1$ for the full DESI survey. This exacerbates the trends in the uncertainties, and so we conclude that our fiducial choice of fitting scales $0.02 \kMpc < k < 0.30 \kMpc$ is reasonable --- it is not worth extending the fit to smaller scales for DESI DR1 or its upcoming DESI 2024 results. But, as we see no evidence of additional bias, this may be worth testing for the full DESI data.

For the correlation function, we find that the results are remarkably stable to the choice of scale cuts, with a small increase in bias but a decrease in error as $s_{\mathrm{min}}$ is reduced. The bias reaches $0.1\%$ in $\alpha_{\mathrm{iso}}$ for $s_{\mathrm{min}} < 28\Mpc$ across the range of $s_{\mathrm{max}}$ we test, and $\sim0.15\%$ in $\alpha_{\mathrm{ap}}$ when a maximum fitting scale $s_{\mathrm{max}} < 130 \Mpc$ is also applied. When the upper fitting range is reduced to this, there is enough freedom in the quadrupole model (particularly in its amplitude/BAO damping) that one starts to affect the BAO information if additional scales are not included on either side of the BAO feature with which to constrain the broadband model. Beyond this, any choice of fitting scale seems as good as any other, with $<10\%$ relative change in the uncertainties across all choices, so we opt to fix to our fiducial choice of $50 \Mpc < s < 150 \Mpc$.

\subsection{Impact of de-wiggling methodology}
\label{ssec:pnw}

\begin{figure}
 \includegraphics[width=0.5\textwidth]{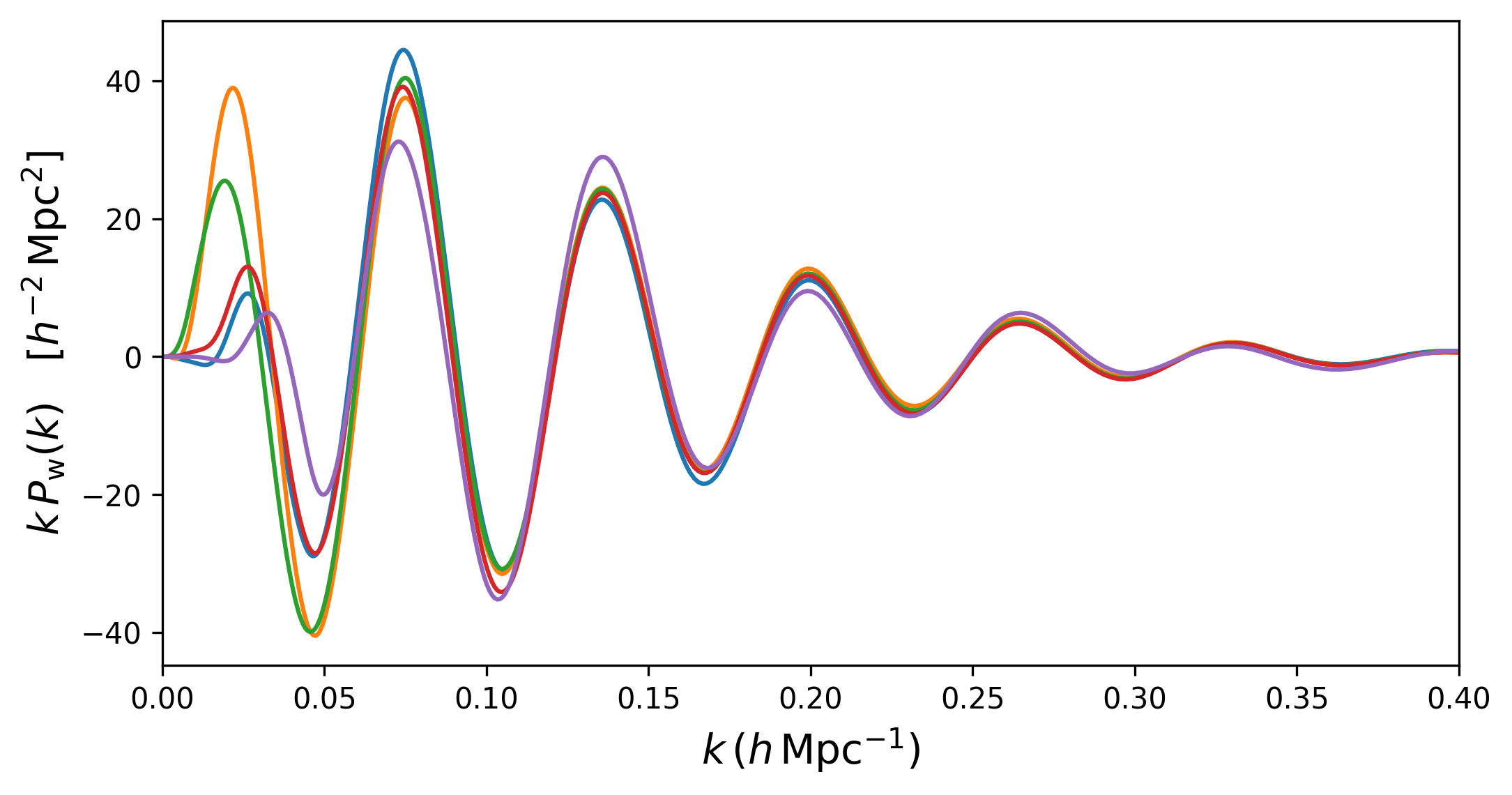}
 \includegraphics[width=0.5\textwidth]{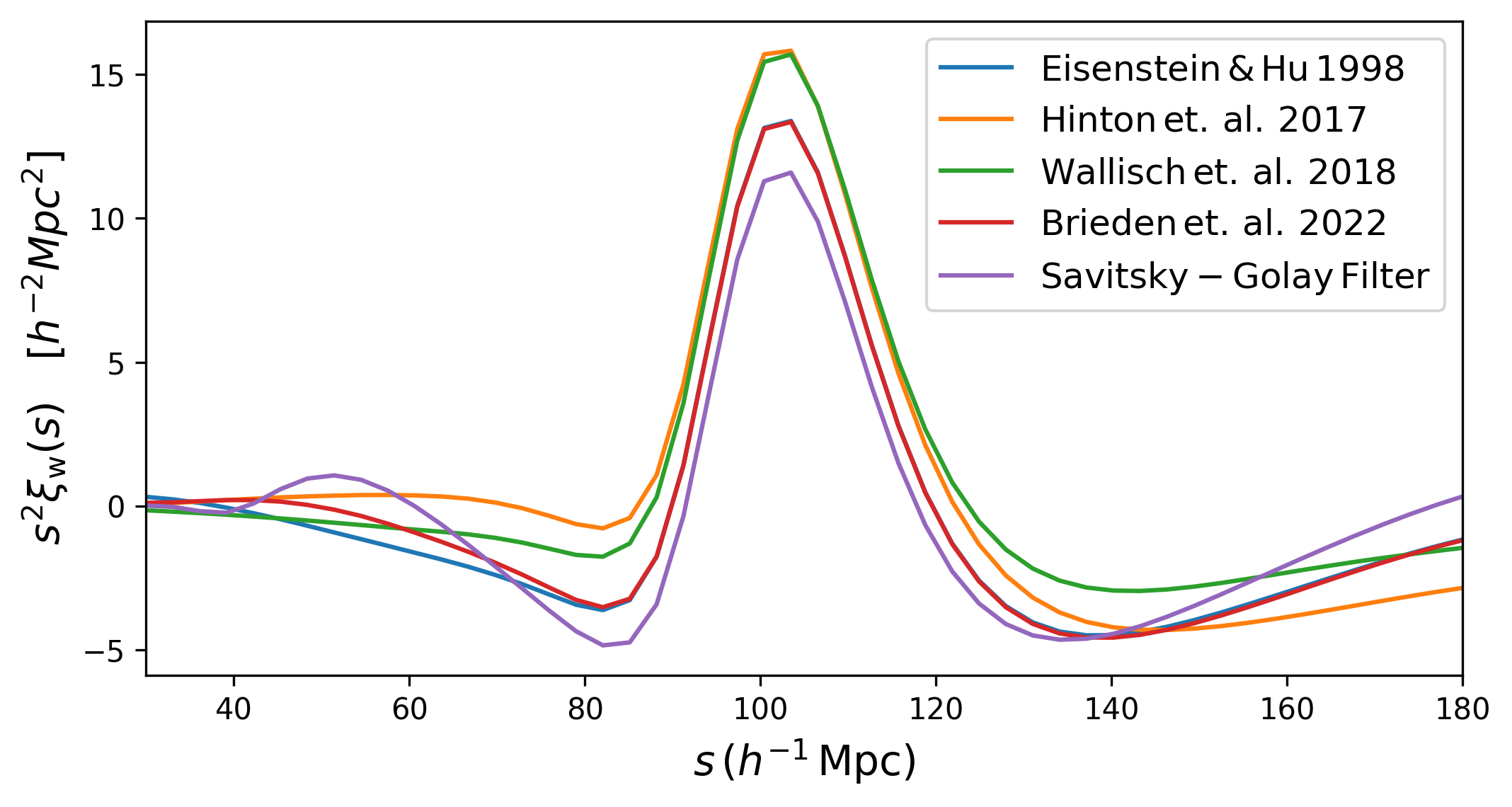}
 \caption{A comparison of the BAO template extracted using a variety of different literature approaches applied to the same linear power spectrum.}
 \label{fig:smoothtypes}
\end{figure}

When performing a BAO fit, a choice must be made of how to isolate the wiggle and no-wiggle components in the linear power spectrum. A number of different algorithms/approaches for this have been developed; using fitting formulae \citep{EH1998}, filters designed for signal processing (such as the Savitsky-Golay filter, B-splines, or a Gaussian filter as used in \citealt{Hinton2017}), or methods that numerically search for changes in the gradient in the power spectrum or its Fourier transform \citep{Wallisch2018, Brieden2022}. As shown in Fig.~\ref{fig:smoothtypes}, the wiggle component of the model arising from these different choices of algorithm can differ appreciably and, without a clear theoretical reason to prefer one over the other, the impact of this choice must be explored.

\begin{figure}
    \centering
    \includegraphics[width=0.5\textwidth]{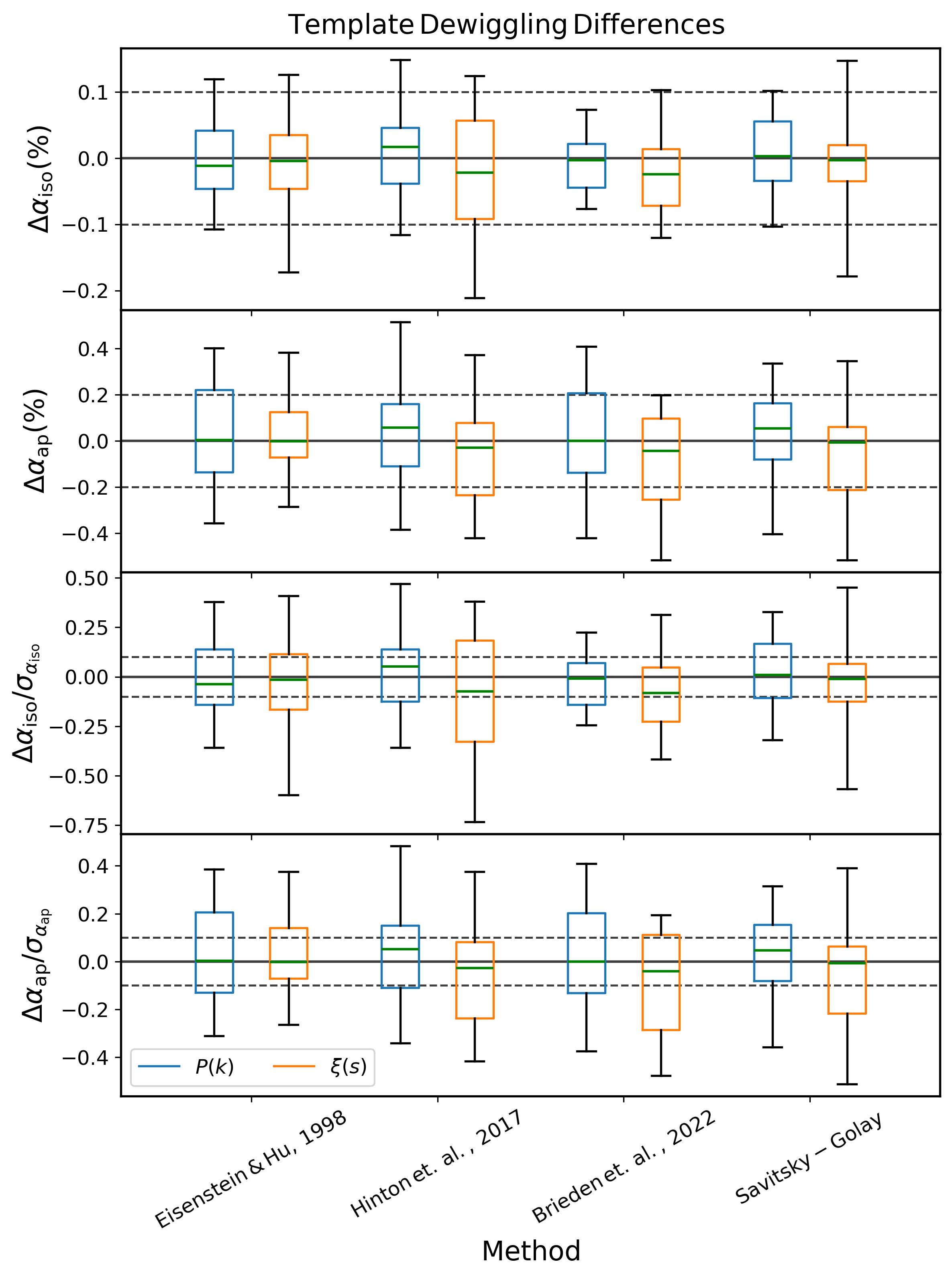}
    \caption{As for Figure~\ref{fig:fog_test}, but showing the differences in the BAO constraints between our 25 mock realisations using different BAO template extraction methods. In all cases, the template cosmology and linear power spectrum are the same and the differences are taken relative to our fiducial \protect{\citep{Wallisch2018}} method.}
    \label{fig:template_test}
\end{figure}

To do this, we follow the same procedure used previously and fit our 25 individual LRG mock realisations using each of the different template extraction methods. We then take the mock-to-mock difference between the constraints relative to our fiducial \cite{Wallisch2018} method. The results are shown in Fig~\ref{fig:template_test}. Despite the clear differences in the templates, the systematic difference in the BAO constraints is $<0.02\%$ ($<0.09\%$) for $\alpha_{\mathrm{iso}}$ ($\alpha_{\mathrm{ap}}$). However, as with the case where we investigate the impact of including the FoG damping in both the smooth and wiggle model components, individual realisations can show larger statistical variations when the template is varied --- up to $0.2\%$ ($0.75\sigma$) for $\alpha_{\mathrm{iso}}$, and $0.5\%$ ($0.5\sigma$) for $\alpha_{\mathrm{ap}}$. The correlation function appears slightly more sensitive to these variations than the power spectrum, consistent with the larger variations between templates over our fitting range seen in the correlation function of Fig.~\ref{fig:smoothtypes}.

\subsection{Impact of dilating the smooth model component or not}
\label{sec:dilate}

BAO fitting constraints arise from dilating the scales in the two-point clustering signal by the $\alpha$ parameters and then comparing the dilated model (Eq.\ref{eqn:BAO_model}) to the data. In our fiducial set-up, we dilate only the wiggle term $C(k,\mu)P_{w}(k)$, evaluating the remainder of the model at the undilated scales/angles $k$ and $\mu$. This ensures that constraints on the $\alpha$'s arise only from the wiggle-component. However, one could also apply the dilation to the smooth component of the model. Previous results in the literature have often taken this latter approach (e.g., \citealt{Beutler17,Ross17,Gil-Marin2020,Bautista21}).

Following the same approach as above, fitting the 25 individual LRG mocks with both approaches and looking at the distribution of differences in the BAO constraints, we find that the choice of whether or not to dilate the smooth model component in addition to the wiggles results in a systematic error of $<0.02\%$ ($<0.09\%$) for $\alpha_{\mathrm{iso}}$ ($\alpha_{\mathrm{ap}}$). The largest individual differences are $\sim0.15\%$ ($0.5\sigma$) for $\alpha_{\mathrm{iso}}$, and $0.75\%$ ($0.75\sigma$) for $\alpha_{\mathrm{ap}}$.

\subsection{Impact of different methods for transforming from the power spectrum to the correlation function}
\label{sec:pk2xi}

In this work, we provide a fully consistent/equivalent model for the correlation function and power spectrum, treating the former purely as the Hankel transform of the latter, and applying all our modelling choices and BAO dilation to the power spectrum. The new spline broadband method also ensures a consistent modelling between the two. However, some previous analyses \citep{Anderson14,Ross17} instead transformed the undilated power spectra, without galaxy bias, and applied both the BAO dilation and galaxy bias terms in configuration space. They also used two different galaxy bias terms for the two correlation function multipoles rather than including a physical angular-dependent RSD with free parameter $f$ or $\beta$. \edit{Eqs.~13-17 and accompanying text in \cite{Ross17} describe this process.}

\edit{We estimate the impact of the differences between these two methods by implementing both and fitting the individual LRG realisations using the spline-based broadband. W}e find a negligible systematic difference in $\alpha_{\mathrm{iso}}$, but a detection of a small difference of $0.12\pm0.03\%$ in $\alpha_{\mathrm{ap}}$. The fact that the two approaches differ mainly in their modelling of the quadrupole and the difference occurs in $\alpha_{\mathrm{ap}}$ indicates the presence of a real difference, not just noise. Although we believe that our new approach is preferable, we hence accommodate for this difference in our total systematic error on $\alpha_{\mathrm{ap}}$. On individual realisations, we find that the largest difference between the two methods is $0.04\%$ ($0.15\sigma$) for $\alpha_{\mathrm{iso}}$, and $0.4\%$ ($0.4\sigma$) for $\alpha_{\mathrm{ap}}$. Despite the small detection of systematic differences when averaging over all 25 realisations, this is consistent with the statistical differences on individual realisations seen from other modelling choices.

\subsection{Impact of different parameter sampling methods}

The last choice we investigate is less about the evaluation of the modelling for the BAO, but rather how the constraints on the $\alpha$'s are obtained numerically. \textsc{barry} exposes a number of well-known sampling algorithms with which one can fit the data and recover posterior samples (and possibly evidences) on the dilation parameters --- these include the publicly available packages \textsc{emcee}, \textsc{dynesty}, \text{zeus} and \textsc{nautilus} \citep{Foreman-Mackey2013, Speagle2020, Karamanis2021, Lange2023}, each with automated convergence checking.

\begin{figure}
    \centering
    \includegraphics[width=0.5\textwidth]{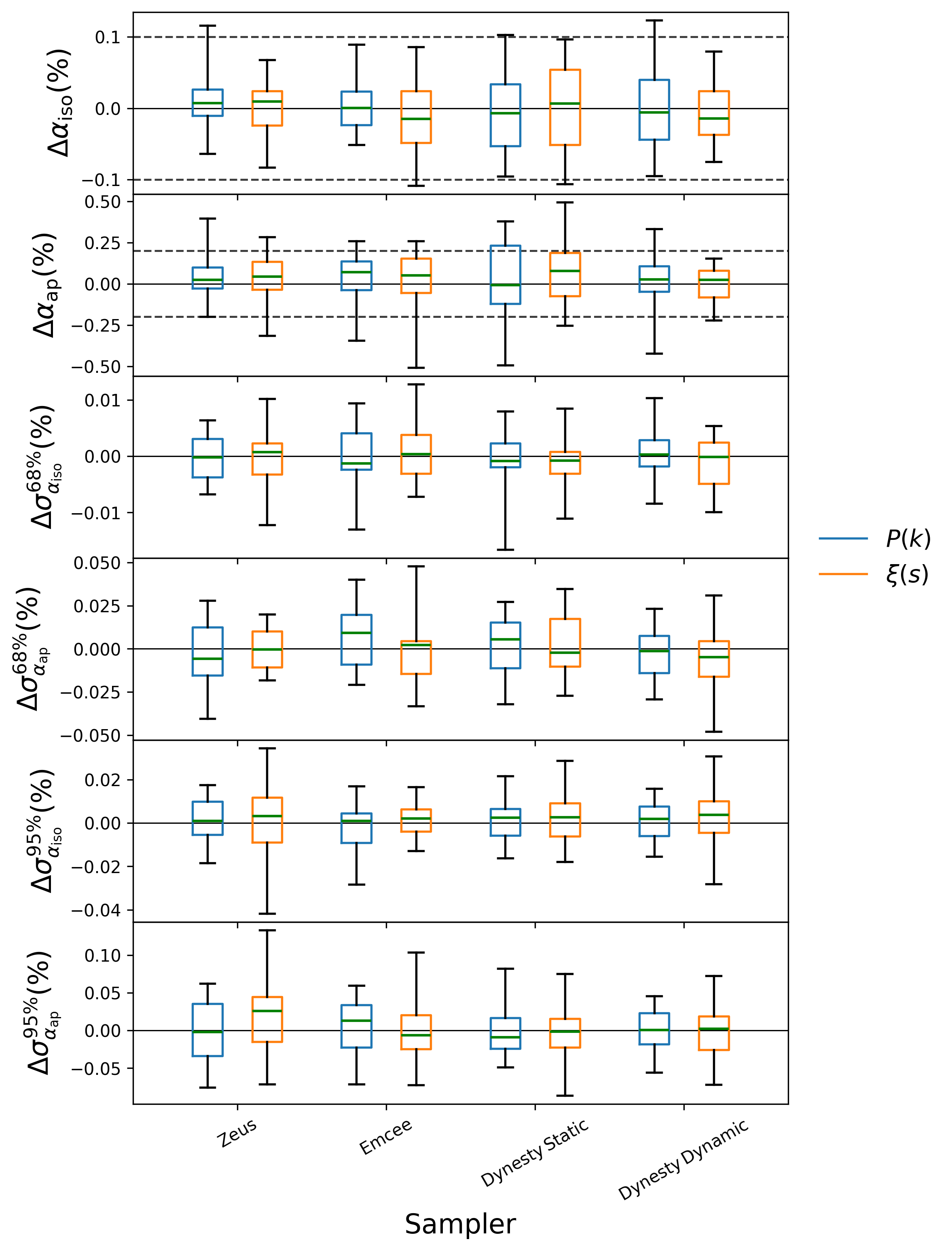}
    \caption{Difference in the BAO constraints between our 25 mock realisations fit with different MCMC/Nested samplers, relative to those obtained using \textsc{nautilus}. Each box-and-whisker plot shows the median difference over the individual mocks (\edit{green} line) the interquartile range (blue/orange boxes for power spectrum/correlation function respectively) and the largest individual differences (black error bars). The top two panels show the percentage differences between the same realisation, the middle two panels show the differences between the width of the 68\% confidence intervals (i.e., in the $1\sigma$ bounds), and the bottom two panels show the same for the 95\% confidence intervals. The horizontal lines are to guide the eye and are placed at 0.1 for $\Delta \alpha_{\mathrm{iso}}$ and $0.2\%$ for $\Delta \alpha_{\mathrm{ap}}$.}
    \label{fig:samplers_test}
\end{figure}

Fig.~\ref{fig:samplers_test} shows the marginalised posterior means on the $\alpha$'s for our individual realisations for each of the samplers we test, relative to those obtained using \textsc{nautilus}. We also investigate the width of the $68\%$ and $95\%$ confidence intervals on the $\alpha$'s. Given these samplers are all well tested and use reasonable automated convergence checks we do not expect and do not see any systematic differences. What is interesting to quantify however is that there is statistical differences between the posterior means for individual realisations depending on the choice of sampler --- up to $0.1\%$ and $0.5\%$ on $\alpha_{\mathrm{iso}}$ and $\alpha_{\mathrm{ap}}$ respectively. In all cases, we see much better convergence in the $68\%$ and $95\%$ confidence intervals than the posterior mean, consistent with statistical sampling noise around the peak of the posterior.

We measured the speed to convergence of these various samplers across the 25 realisations and found that \textsc{nautilus} performed significantly faster than other methods. It also has the benefit of returning the Bayesian evidence as well as posterior samples. For these reasons, we adopted \textsc{nautilus} as our default method, including in this paper.

\subsection{Summary of methodology systematics and differences between our new method and the (e)BOSS methodology}
Overall, our tests in this Section have shown that BAO measurements are remarkably robust to model choices not specified by theory. We find that, if our suggested best-practices (summarised further below) are followed, we expect a systematic error of no more than $0.05\%$ in $\alpha_{\mathrm{iso}}$ and $0.12\%$ in $\alpha_{\mathrm{ap}}$ to arise from any modelling choices. For clarity compared to past studies and as a guide for the future, we now reiterate our best-practices, highlighting differences with previous literature:

\label{sec:bossdiff}
\begin{enumerate}
    \item{For BAO reconstruction, previous works \citep{Xu12,Anderson12,Anderson14,Alam17,Alam21} largely followed \textbf{RecIso} convention. In contrast, we have demonstrated a clear theoretical preference for the \textbf{RecSym} reconstruction convention \edit{(see Section~\ref{sec:reconstandard} for a description of these two conventions, and Eqs.~\ref{eqn:recsym} and~\ref{eqn:propagator} for the corresponding \textbf{RecSym} propgators)}. Our companion papers \citet{OptimalRecon1} and \citet{OptimalRecon2} further detail how this can be applied optimally in the context of DESI.}
    \item{We use the spline-based broadband model validated in \S\ref{sec:spline}. Previous analyses used polynomials (of slightly varying degrees and functional forms) to fit the broadband. We find the spline method has better theoretical motivation, and to be less arbitrary in its choice of free parameters. \edit{For our default fitting of the monopole and quadrupole, between the ranges of $k=0.02-0.30$\hompc~and $s=50-150$\mpcoh~and with a spline-width $\Delta=0.06$\hompc, this gives 14 free broadband parameters for the power spectrum and 6 for the correlation function. We again clarify that the broadband between the two has the same theoretical basis --- the difference in number of parameters stems only from the behaviour of the Hankel transform and the different fitting ranges used for the two sets of clustering measurements.}}
    \item{We recommend choosing the BAO damping parameters based on theoretical calculations and allowing these to vary within reasonably tight priors ($1-2$\mpcoh). \edit{For DESI, we generally use Gaussian priors of width $1$\mpcoh~ for $\Sigma_{\perp}$ and $2$\mpcoh~ for $\Sigma_{||}$ and $\Sigma_{s}$, corresponding to approximately $20\%$ variations from their expected values. Central values for these priors are chosen based on fits to simulated data and theoretical considerations. More info on the priors adopted for our fiducial data analysis can be found in \cite{KP4}.} We found that completely fixing the BAO damping has the potential to lead to biased $\alpha$'s if the wrong values are chosen, while allowing them to vary freely can also cause bias, and generally weak constraints, in the case of noisy data. Accounting for shot-noise and non-linearities is able to give robust analytic predictions. Previous BOSS/eBOSS analyses fixed these based on fits to simulations (although found consistent results compared to when these were allowed to freely vary).}
    \item{We apply the Finger-of-God damping to \textit{only} the smooth model components, allowing for no interaction with the BAO wiggles. This makes interpreting the impact of the BAO damping parameters simpler. Previous models applied the FoG term equally to both the wiggle and no-wiggle components.}
    \item{The DESI BAO template is constructed following the method of \cite{Wallisch2018} Previous BOSS and eBOSS analyses have used varying methods for extracting the template (i.e., \citealt{EH1998,Kirkby2013}). Our tests do not point to a single correct choice, but we find a maximum systematic error associated with this choice of only $0.02\%$ and $0.09\%$ for $\alpha_{\mathrm{iso}}$ and $\alpha_{\mathrm{ap}}$ respectively when following our fiducial fitting methodology.}
    \item{We dilate only the BAO wiggles by the $\alpha$'s. Previous analyses generally dilated both the smooth and wiggle components of the model. We believe the new method to be preferable as it guarantees only information from the BAO is captured. Given the large amount of freedom in the broadband used in previous analyses, we do not expect any information was included from the full-shape of the clustering with the previous method, but our model choice ensures this remains true even if the broadband model is less flexible.}
    \item{We give both the galaxy bias $b$ and RSD term $\beta$ flat priors, finding that difficulties determining the appropriate central value of other prior choices can lead to poor fits on (a small number) of individual realisations. Previous analyses typically used more informative (but reasonably wide) Gaussian priors on one or both of these.}
    \item{We provide a fully consistent/equivalent model for the correlation function and power spectrum, treating the former purely as the Hankel transform of the latter, and applying all our modelling choices and BAO dilation to the power spectrum. BOSS analyses instead transformed the undilated power spectra, without galaxy bias, and applied both the BAO dilation and galaxy bias terms in configuration space. They also used two different galaxy bias terms for the correlation function multipoles rather than the RSD parameter $\beta$. We find the new, consistent, modelling method to be preferable, and find $0.12\%$ of systematic difference in $\alpha_{\mathrm{ap}}$ between the two methods.}
\end{enumerate}

\section{Conclusions}
\label{sec:conclusions}

Baryon acoustic oscillations (BAO) are one of the premier probes of the cosmic expansion history, especially at high redshifts, providing percent-level and below measurements of cosmological distances up to $z=2$. Since BAO are an oscillatory signal in Fourier space, with a distinct and fast oscillation frequency set by the sound horizon $r_d$ at the baryon drag epoch, its imprint on large-scale structure cannot easily be reproduced by either survey systematics or nonlinearities, making the measured BAO scale remarkably robust. Indeed, in the galaxy power spectrum, the latter typically induces shifts of no more than a percent, shifts which are further reduced by reconstruction. On the other hand, modern spectroscopic surveys, both planned and currently online, promise to probe galaxy clustering with larger volumes and greater number densities than ever before, leading to BAO measurements with statistical power verging upon the bias naively induced by these systematic effects. Indeed, the Dark Energy Spectroscopic Instrument (DESI), which will soon be releasing its first data release (DR1) is projected to yield a cumulative $0.2\%$ measurement of the BAO scale when its five-year run is completed.

Given the above, our goal in this paper has been to perform a thorough accounting of the theory and modeling systematics incurred in measuring the \textit{nonlinear} BAO in galaxy clustering. Our approach has been two-fold: first, we reviewed the theoretical literature modeling the BAO signal in galaxies, filling in holes and updating results where necessary, in order to both estimate the shift in the BAO scale due to nonlinearities and suggest an appropriate fitting form for the observed signal. Secondly, we test these methods on a set of high-precision mocks whose sample-variance has been reduced using Zeldovich control variates, paying particular attention to systematic effects of numerical choices not uniquely specified by the theory.

In the first part of the paper we (re)examine the perturbation theory (PT) of the BAO, particularly as pertains to the nonlinear shift of the nonlinear BAO peak and its damping by long-wavelength displacements. Two of the largest contributions to the former are (a) a nonlinear ``contraction'' of the BAO due to density modes larger than $r_d$, whose effect in redshift space we derive and (b) imprints of relative baryon-dark matter perturbations on galaxy clustering. The former effect is substantially larger in redshift space, where spectroscopic surveys naturally operate, than previous calculations in real-space suggest, but is significantly diminished after reconstruction, up to an error induced by a mismatch of the fiducial and true matter clustering amplitude $\Delta \sigma_8^2 / \sigma_8^2$. The impact of these two nonlinear effects can computed as a function of bias and redshift and added to the theoretical error budget of BAO measurements. Post reconstruction, assuming that the matter clustering on large scales $\propto \sigma_8$ is known to $5\%$, effect (a) leads to shifts in the BAO scale of no more than $0.1\%$ in the regimes relevant for DESI, while effect (b) leads to biases of less than $0.05\%$.

While it has long been known that reconstruction reduces the nonlinear damping of the BAO in galaxy clustering due to relatively long-wavelength modes, there has been significant confusion in the literature about the appropriate way to model their effect analytically, with many works suggesting a complicated scenario where the displaced galaxies and randoms have different damping scales (e.g. \citealt{Ding18,Chen19b}). Here we show that in the \textbf{RecSym} scheme, where both galaxies and randoms are moved by the same redshift-space displacement, this is in fact not the case; the long-modes in fact produce a simple and unified exponential damping across both the shifted randoms and galaxies. On the other hand, we show that the \textbf{RecIso} scheme adopted by previous galaxy surveys like BOSS and eBOSS results in a complicated damping form where the effect of modes longer than the BAO are not controlled, leading to non-negligible shifts in the measured BAO in the process. For this reason, the \textbf{RecSym} scheme is the preferred one, adopted for DESI Y1 and which we recommend for future BAO analyses, \edit{though we note that in practice the differences are relatively small, especially at the statistical level of DESI year one and existing surveys, and we refer the reader to the accompanying refs.~\cite{OptimalRecon1,OptimalRecon2} for further validation on simulations.} In addition, in the process of preparing this paper, we learned that \citet{Sugiyama24} had reached a similar conclusion on the cancellation of long-wavelength modes in reconstruction schemes where galaxies and randoms are moved by the same reconstructed displacement. Where our paper focused on deriving the IR-resummation of the reconstructed BAO and the cancellation of the mode-coupling induced phase shift, \citet{Sugiyama24} instead performed a more detailed and general look at the cancellation of long-wavelength displacements in the power spectrum as a whole. We refer readers more interested in the latter aspect to that work, which we have coordinated to appear at a similar time  with this one and which should be rather complementary.

In the second part of the paper, we investigate the impact of a number of other modelling choices, using numerical fits to a set of highly precise simulations. We arrive at a simple set of best-practices for how to model the BAO, enumerated in \S\ref{sec:bossdiff}. We demonstrate that if these are followed we expect a conservative systematic \textit{modelling} error budget of $0.05\%$ and $0.12\%$ on $\alpha_{\mathrm{iso}}$ and $\alpha_{\mathrm{ap}}$ respectively. Any larger deviations from the expected $\alpha$ values can be attributed to measurement noise. As will be demonstrated in upcoming publications, analysis of the DESI Y1 data follows these best-practices.

Table~\ref{tab:summary} summarises our systematic error contributions from both theory and modelling. Combining our two prongs of investigation we find that the combined systematics in BAO fitting can be controlled, conservatively, to within $0.1\%$ for $\alpha_{\mathrm{iso}}$ and $0.2\%$ for $\alpha_{\mathrm{ap}}$. We arrive at these numbers by summing in quadrature all the \textit{detected} systematics in Table~\ref{tab:summary} and rounding up to account for the unresolved systematics for which we have only upper limits. These are the values that will be adopted for modelling and theory systematics in the DESI Y1 analysis and are substantially less than the aggregate $0.0 < z < 1.1$ precision on the BAO expected from this data. They are also more than a factor of 2 less than the expected aggregate BAO precision for the full DESI sample, indicating that even with this exceptional dataset, we expect to remain in a regime where the uncertainties are dominated by statistics rather than systematics.

\begin{table}
\centering
\caption{Different contributions to the DESI BAO theory and modelling systematic error budget for $\alpha_{\mathrm{iso}}$ and $\alpha_{\mathrm{ap}}$ considered in this work. The section column denotes where the discussion of this contribution primarily occurs. In the theory cases, where the exact contributions depend on the nature of the tracers (galaxy bias, redshift, etc,), we have estimated a conservative value considering all DESI tracers/redshift bins.}
\begin{tabular}{cccc}
\hline\hline
Name/Description & Section & $\sigma_{\alpha_{\mathrm{iso}}}$ & $\sigma_{\alpha_{\mathrm{ap}}}$ \\\hline
Non-linear mode-coupling & \S\ref{sec:nlpre};  \S\ref{sec:nlrecon} & $<0.1\%$ &
$<0.1\%$ \\
Relative velocity effects & \S\ref{sec:nlpre};  \S\ref{sec:nlrecon} & $<0.05\%$ &
$<0.05\%$ \\
Broadband modelling & \S\ref{sec:damping} & $<0.02\%$ &
$0.11\%$ \\
BAO wiggle extraction & \S\ref{ssec:pnw} & $<0.02\%$ &
$<0.09\%$ \\
Dilating smooth vs. wiggle & \S\ref{sec:dilate} & $<0.02\%$ &
$<0.09\%$ \\
Modelling $\xi(s)$ from $P(k)$ & \S\ref{sec:pk2xi} & $<0.01\%$ &
$0.12\%$ \\
\hline
\textbf{Combined} & \S\ref{sec:conclusions} & $ 0.1\%$ &
$ 0.2\%$ \\
\hline\hline
\label{tab:summary}
\end{tabular}
\end{table}

Looking to the future, it is clear that in order to further validate modeling systematics for the BAO beyond that which has been shown here, a method to test our models beyond running large numbers of N-body simulations will have to be found. Firstly, the statistical errors even from fitting sample-variance reduced DESI mocks in this work prohibit us from clearly identifying (i.e., at $\ge 3\sigma$) systematic errors at levels below $\sim0.1\%$. Producing more simulations with the required precision and fidelity to overcome this will be rather costly. Secondly, we are in a realm of accuracy where the potential errors induced by galaxy clustering physics that cannot be easily introduced into N-body simulations are now important and, in addition, numerical simulations struggle due to approximations in the initial condition generation, finite computational resources and numerical approximations made during the evolution and post-processing of the simulations themselves \citep{Angulo22,Grove2022,Ding2022}. While sufficient for DESI, addressing this is of the utmost importance if we aim to determine the limiting systematic precision with post-DESI BAO measurements.

\section*{Acknowledgements}

We thank Naonori Sugiyama for useful discussions, particularly on the perturbative treatment of infrared effects in reconstruction. We thank Alejandro Aviles and Florian Beutler for graciously agreeing to serve on the internal DESI review committee and providing helpful feedback and additionally thank Will Percival and Sesh Nadathur for useful comments.

SC acknowledges the support of the National Science Foundation at the Institute for Advanced Study. CH acknowledges support from the Australian Government through the Australian Research Council’s Laureate Fellowship (project FL180100168) and Discovery Project (project DP20220101395) funding schemes. H-JS acknowledges support from the U.S. Department of Energy, Office of Science, Office of High Energy Physics under grant No. DE-SC0019091 and No. DE-SC0023241. H-JS also acknowledges support from Lawrence Berkeley National Laboratory and the Director, Office of Science, Office of High Energy Physics of the U.S. Department of Energy under Contract No. DE-AC02-05CH1123 during the sabbatical visit. The author list was produced using the publicly available \texttt{mkauthlist}.\footnote{\url{https://github.com/DarkEnergySurvey/mkauthlist}}

This material is based upon work supported by the U.S. Department of Energy (DOE), Office of Science, Office of High-Energy Physics, under Contract No. DE–AC02–05CH11231, and by the National Energy Research Scientific Computing Center, a DOE Office of Science User Facility under the same contract. Additional support for DESI was provided by the U.S. National Science Foundation (NSF), Division of Astronomical Sciences under Contract No. AST-0950945 to the NSF’s National Optical-Infrared Astronomy Research Laboratory; the Science and Technology Facilities Council of the United Kingdom; the Gordon and Betty Moore Foundation; the Heising-Simons Foundation; the French Alternative Energies and Atomic Energy Commission (CEA); the National Council of Science and Technology of Mexico (CONACYT); the Ministry of Science and Innovation of Spain (MICINN), and by the DESI Member Institutions: \url{https://www.desi.lbl.gov/collaborating-institutions}. Any opinions, findings, and conclusions or recommendations expressed in this material are those of the author(s) and do not necessarily reflect the views of the U. S. National Science Foundation, the U. S. Department of Energy, or any of the listed funding agencies.

The authors are honored to be permitted to conduct scientific research on Iolkam Du’ag (Kitt Peak), a mountain with particular significance to the Tohono O’odham Nation.

\section*{Data Availability}

Data from the plots in this paper are available on Zenodo (\url{https://zenodo.org/records/10685759}) as part of DESI's Data Management Plan.



\bibliographystyle{mnras}
\bibliography{main2} 




\appendix

\section{Redshift Evolution}
\label{app:zevol}

The measured power spectrum (or correlation function) is a weighted average over the entire redshift bin, given by
\begin{equation}
    \hat{P}(\bk) = \sum_i w_i P(\bk(z_i), z_i).
\end{equation}
The corresponding wavenumber $\bk$ varies with redshift since, if the fiducial and true cosmologies do not exactly match, the AP parameters translating between distances therein will evolve with time (see Eq.~\ref{eqn:kalpha}). Within each redshift bin, the measured power spectrum is primarily sensitive to clustering at the effective redshift $z_{\rm eff} = \sum_i w_i z_i$ at which linear trends in the time evolution cancel, i.e.
\begin{equation}
    \hat{P}(\bk) = P(\bk(z_{\rm eff})) + \frac{1}{2} \langle (z - z_{\rm eff})^2 \rangle \Big( \frac{d^2 P}{dz^2} \Big)_{z=z_{\rm eff}} + ... 
\end{equation}
where angular brackets denote the weighted mean so that the leading correction is proportional to the variance of the redshift distribution $\sigma_z^2$. This motivates the interpretation of the measured AP parameters from BAO  as distances at $z_{\rm eff}$ with a theoretical error set by the width of the redshift bin.

Let us try to gain some heuristic understanding of the kind of errors induced in the AP parameters by the finiteness of $\sigma_z$. All else being equal we can write $P = P(\apar, \aperp)$ such that for example
\begin{equation*}
    \frac{dP}{dz} = \frac{\partial P}{\partial \apar} \frac{d\apar}{dz} + \frac{\partial P}{\partial \aperp} \frac{d\aperp}{dz}.
\end{equation*}
The derivatives $\partial P / \partial \alpha_{\parallel,\perp}$ are precisely the templates for the AP parameters and, if the incorrect effective redshift were used and the linear redshift evolution not entirely cancelled, the measured distances would be off by $\Delta \alpha_{\parallel,\perp} = (d \alpha_{\parallel,\perp}/dz) \Delta z$. This would clearly be undesirable since $\Delta z \sim 0.1$ and, within even the baseline $\Lambda$CDM model we might expect $d \alpha_{\parallel,\perp}/dz \sim \Delta \Omega_m / \Omega_m \sim 0.1$, leading to percent-level systematic errors.

We can repeat the same analysis for the case where $z_{\rm eff}$ \textit{were} correctly determined, such that the leading error is given by the second derivative. Let us further assume that, while the first derivatives lead to phase shifts in the BAO wiggles, the second derivatives lead to only in-phase corrections that do not bias distance measurements.\footnote{Roughly speaking we can imagine $P_w \sim \cos(\alpha k r_d)$ such that the second derivative is again a cosine, now multiplied by $k^2$.} Within this approximation we have that
\begin{equation*}
    \frac{d^2P}{dz^2} \sim \frac{\partial P}{\partial \apar} \frac{d^2\apar}{dz^2} + \frac{\partial P}{\partial \aperp} \frac{d^2\aperp}{dz^2}
\end{equation*}
which we can again directly translate into an error on the AP parameters as
\begin{equation}
    \Delta \alpha_{\parallel, \perp} = \frac{1}{2} \Big( \frac{d^2\alpha_{\parallel, \perp}}{dz^2} \Big)_{z=z_{\rm eff}} \sigma_z^2
\end{equation}
For $\sigma_z \sim 0.1$ and for cosmologies wherein distances are within 10$\%$ of the fiducial one this translates to errors of less than $0.1\%$, but even in the case of larger redshift bins or models with more freedom, this shift can be exactly computed, since $\sigma_z$ depends only on the redshift distribution and the second derivative can be computed given an cosmology, and included when performing cosmological inference.

\section{Wide-Angle Effects}
\label{app:wide_angle}

In the main body of the text we have operated almost exclusively within the plane-parallel approximation where the line-of-sight (LOS) can be treated as a constant vector $\hat{n}$. This approximation is an excellent one for most of the scales on which the BAO is measured, with corrections scaling as negative powers $x =k \chi$ where $\chi$ is the comoving distance \citep{Reimberg16,Castorina18}. These corrections to the plane parallel limit are due to (a) the physical LOS along which galaxies are redshifted being not parallel but radial (b) the radial evolution of the galaxy selection function $d(\chi^2 \bar{n})/d\ln \chi$ and (c) anisotropies incurred by the numerical implementation of the power spectrum estimator. The leading effect due to (c) is a practical rather than physical one and can be absorbed into the window function \citep{Castorina18,Beutler19a} --- since we adopt this formalism in DESI we will focus on the first two effects \citep{KP3}, whose leading order contributions are proportional to $x^{-2}$.

Since these wide-angle effects are on large scales we can evaluate them assuming the underlying dynamics are linear, in which case the first effect (a) gives \citep{Castorina18,Beutler19a}
\begin{align}
    \xi_0(s,\chi) &\supset -\left( \frac{4 f^2}{45}  \xi_0^0 + \frac{f(9 b_1 + f)}{45} \xi^2_0 \right)  \left(\frac{s}{\chi} \right)^2 \nonumber \\
    \xi_2(s,\chi) &\supset \left( \frac{4 f^2}{45}  \xi^0_0 + \frac{f(189 b_1 + 53f)}{441} \xi^2_0 - \frac{4 f^2}{245}  \xi^4_0 \right) \left(\frac{s}{\chi} \right)^2 \nonumber
\end{align}
while the selection function effect gives in addition
\begin{align}
    \xi_0(s,\chi) & \supset \frac{4 f^2}{3 \chi^2} \xi^0_{-2} + \frac{2f(b_1-f)}{3\chi} \xi^1_{-1} \left(\frac{s}{\chi} \right)^2 \nonumber \\
    \xi_2(s,\chi) & \supset -\frac{8 f^2}{3 \chi^2} \xi^2_{-2} - \frac{8 f (5 b_1 + f)}{15 \chi} \xi^1_{-1} \left(\frac{s}{\chi} \right) + \frac{4 f^2}{5 \chi} \xi^3_{-1} \left(\frac{s}{\chi}  \right)
\end{align}
in the limit of constant number density $\bar{n}$.\footnote{We take this limit in order to provide a rough estimate the size of the effect on BAO measurements, but note that for a spectroscopic survey like DESI the redshift dependence of $\bar{n}$ can be measured to very high precision so the actual contributions could be calculated very precisely if desired.} In the above we have defined the generalized correlation functions
\begin{equation}
    \xi^\ell_n = \int \frac{dk\ k^2}{2\pi^2}\ k^n P(k) j_\ell(ks) \quad .
\end{equation}
The contributions to the power spectrum multipoles are related to the correlation function expressions above via the usual Hankel transforms. For a realistic survey, these effects have to be averaged over line-of-sight vectors $\Vec{\chi}$, though for the exploratory purposes of this section, we will assume a fixed $\chi = \chi(z)$. 

In order to estimate the bias these additional contributions cause in standard BAO fits we perform a simple Fisher forecast as we did for nonlinear effects in \S\ref{sec:nonlinear}. These effects are largest at low redshift, where $k \chi$ is smallest, and when the galaxy bias is low. For this reason, let us consider the case of a sample-variance limited survey with $b_1 = 1$ at $z = 0.1$, i.e.\ the lowest redshift boundary for the BGS sample. Even in this case, the shifts to both $\alpha_{\parallel, \perp}$ are uniformly less than $0.05\%$ using our fiducial setup with $k_{\rm min} = 0.02 \kMpc$. The biggest shifts are due to the selection function contribution since the odd Bessel functions are out of phase with the even ones. Nonetheless, while the wide-angle effects can reach sizes of up to several percent of the total signal at $k_{\rm min}$ their sizes are not amplified by the large BAO scale as is the nonlinear phase shift derived in \S\ref{sec:nonlinear}, and furthermore since they roughly decay with wavenumber as $k^{-2}$ their overall shape is very different from a shift in the BAO scale $\propto k P_w'(k)$.

Let us comment briefly on some other large-scale effects on the observed galaxy 2-point function not covered in the main text, such as unequal time and relativistic effects (see e.g.\ \citealt{Grimm20,Castorina22,Raccanelli23} for recent examples).  In general, these effects are suppressed on subhorizon scales like the wide-angle effects above and, moreover, do not have the exact shape of shifts in the BAO scale. For example, \citet{Raccanelli23} recently showed that unequal time effects can contribute a term proportional to $k P'(k)$; however, this term is suppressed on large scales and is proportional to an additional factor of $(H(z)/k)^2$ thus it contributes negligibly to the measured BAO scale. As we do not expect any of the large-scale effects on the galaxy power spectrum to significantly bias DESI BAO measurements, we leave their detailed analysis to future work.

\section{The Nonlinear BAO Shift in Redshift Space}
\label{app:rsd_bao_shift}

In order to compute the nonlinear BAO shift in redshift space in Eulerian perturbation theory we need to compute the galaxy density to second-order
\begin{equation}
    \delta_{g,s}^{(2)} = \int_\bp Z_2(\bp,\bk-\bp) \delta_{\rm lin}(\bp) \delta_{\rm lin}(\bk-\bp) \nonumber 
\end{equation}
where the second-order redshift space kernel is given in Fourier space as (see e.g. \citealt{Bernardeau02})
\begin{align}
    Z_2(\bp_1,\bp_2) &= \frac12 b_2 + b_s \left( \frac{(\bp_1 \cdot \bp_2)^2}{p_1^2 p_2^2} - \frac13 \right) \nonumber \\
    &+ b_1 F_2(\bp_1, \bp_2) + f \mu^2 G_2(\bp_1,\bp_2) \nonumber \\
    &+ \frac{f k \mu}{2} \left( \frac{\hn \cdot \bp_1}{p_1^2} (b_1 + f\mu_2^2) + \frac{\hn \cdot \bp_2}{p_2^2} (b_1 + f\mu_1^2) \right)
\end{align}
and we have defined $\mu_{n} = \hn \cdot \hat{p}_n$ and $\hk = \bp_1 + \bp_2$, with $\mu = \hn \cdot \hk$. We have also used the standard definitions of the matter density and velocity divergence kernels for $F_2$ and $G_2$ and will return to the cubic kernels $Z_3$ when discussing BAO damping below.

The largest out-of-phase contributions to the BAO in the galaxy power spectrum are sourced by the square of the quadratic contribution above \citep{Crocce08,Padmanabhan09b,Sherwin12}
\begin{equation*}
    P^{(22)}_{g,s}(\bk) = 2\int_\bp Z_2(\bp,\bk-\bp)^2 \Plin(\bp) \Plin(\bk-\bp).
\end{equation*}
Specifically, this term contains contributions from modes longer than the BAO proportional to a shift, i.e.\ logarithmic derivative, of the BAO template, e.g.
\begin{equation}
    \int_{|\bp|\lesssim 1/r_d} \left(\frac{\bk \cdot \bp}{p^2} \right) P_{\rm lin}(p) \left(-\frac{\bk \cdot \bp}{k} \right)\ P_w'(k)
\end{equation}
where the first factor comes from the displacement $\Psi \cdot \nabla \delta$ and the second one comes from expanding $P(|\bk-\bq|)$ for small $\bq$. The leading terms will be those that scale as $p^0$ in the integrand like the one above. The bound on $\bp$ comes the fact that $P(|\bk-\bq|)$ begins to oscillate for larger $\bq$. Note that there are in fact two long-wavelength regions in the 22 integral since both $\bp$ and $\bk-\bp$ can be small. Additionally, the inclusion of redshift-space distortions can introduce powers of the line-of-sight angle $\mu$ into the above integral since displacements are amplified along $\hn$.

In order to compute all such terms to one-loop order in the redshift-space galaxy power spectrum we use \textsc{Mathematica}, setting up our calculation in the spherical coordinate system
\begin{align}
    \hat{k} &= \left\{0, 0, 1 \right\}, \quad \hat{n} = \left\{\sqrt{1-\mu^2}, 0, \mu \right\} \nonumber \\
    \hat{p} &= \left\{\sqrt{1-\nu^2} \cos\phi, \sqrt{1-\nu^2}\sin\phi, \nu \right\}.
    \label{eqn:mathematica_coords}
\end{align}
such that the $\phi$ and $\nu$ dependence can be performed analytically in the small $\bp$ limit. Specifically, we Taylor-series expand $Z_2^2$ and $\Plin(\bk-\bp)$ in $p = |\bp|$, perform the angular integrals in $\nu, \phi$, and compute the derivative of this integral with respect to $\Plin'(k)$, leading to Equation~\ref{eqn:bao_shift}. 

\section{Modeling Reconstruction in Eulerian Perturbation Theory}
\label{app:ept_recon}

In Appendix \ref{app:rsd_bao_shift} we outlined the calculation of the nonlinear BAO shift in the galaxy power spectrum. This shift is, however, known to be significantly reduced by reconstruction, which we will show analytically using perturbation theory in this appendix.

\subsection{Post-Reconstruction BAO Shift in Real Space}
\label{app:recon_shift}

We begin by considering the case of matter in real space. The smoothed matter density yields displacements $\Psi^{(n)}_{r}(\bx) = \nabla^{-1} (\mathcal{S} \ast \delta^{(n)}_m(\bx))$ which can be used to compute post-reconstruction densities as
\begin{align}
    (2\pi)^3 \delta_D(\bk) + \delta_a(\bk) &= \int d^3\bx e^{-i\bk\cdot\bx - i\bk\cdot\Psi_r(\bx) } \left( 1 + \delta_a(\bx) \right) \nonumber \\
    &= \sum_{n=0}^\infty \frac{(-i)^n}{n!} \text{FT}\left[ (\bk\cdot\Psi_r)^n (1 + \delta_a) \right](\bk).
    \label{eqn:postrecon_kernels}
\end{align}
Here $\delta_a(\bx)$ is the pre-reconstruction density, e.g. $\delta_d(\bx) = \delta_m(\bx)$ and $\delta_s(\bx) = 0$. We then have at second order \citep{Hikage17}
\begin{align}
    K^{(2)}_d(\bp_1,\bp_2) = (1 - \mathcal{S}_\bk) F_2(\bp_1,\bp_2) &- \frac12 \left( \frac{\bk\cdot\bp_1}{p_1^2} \mathcal{S}_{\bp_1} + \frac{\bk\cdot\bp_2}{p_2^2} \mathcal{S}_{\bp_2} \right) \nonumber \\
    &+ \frac12 \left( \frac{\bk\cdot\bp_1}{p_1^2} \right) \left( \frac{\bk\cdot\bp_2}{p_2^2} \right) \mathcal{S}_{\bp_1} \mathcal{S}_{\bp_2} \nonumber
\end{align}
\begin{align}
    K^{(2)}_s(\bp_1,\bp_2) = - \mathcal{S}_\bk F_2(\bp_1,\bp_2) + \frac12 \left( \frac{\bk\cdot\bp_1}{p_1^2} \right) \left( \frac{\bk\cdot\bp_2}{p_2^2} \right) \mathcal{S}_{\bp_1} \mathcal{S}_{\bp_2}
\end{align}
where we have used the shorthand $\mathcal{S}_\bp = \mathcal{S}(\bp)$.

We can now proceed to compute the nonlinear BAO shift as in the pre-reconstruction case with these kernels by considering the products $K^{(2)}_a(\bp, \bk-\bp) K^{(2)}_b(\bp, \bk-\bp)$. If the smoothing scale of $\mathcal{S}$ is significantly smaller than the BAO scale $r_d$ then for the long-wavelength modes $\bp$ contributing to the shift we can use $\mathcal{S}_\bp \approx 1$ and $\mathcal{S}_{\bk-\bp} \approx \mathcal{S}_\bk$. For example, performing this procedure for the autospectrum of the shifted randoms yields
\begin{align}
    P_{ss} &\supset k P_w'(k) \int_\bp \left( -\frac13 \mathcal{S}_\bp^2 \mathcal{S}_{\bk-\bp}^2 + \frac{82}{105} \mathcal{S}_\bp \mathcal{S}_{\bk-\bp} \mathcal{S}_{\bk} - \frac{47}{105} \mathcal{S}_{\bk}^2 \right) \Plin(\bp) \nonumber \\
    &\approx k P_w'(k) \int_\bp \left( -\frac13 \mathcal{S}_{\bk}^2 + \frac{82}{105} \mathcal{S}_{\bk}^2 - \frac{47}{105} \mathcal{S}_{\bk}^2 \right) \Plin(\bp) = 0,
\end{align}
i.e.\ a vanishing contribution to the BAO phase shift. Computing this contribution for the $dd$ and $ds$ spectra similarly yields that they vanish. 

We can also repeat the above calculations for galaxies rather than matter. In this case we have instead $\Psi^{(n)}_{r}(\bx) = \nabla^{-1} (\mathcal{S} \ast \delta^{(n)}_g(\bx)/b_1)$ and $\delta_d(\bx) = \delta_g(\bx)$, which we can nonetheless plug back into Equation~\ref{eqn:postrecon_kernels} to derive $K^{(2)}_{d,s}$ for post-reconstruction galaxies. In this case, an equivalent calculation to the above yields again that the nonlinear shift is exactly cancelled by the contributions introduced by reconstruction. This is true regardless of the details of the nonlinear biasing since the nonlinear shift is a product of the bias and the quadratic shift term $\Psi \cdot \nabla \delta$; the latter is always removed by reconstruction given that the correct $b_1$ is used. Indeed, the above calculation can be redone when
\begin{equation}
    \Psi_r = \nabla^{-1}\left( \frac{ \mathcal{S} \ast \delta_g(\bx)}{b_1^{\rm fid}}  \right)
\end{equation}
allowing for $b_1^{\rm fid} \neq b_1$. Doing so yields a shift
\begin{equation}
    \Delta \alpha \propto \left(\frac{b_1^{\rm fid}}{b_1} - 1 \right) \sigma^2_{s}
\end{equation}
where the proportionality constant is of the same order as the usual nonlinear shift pre-reconstruction, i.e.\ the residual shift in the BAO post-reconstruction is proportional to how well the ``true'' matter overdensity can be estimated to produce the displacements. It is worth noting that the post-reconstruction shift left over due to incorrect assumptions will generally be in addition scale dependent and may not be entirely degenerate with BAO measurements, so that our estimates here should be taken to be conservative.

\subsection{Post-Reconstruction BAO Damping in Real Space}
\label{app:recon_damping}

We can also extend the perturbation theory analysis above to look at the damping of the BAO after reconstruction. More specifically, we can use 1-loop perturbation theory to examine what long-wavelength contributions get enhanced by the BAO such that they need to be resummed.

Before we do so it is useful to recall what happens pre-reconstruction. In this case we have for example in the matter power spectrum that \citep{Bernardeau02}
\begin{align}
    P_{mm}^{\rm loop}(\bk) = \int_\bp 2 &F_2(\bp,\bk-\bp)^2 \Plin(p) \Plin(|\bk-\bp|) \nonumber \\
    &\quad \quad + 6 F_3(\bk,\bp,-\bp) \Plin(p) \Plin(k). \nonumber
\end{align}
Both $F_2^2$ and $F_3$ have poles proportional to $1/p_n^2$ in the long-wavelength limit where one of their arguments $p_n \rightarrow 0$, such that their individual contributions will be quite sensitive to these IR contributions. However, if $\Plin$ is a smooth function we can approximate $\Plin(|\bk-\bp|) \approx \Plin(k)$ to leading order, and with some algebra it is possible to show that the poles at $\bp, \bk-\bp \rightarrow 0$ in the $F_2^2$ piece cancel the $\bp \rightarrow 0$ pole in the $F_3$ piece. On the other hand, if the power spectrum contains a non-smooth component such as the BAO ($\sim \sin(r_d k)$) then the derivative of $\Plin$ with respect to $k$ is enhanced by the oscillation frequency $r_d$. In this case, the infrared contribution can be written as \citep{Baldauf15,Senatore18}
\begin{align}
    P_{mm}^{\rm loop} &\supset -\int_{\bp}^{p \ll k} \frac{(\bk \cdot \bp)^2}{p^4} \Plin(p) \left( P_w(k) - P_w(|\bk-\bp|)  \right) \nonumber \\
    &= - \int d^3\bx\ e^{-i\bk\cdot\bx} \ \xi_w(\bx) \ \int_{\bp, p \ll k} \frac{(\bk \cdot \bp)^2)}{p^4} \Plin(p) \left( 1 - e^{i\bp \cdot \bx} \right) \nonumber \\
    &\approx - \frac12 k_i k_j \int d^3\bx\ e^{-i\bk\cdot\bx} \ \xi_w(\bx) \left[ A_{ij}(\bx) \right]_{x = r_d} \nonumber \\
    &\approx -\frac12 k^2 \Sigma^2_{\rm NL} P_w(k).
\end{align}
where $\xi_w$ is the Fourier transform of the $P_w$ and we have used that $\xi_w$ is localized at $r_d$ to perform a saddle point approximation between the second and third lines. This is none other than the 1-loop contribution to the BAO damping $e^{-\frac12 k^2 \Sigma^2_{\rm NL}} P_w$ and uses the same saddle-point approximation in LPT described in \S\ref{sec:nonlinear}, with the result that the BAO is damped by (numerically) large contributions from long-wavelength modes.

We can perform the same calculation post-reconstruction. For brevity, we will only consider the case of the shifted randoms in matter reconstruction in order to highlight why our result in \S\ref{sec:recon} differs from the existing literature---the calculation for shifted galaxies etc. is entirely analogous. In order to do so we need to compute the third-order contribution in Equation~\ref{eqn:postrecon_kernels}
\begin{align*}
    K^{(3)}_s(\bp_1, \bp_2, \bp_3) &= - \mathcal{S}_\bk F_3(\bp_1,\bp_2,\bp_3) \nonumber \\
    &+ \frac13 \left( \frac{(\bk\cdot\bp_1)(\bk\cdot\bp_{23})}{p_1^2 p_{23}^2} \mathcal{S}_{1} \mathcal{S}_{23} + {\rm et.\ cycl.}  \right) \nonumber \\
    &- \frac16 \left(\frac{(\bk\cdot\bp_1)(\bk\cdot\bp_2)(\bk\cdot\bp_3)}{p_1^2 p_2^2 p_3^2}\right) \mathcal{S}_1 \mathcal{S}_2 \mathcal{S}_3
\end{align*}
where we have used numerical indices to indicate arguments by $\bp_n$ and used the shorthand $\bp_{nm} = \bp_n + \bp_m$. Plugging this into the 1-loop power spectrum then yields
\begin{align}
    &P^{\rm loop}_{ss} = \int_\bp 2 K^{(2)}_s(\bp,\bk-\bp)^2 \Plin(p) \Plin(|\bk-\bp|) \nonumber \\
    &\quad \quad \quad \quad \quad \quad + 6 K^{(3)}_s(\bk,\bp,-\bp) \Plin(p) \Plin(k) \nonumber \\
    &\supset - \mathcal{S}_\bk^2 \int_{\bp}^{p \ll k} \frac{(\bk \cdot \bp)^2}{p^4} (1 - \mathcal{S}_\bp)^2 \Plin(p) \left( P_w(k) - P_w(|\bk-\bp|)  \right).
\end{align}
This is identical to the integral except that (1) there is a factor of $\mathcal{S}_\bk^2$ in front, since $P_{ss}$ is given by $\mathcal{S}^2 \Plin$ at linear order and more (2) the integrand is suppressed by $(1 - \mathcal{S}_\bp)^2$ internally. This is because, as described in \S\ref{sec:recon}, the shifted \textit{field} is in fact displaced by bulk motions given to leading order by $(1 - \mathcal{S}) \ast \Psi$ even though the randoms themselves are only moved by $-\mathcal{S} \ast \Psi$. This is also reflected by the fact that the nonlinear BAO shift derived above cancels independently in $P_{ss}$ by virtue of the $\Psi \cdot \nabla \delta$ contribution cancelling there as well. Similar calculations show that the same result occurs in both $P_{dd,ds}$ as well, such that all of the pieces of the reconstructed power spectrum have the same BAO damping. The case for galaxies is also entirely similar.

\subsection{Nonlinear BAO in Redshift Space}
\label{app:recon_rsd}

Finally, let us extend our treatment of the nonlinear BAO post-reconstruction at 1-loop in EPT to include redshift-space distortions. The order $n$ contribution to the reconstructed displacement is given by
\begin{equation}
    \Psi_{r,s}^{(n)}(\bk) = - \left( \frac{i \bk}{k^2} \right) \left( \frac{\delta^{(n)}_{g,s}(\bk)}{b_1 + f\mu^2} \right).
\end{equation}
In \textbf{RecSym} both the galaxies and randoms are shifted by this displacement \textit{multiplied by} the redshift-space transformation matrix $R_{ij} = \delta_{ij} + f \hat{n}_i \hat{n}_j$. Let us again consider the case of matter such that $b_1 = 1$ for simplicity. In this case, the quadratic kernel for the shifted randoms is now
\begin{align}
    &Z^{(2)}_s(\bp_1,\bp_2) = (- i k_i) R_{ij} \left( - \left( \frac{i k_j}{k^2} \right) \left( \frac{Z^{(2)}(\bp_1,\bp_2)}{1 + f\mu^2} \right) \right) \mathcal{S}_\bk \nonumber \\
    &- \frac12 \left(k_i R_{ij} \left( \frac{-i p_{1,j}}{p_1^2} \frac{Z_1(\bp_1)}{1 + f\mu_1^2} \right) \right) \left(k_i R_{ij} \left( \frac{-i p_{2,j}}{p_2^2} \frac{Z_1(\bp_2)}{1 + f\mu_2^2} \right) \right) \mathcal{S}_1 \mathcal{S}_2
\end{align}
where as usual $\bk = \bp_1 + \bp_2$ and $\mu_n = \hn \cdot \hat{p}_n$. We have included the dependence on $Z_1$ explicitly but note that $Z_1(\bp_n)/(1 + f\mu_n^2) = 1$. Repeating the real-space calculation using the coordinate system of Equation~\ref{eqn:mathematica_coords}, and explicitly integrating over the angular dependence of the long mode $\bp$, it is straightforward, if laborious, to show that the nonlinear shift proportional to $\sigma_{s}^2 k P_w'(k)$ vanishes. In addition, we can also see that the arguments about the damping scale being set by $(1 - \mathcal{S}) \ast \Psi$ also hold in redshift space since the relevant (resummed) contribution comes from powers of the redshift-space displacement sourced by $\mathcal{S}_{\bp_n} \delta^{(1)}(\bp_n) / (b_1 + f\mu_n^2)$. These smooth displacements cancel the exponentiated Zeldovich displacements on large scales at all orders, such that their total contribution can be resummed as claimed in the main text. The case of displaced galaxies is entirely analogous. 

Finally, let us repeat the calculation for \textbf{RecIso}. In \textbf{RecIso} the particles are shifted by the displacement without multiplying by $R_{ij}$, and we can reproduce this in our calculations by taking $R_{ij} \rightarrow \delta_{ij}$ in the above. Performing the same calculation yields a BAO shift in \textbf{RecIso} of
\begin{equation*}
    \left( \frac{7 f^3 \left(4-9 \mu ^2\right) \mu ^4+f^2 \left(14-73 \mu ^2\right) \mu ^2-12 f \mu ^2}{105
   \left(1 + f \mu ^2\right)} \right) \sigma_{s}^2 W^2(k) k P_w'(k)
\end{equation*}
representing a nonzero shift unlike \textbf{RecSym}.

\section{Hankel Transform of Broadband Spline Model}
The two relevant Hankel-transformed spline coefficients for the correlation function were evaluated to  
\label{app:corrspline}
\begin{align}
    B_{2,0}(x) = \frac{2}{x^6}\Big( &x^3 \text{Si}(x)-2 x^3 \text{Si}(2 x) \nonumber \\
    &+x^2 \cos (x)-x^2 \cos (2 x) 
    -x \sin (x)-16 \cos (x) \nonumber \\
    &+4 \cos
   (2 x)+x \sin (x) \cos (x)+12\Big) \nonumber
\end{align}
\begin{align}
    B_{2,1}(x) = -\frac{1}{2x^3} \Big( & 6 x^3 \text{Si}(x)-32 x^3 \text{Si}(2 x)+27 x^3 \text{Si}(3 x) \nonumber \\
    &+8 x^2+6 x^2 \cos (x)-16 x^2 \cos (2 x)+9
   x^2 \cos (3 x) \nonumber \\
   &-6 x \sin (x)+8 x \sin (2 x)-3 x \sin (3 x) \nonumber \\
   &-96 \cos (x)+64 \cos (2 x) -16 \cos (3 x)+48 \Big)
\end{align}
where $\text{Si}(x)$ is the sine integral $\int_0^z dt \sin(t)/t$.

\section{Analytic Marginalisation of the Broadband} 
\label{app:marg}

Our full model for the BAO contains a large number of nuisance parameters designed to recover unbiased constraints on the two parameters we actually care about ($\alpha_{\mathrm{iso}}$ and $\alpha_{\mathrm{ap}}$) in the presence of other physics that affects the shape of the galaxy clustering, survey selection effects and measurement noise.

In our base model for the galaxy clustering however, most of these nuisance parameters enter at linear order, and so can be marginalised over analytically at the level of the likelihood without needing to be sampled, expensively and numerically, as part of an MCMC or Nested sampling algorithm \citep{Bridle2002,Taylor10,d'Amico2020}. This is useful as we are typically not interested in the constraints on these other parameters. 

To demonstrate how this works, we rewrite the multipole version of our clustering model from Eq.~\ref{eqn:BAO_model} into two parts, where the first combines the wiggle and no-wiggle model components, and the latter is the broadband, making clear its dependence on the linear nuisance parameters $a_{\ell,n}$
\begin{equation}
    P_{gg,\ell}(k) =  \tilde{P}_{gg,\ell}(k)+ \sum_{n} a_{\ell,n}\frac{\partial\mathcal{D}_{\ell}(k)}{\partial a_{\ell,n}}.
    \label{eqn:BAO_model2}
\end{equation}
When performing a BAO fit, this model is then convolved with the survey window function (matrix $\boldsymbol{\mathsf{W}}_{\ell\ell'}$) to give a new vector $\boldsymbol{P}^{W}_{\ell}=\boldsymbol{\mathsf{W}}_{\ell\ell'} \boldsymbol{P}_{gg,\ell'} = \boldsymbol{\tilde{P}}^{W}_{\ell} + \sum_{n}a_{\ell,n}\mathcal{\boldsymbol{D}}^{W}_{\ell,n}$ where we have still kept the windowed model in two parts, those that do and do not depend on the linear order nuisance parameters respectively. The matrix $D^W_{\ell,n}$ then gives a set of purely linear templates in the power spectrum model. These templates can in general depend on the nonlinear degrees of freedom (e.g. bias and dilation parameters) though they do not in our particular case.

Concatenating the different multipoles, we then take the common approach of using a Gaussian likelihood to describe our data ($\hat{\boldsymbol{P}}$) given the model with parameters $\boldsymbol{\theta}$,
\begin{equation}
\mathcal{L}(\hat{\boldsymbol{P}} | \boldsymbol{\theta}) \propto \mathrm{Exp}\biggl[-\frac{1}{2}[\hat{\boldsymbol{P}}-\boldsymbol{P}^{W}]
\boldsymbol{\mathsf{C}}^{-1}[\hat{\boldsymbol{P}}-\boldsymbol{P}^{W}]^{T}\biggl]. \label{eq:like}
\end{equation}
The act of marginalising over the nuisance parameters, before or after sampling them, is akin to performing the integral 
\begin{equation}
\mathcal{L}(\hat{\boldsymbol{P}} | \boldsymbol{\Omega}) \propto \int d\vec{\boldsymbol{a}}\,\mathcal{L}(\hat{\boldsymbol{P}} | \boldsymbol{\theta}) P(\vec{\boldsymbol{a}})
\end{equation}
where $P(\vec{\boldsymbol{a}})$ are the prior probabilities for the various $a_{\ell,n}$ and $\boldsymbol{\Omega}$ denotes all the other free parameters of the model. When using the Gaussian likelihood in Eq.~\ref{eq:like} and a flat (unbounded) prior for $P(\vec{\boldsymbol{a}})$, this integral has an exact analytic solution,
\begin{equation}
\mathcal{L}(\hat{\boldsymbol{P}} | \boldsymbol{\Omega}) \propto \frac{1}{|\boldsymbol{\mathsf{F}}|^{1/2}}\mathrm{Exp}\biggl[-\frac{1}{2}[\hat{\boldsymbol{P}}-\boldsymbol{\tilde{P}}^{W}]
\boldsymbol{\mathsf{\tilde{C}}}^{-1}[\hat{\boldsymbol{P}}-\boldsymbol{\tilde{P}}^{W}]^{T}\biggl], \label{eq:like2}
\end{equation}
where 
\begin{equation}
\boldsymbol{\mathsf{F}} = \boldsymbol{\mathsf{D}}^{W}\boldsymbol{\mathsf{C}}^{-1}\boldsymbol{\mathsf{D}}^{W,T}
\end{equation}
is a square matrix with length equal to the number of nuisance parameters, and
\begin{equation}
\boldsymbol{\mathsf{\tilde{C}}}^{-1} = \boldsymbol{\mathsf{C}}^{-1} - \boldsymbol{\mathsf{C}}^{-1}\boldsymbol{\mathsf{D}}^{W,T}\boldsymbol{\mathsf{F}}^{-1}
\boldsymbol{\mathsf{D}}^{W}
\boldsymbol{\mathsf{C}}^{-1}.
\end{equation}
This has the interpretation that the full covariance is given by the data covariance plus a theory covariance, such that the total precision matrix is reduced. Similar analytic solutions can be obtained for the cases where $P(\vec{\boldsymbol{a}})$ takes a Gaussian, in which cases one simply replaces
\begin{align}
    \boldsymbol{\mathsf{F}} & \rightarrow \boldsymbol{\mathsf{F}} + \boldsymbol{\mathsf{\Sigma}}
\end{align}
where $\boldsymbol{\mathsf{\Sigma}}$ is the covariance matrix for the Gaussian prior on $\vec{\boldsymbol{a}}$.

Given the above expressions, one can also identify the best-fit values of the linear order nuisance parameters given the remainder of the model $\boldsymbol{\tilde{P}}^{W}$,
\begin{equation}
\boldsymbol{a}_{\mathrm{best-fit}} =  \boldsymbol{\mathsf{F}}^{-1}
\boldsymbol{\mathsf{D}}^{W}\boldsymbol{\mathsf{\tilde{C}}}^{-1}[\hat{\boldsymbol{P}}-\boldsymbol{\tilde{P}}^{W}]^{T}.
\end{equation}
The improvement in $\chi^2$ obtained by using these best fit parameters is equal to the reduction in the precision matrix above.

If the linear templates in $\boldsymbol{\mathsf{D}}^W$ are independent of the nonlinear degrees of freedom, as is the case for the BAO broadband, an additional simplification occurs. In this case $\boldsymbol{\mathsf{F}}$ and  $\boldsymbol{\mathsf{\tilde{C}}}^{-1}$ can be precomputed and so evaluation of the likelihood takes no additional time compared to the unmarginalised case.  Combining with the above discussion this allows one to also take the ``partially'' marginalised approach --- evaluating the best-fit broadband parameters at each sampling iteration, multiplying by $\boldsymbol{\mathsf{D}}^{W}$ and adding them to the rest of the model to evaluate the standard $\chi^{2}$. This is the approach taken in some past literature \citep{Ross17,Bautista21} and gives the same marginalised posterior on the parameters of interest in this special case. However, as can be seen from the description above, it requires more numerical operations per likelihood evaluation and so we argue the direct marginalisation is preferable.

In either case, the benefits to sampling and run-time are substantial when including analytic marginalisation. For our default model, when fitting the LRG mocks the number of free parameters in the sampling is reduced from 21 to 7. As such, we adopt full analytic marginalisation over the broadband parameters as our default.

Analytic marginalisation can be carried out similarly for the correlation function, and further inspection shows that the galaxy bias (squared) could also be analytically marginalised over if desired. However, we do not include this latter option in our default methodology as tests on noisier data suggest that allowing the squared galaxy bias to go negative (as must be allowed in the case of a flat, unbounded prior) can give incorrect fits to the BAO in the correlation function, where the `dip' in the correlation function on scales smaller than the BAO is mistaken for (a flipped version of) the BAO itself. The method of analytic marginalisation does not lend itself to mixed priors, and so an alternative for further investigation would be to investigate Gaussian priors on the galaxy bias and broadband terms.

\section{Affiliations}
\label{app:affiliations}

$^{1}$ Institute for Advanced Study, 1 Einstein Drive, Princeton, NJ 08540, USA\\
$^{2}$ School of Mathematics and Physics, University of Queensland, 4072, Australia\\
$^{3}$ Department of Physics, University of California, Berkeley, 366 LeConte Hall MC 7300, Berkeley, CA 94720-7300, USA\\
$^{4}$ Lawrence Berkeley National Laboratory, 1 Cyclotron Road, Berkeley, CA 94720, USA\\
$^{5}$ Center for Cosmology and AstroParticle Physics, The Ohio State University, 191 West Woodruff Avenue, Columbus, OH 43210, USA\\
$^{6}$ Department of Physics \& Astronomy, Ohio University, Athens, OH 45701, USA\\
$^{7}$ Physics Department, Yale University, P.O. Box 208120, New Haven, CT 06511, USA\\
$^{8}$ Physics Dept., Boston University, 590 Commonwealth Avenue, Boston, MA 02215, USA\\
$^{9}$ Tata Institute of Fundamental Research, Homi Bhabha Road, Mumbai 400005, India\\
$^{10}$ University of Michigan, Ann Arbor, MI 48109, USA\\
$^{11}$ Leinweber Center for Theoretical Physics, University of Michigan, 450 Church Street, Ann Arbor, Michigan 48109-1040, USA\\
$^{12}$ NSF NOIRLab, 950 N. Cherry Ave., Tucson, AZ 85719, USA\\
$^{13}$ Department of Physics \& Astronomy, University College London, Gower Street, London, WC1E 6BT, UK\\
$^{14}$ Institute for Computational Cosmology, Department of Physics, Durham University, South Road, Durham DH1 3LE, UK\\
$^{15}$ Department of Physics and Astronomy, The University of Utah, 115 South 1400 East, Salt Lake City, UT 84112, USA\\
$^{16}$ Instituto de F\'{\i}sica, Universidad Nacional Aut\'{o}noma de M\'{e}xico,  Cd. de M\'{e}xico  C.P. 04510,  M\'{e}xico\\
$^{17}$ Department of Astronomy, School of Physics and Astronomy, Shanghai Jiao Tong University, Shanghai 200240, China\\
$^{18}$ University of California, Berkeley, 110 Sproul Hall \#5800 Berkeley, CA 94720, USA\\
$^{19}$ Institut de F\'{i}sica dÕAltes Energies (IFAE), The Barcelona Institute of Science and Technology, Campus UAB, 08193 Bellaterra Barcelona, Spain\\
$^{20}$ Ecole Polytechnique F\'{e}d\'{e}rale de Lausanne, CH-1015 Lausanne, Switzerland\\
$^{21}$ Departamento de F\'isica, Universidad de los Andes, Cra. 1 No. 18A-10, Edificio Ip, CP 111711, Bogot\'a, Colombia\\
$^{22}$ Observatorio Astron\'omico, Universidad de los Andes, Cra. 1 No. 18A-10, Edificio H, CP 111711 Bogot\'a, Colombia\\
$^{23}$ Center for Astrophysics $|$ Harvard \& Smithsonian, 60 Garden Street, Cambridge, MA 02138, USA\\
$^{24}$ Department of Physics, The University of Texas at Dallas, Richardson, TX 75080, USA\\
$^{25}$ Institut d'Estudis Espacials de Catalunya (IEEC), 08034 Barcelona, Spain\\
$^{26}$ Institute of Cosmology and Gravitation, University of Portsmouth, Dennis Sciama Building, Portsmouth, PO1 3FX, UK\\
$^{27}$ Institute of Space Sciences, ICE-CSIC, Campus UAB, Carrer de Can Magrans s/n, 08913 Bellaterra, Barcelona, Spain\\
$^{28}$ Department of Physics, The Ohio State University, 191 West Woodruff Avenue, Columbus, OH 43210, USA\\
$^{29}$ The Ohio State University, Columbus, 43210 OH, USA\\
$^{30}$ Departament de F\'{i}sica, Serra H\'{u}nter, Universitat Aut\`{o}noma de Barcelona, 08193 Bellaterra (Barcelona), Spain\\
$^{31}$ Laboratoire de Physique Subatomique et de Cosmologie, 53 Avenue des Martyrs, 38000 Grenoble, France\\
$^{32}$ Instituci\'{o} Catalana de Recerca i Estudis Avan\c{c}ats, Passeig de Llu\'{\i}s Companys, 23, 08010 Barcelona, Spain\\
$^{33}$ Department of Physics and Astronomy, University of Waterloo, 200 University Ave W, Waterloo, ON N2L 3G1, Canada\\
$^{34}$ Waterloo Centre for Astrophysics, University of Waterloo, 200 University Ave W, Waterloo, ON N2L 3G1, Canada\\
$^{35}$ IRFU, CEA, Universit\'{e} Paris-Saclay, F-91191 Gif-sur-Yvette, France\\
$^{36}$ Perimeter Institute for Theoretical Physics, 31 Caroline St. North, Waterloo, ON N2L 2Y5, Canada\\
$^{37}$ Max Planck Institute for Extraterrestrial Physics, Gie\ss enbachstra\ss e 1, 85748 Garching, Germany\\
$^{38}$ Instituto de Astrof\'{i}sica de Andaluc\'{i}a (CSIC), Glorieta de la Astronom\'{i}a, s/n, E-18008 Granada, Spain\\
$^{39}$ Department of Physics, Kansas State University, 116 Cardwell Hall, Manhattan, KS 66506, USA\\
$^{40}$ Department of Physics and Astronomy, Sejong University, Seoul, 143-747, Korea\\
$^{41}$ Centre for Astrophysics \& Supercomputing, Swinburne University of Technology, P.O. Box 218, Hawthorn, VIC 3122, Australia\\
$^{42}$ CIEMAT, Avenida Complutense 40, E-28040 Madrid, Spain\\
$^{43}$ SLAC National Accelerator Laboratory, Menlo Park, CA 94305, USA\\
$^{44}$ National Astronomical Observatories, Chinese Academy of Sciences, A20 Datun Rd., Chaoyang District, Beijing, 100012, P.R. China\\


\bsp	
\label{lastpage}
\end{document}